\documentclass[aps,
prd,
amsmath,amssymb,
superscriptaddress,
footinbib,
onecolumn,
floatfix,
11pt
]{revtex4-2}

\usepackage[english]{babel}
\usepackage[utf8]{inputenc}
\usepackage[T1]{fontenc}
\usepackage{graphicx}
\usepackage{subfigure}
\usepackage{mathrsfs}
\usepackage{bbm}
\usepackage{bm}
\usepackage{enumerate}
\usepackage[dvipsnames]{xcolor}
\usepackage{float}
\usepackage{subdepth}
\usepackage{booktabs}
\usepackage[math]{cellspace}
\usepackage{physics}
\usepackage{multirow}

\usepackage[colorlinks,
linkcolor=BrickRed,
citecolor=MidnightBlue,
urlcolor=MidnightBlue,
bookmarks=true,
bookmarksopen=true,
bookmarksnumbered=true,
]{hyperref}

\graphicspath{{figures/}}

\newcommand{\bmat}{\left ( \begin{array}{cc}}
\newcommand{\emat}{ \end{array}\right )}

\renewcommand\epsilon\varepsilon
\renewcommand\phi\varphi

\newcommand\be{\begin{eqnarray}}
\newcommand\ee{\end{eqnarray}}

\newcommand\eref[1]{(\ref{#1})}

\newcounter{jvcc}
\newcounter{amgg}
\newcounter{ls}

\renewcommand*\d{\mathop{}\!\mathrm{d}}

\newcommand{\av}[1]{\left\langle#1\right\rangle}

\renewcommand{\i}{\mathrm{i}}
\renewcommand{\(}{\left(}
\renewcommand{\)}{\right)}
\renewcommand{\mod}{\,\mathrm{mod}\,}
\newcommand{\id}{\mathbbm{1}}

\newcommand{\sH}{\mathcal{H}}
\newcommand{\sC}{\mathcal{C}}
\newcommand{\sT}{\mathcal{T}}
\newcommand{\sP}{\mathcal{P}}
\newcommand{\sR}{\mathcal{R}}
\newcommand{\sS}{\mathcal{S}}

\newcommand{\cH}{\mathcal{H}}
\newcommand{\diag}{\mathrm{diag}}
\newcommand{\adiag}{\mathrm{antidiag}}
\newcommand{\sN}{\mathcal{N}}

\newcommand{\matAIIId}{
	\begin{pmatrix}  & A \\ B &  \end{pmatrix}
}
\newcommand{\matAIId}{
	\begin{pmatrix} A & B \\ C & A^\top \end{pmatrix},\ \begin{cases}B=-B^\top\\C=-C^\top\end{cases}
}
\newcommand{\matC}{
	\begin{pmatrix} A & B \\ C & -A^\top \end{pmatrix},\ \begin{cases}B=B^\top\\C=C^\top\end{cases}
}
\newcommand{\matAIdp}{
	\begin{pmatrix} & A \\ A^\top & \end{pmatrix}
}
\newcommand{\matAIIdp}{
	\begin{pmatrix}
		&         & A & B \\
		&         & C & D \\
		D^\top & -B^\top &   &   \\
		-C^\top &  A^\top &   &
	\end{pmatrix}
}
\newcommand{\matAIdm}{
	\begin{pmatrix} & A \\ B & \end{pmatrix},\ \begin{cases}A=A^\top\\B=B^\top\end{cases}
}
\newcommand{\matAIIdm}{
	\begin{pmatrix} & A \\ B & \end{pmatrix},\ \begin{cases}A=-A^\top\\B=-B^\top\end{cases}
}

\newcommand{\matAIIIm}{
	\begin{pmatrix} & A \\ B & \end{pmatrix},\ \begin{cases}A=A^\dagger\\B=B^\dagger\end{cases}
}

\newcommand{\matBDIdmp}{
\begin{pmatrix} & A \\ B & \end{pmatrix},\ \begin{cases}
	A=A^\dagger=A^\top=A^*\\B=B^\dagger=B^\top=B^*
\end{cases}
}

\newcommand{\matCIIdmp}{
	\begin{pmatrix}  
	&     & A    & B   \\
	&     & -B^* & A^* \\
	C    & D   &      &     \\
	-D^* & C^* &      &
\end{pmatrix},\ 
\begin{cases}
	A=A^\dagger \\B=-B^\top \\ C=C^\dagger \\ D=-D^\top
\end{cases}
}

\newcommand{\matCIIpm}{
	\begin{pmatrix}  
		&     & A    & B   \\
		&     & B^* & -A^* \\
		C    & D   &      &     \\
		D^* & -C^* &      &
	\end{pmatrix},\ 
	\begin{cases}
			A=A^\dagger \\B=B^\top \\ C=C^\dagger \\ D=D^\top
	\end{cases}
}

\newcommand{\matBDIpm}{
	\begin{pmatrix} & A \\ B & \end{pmatrix},\ \begin{cases}A=A^\dagger=-A^\top=-A^*\\B=B^\dagger=-B^\top=-B^*\end{cases}
}  

\newcommand{\matAII}{
\begin{pmatrix}  
	A    & B   \\
   -B^*  & A^*
\end{pmatrix}	
}

\newcommand{\matAIp}{
	\begin{pmatrix} & A \\ B & \end{pmatrix},\ \begin{cases}
		A=A^*\\B=B^*
	\end{cases}
}

\newcommand{\matAIIp}{
	\begin{pmatrix}  
		&     & A    & B   \\
		&     & -B^* & A^* \\
		C    & D   &      &     \\
		-D^* & C^* &      &
	\end{pmatrix}
}

\newcommand{\matAIm}{
	\begin{pmatrix} & A \\ A^* & \end{pmatrix}
}

\begin{document}

\title{Symmetry Classification and Universality in Non-Hermitian Many-Body Quantum Chaos by the Sachdev-Ye-Kitaev Model}

\author{Antonio M. Garc\'\i a-Garc\'\i a}
\email{amgg@sjtu.edu.cn}
\affiliation{Shanghai Center for Complex Physics,
	School of Physics and Astronomy, Shanghai Jiao Tong
	University, Shanghai 200240, China}

\author{Lucas S\'a}
\email{lucas.seara.sa@tecnico.ulisboa.pt}
\affiliation{CeFEMA, Instituto Superior T\'ecnico, Universidade de Lisboa, Av.\ Rovisco Pais, 1049-001 Lisboa, Portugal}

\author{Jacobus J. M. Verbaarschot}
\email{jacobus.verbaarschot@stonybrook.edu}
\affiliation{Center for Nuclear Theory and Department of Physics and Astronomy, Stony Brook University, Stony Brook, New York 11794, USA}

\date{\today}

\begin{abstract}
  Spectral correlations are a powerful tool to study the dynamics of quantum many-body systems.
  For Hermitian Hamiltonians, quantum chaotic motion is related to random matrix theory spectral correlations.
  Based on recent progress in the application of spectral analysis to non-Hermitian quantum systems, we show that local level statistics, which probe the dynamics around the Heisenberg time, of a non-Hermitian $q$-body Sachdev-Ye-Kitev (nHSYK) model with $N$ Majorana fermions, and its chiral and complex-fermion extensions, are also well described by random matrix theory for $q > 2$, while for $q = 2$, they are given by the equivalent of Poisson statistics. For that comparison, we combine exact diagonalization numerical techniques with analytical results obtained for some of the random matrix spectral observables. Moreover, depending on $q$ and $N$,  we identify $19$ out of the $38$ non-Hermitian universality classes in the nHSYK model, including those corresponding to the tenfold way. In particular, we realize explicitly $14$ out of the $15$ universality classes corresponding to non-pseudo-Hermitian Hamiltonians that involve universal bulk correlations of classes ${\rm AI}^\dagger$ and ${\rm AII}^\dagger$, beyond the Ginibre ensembles. These results provide strong evidence of striking universal features in non-unitary many-body quantum chaos, which in all cases can be captured by nHSYK models with $q > 2$.
\end{abstract}

\maketitle
\newpage

\allowdisplaybreaks[3]

\section{Introduction}
Calculations in strongly interacting quantum systems are generically difficult. The spectrum of the Hamiltonian is arguably one of the least expensive quantities that can be computed numerically. Moreover, the spectrum is basis invariant and that, especially in many-body systems, is a substantial advantage. These facts help us understand the relevance of the Bohigas-Giannoni-Schmit (BGS) conjecture~\cite{bohigas1984}, which states that spectral correlations of quantum chaotic systems at the scale of the mean level spacing are given by the predictions of random matrix theory. It provides a simple, both conceptually and practically, but still very powerful link between the analysis of the spectrum and the quantum dynamics. More importantly, it points to a very robust universality of the late-time quantum dynamics under relatively mild conditions and with relatively well-understood exceptions due to phenomena like Anderson localization or integrability.
Global anti-unitary symmetries, and not dynamical features, label different universality classes, which are rigorously classified using random matrix theory (RMT), the so-called tenfold way \cite{Zirnbauer:1996zz}. Although the Berry-Tabor conjecture~\cite{berry1977} states a similar spectral characterization of integrable systems, its applicability is more restricted because, in some sense, each integrable system is integrable in its own way. 

As an illustration of the aforementioned universality, the study of spectral correlations to characterize quantum motion encompasses multiple disciplines with no direct relation between them. It started 70 years ago in the context of nuclear physics, where Wigner showed~\cite{wigner1951} that short-range spectral correlations of highly excited states of nuclei are well described by random matrix theory. In the 1970s and 1980s, the interest gradually shifted from nuclear physics to the dynamics of single-particle Hamiltonians in both deterministic~\cite{bohigas1984} and random~\cite{altshuler1988} potentials. Later, it was successfully applied~\cite{shuryak1993a,verbaarschot1993a} to describe the level statistics of the lattice QCD Dirac operator in the presence of gauge configurations. The chiral symmetry of the Dirac operator led to the proposal of universality classes in the context of random matrix theory, which were the direct precedent of the tenfold way~\cite{altland1997}. In recent years, it has also found applications in quantum gravity after the proposal~\cite{maldacena2015} that a certain universal bound on chaos is saturated in field theories with a gravity dual. More specifically, it is a bound on the growth of particular out-of-time-order correlation functions in the semiclassical limit. For quantum chaotic systems, the growth is exponential around the Ehrenfest time with a rate controlled by the classical Lyapunov exponent. In the early days of quantum chaos theory~\cite{larkin1969,berman1978}, this exponential growth was broadly employed to characterize quantum chaos in single-particle Hamiltonians. 

More recently, Kitaev~\cite{kitaev2015} showed analytically that a system of $N$ fermions with $q$-body random interactions in zero spatial dimensions, which is now called the Sachdev-Ye-Kitaev (SYK) model, saturates the bound of Ref.~\cite{maldacena2015}, which indicates the existence of a gravity dual. 
Indeed, in the strong-coupling, low-temperature limit, it shares with Jackiw-Teitelboim (JT) gravity~\cite{jackiw1985,teitelboim1983} -- a near-AdS$_2$ background -- the same low-energy effective Schwarzian action. Therefore, the study of the SYK model could reveal features of non-perturbative quantum gravity in low dimensions. Along this line, the observation that level statistics of the SYK model are well described by random matrix theory~\cite{garcia2017,garcia2016,cotler2016} is suggestive that quantum black holes in JT gravity are also quantum chaotic at all but the shortest timescales. Indeed, the relation between SYK and JT gravity is not restricted to black holes, as other gravity configurations like wormholes~\cite{maldacena2018,garcia2021} have been shown to be closely related to two identical SYK models in the low-temperature limit. As temperature increases, a first-order phase transition takes place, which can also be characterized~\cite{garcia2019} by level statistics. 
By tuning $q$ and $N$, and considering also Dirac fermions, it is possible to reproduce all ten universality classes~\cite{you2016,garcia2018a,li2017,kanazawa2017,behrends2019,sun2020,garcia2021d}. Although, in the past, similar models with Dirac fermions have been intensively investigated~\cite{french1970,french1971,mon1975,bohigas1971,bohigas1971a,sachdev1993,kota2014} in nuclear physics, condensed matter, and many-body quantum chaos, it was Kitaev's demonstration that quantum chaos and strong interactions could, to some extent, coexist with analytical tractability, together with its relevance in quantum gravity, that has brought SYK to the research forefront in theoretical physics.

So far, we have restricted our discussion to closed systems where the spectrum is real. However, non-Hermitian~\cite{bender1998} effective descriptions of quantum Hamiltonians appear in a multitude of problems: quantum dissipative systems~\cite{daley2014,sieberer2016} such as cold atoms in dynamical optical potentials, the Euclidean QCD Dirac operator at nonzero chemical potential~\cite{Barbour:1986jf,kanazawa2021}, photons with parity-time symmetry~\cite{feng2017}, the scattering matrix of open quantum systems from quantum dots
\cite{alhassid2000} to compound nuclei~\cite{Verbaarschot:1985jn} and flux lines depinned from columnar defects by a transverse magnetic field in superconductivity~\cite{hatano1996}.
Recently, the application of random matrix theory to generic open quantum systems has also gathered pace. Modeling the generators of driven or dissipative quantum dynamics as random matrices has led to an understanding of the spectral distributions, typical timescales, spectral gaps, and steady-state properties of open systems described by random Lindbladians~\cite{denisov2018,can2019jphysa,can2019prl,akemann2019,sa2020c,wang2020,sommer2021,lange2021,tarnowski2021,xu2019}, quantum channels~\cite{bruzda2009,bruzda2010,sa2020b,kukulski2021}, and Markov generators~\cite{timm2009,tarnowski2021}.

The universality of correlations in non-Hermitian systems has also been the subject of great interest. Spectral correlations of random non-Hermitian matrices of the so-called Ginibre ensembles~\cite{ginibre1965}, defined by the real, complex, or quaternionic nature of the matrix entries, are relatively well understood with explicit results for both the three standard universality classes~\cite{fyodorov2003} and the three chiral classes~\cite{Halasz:1997fc,splittorff2004a,akemann2005a,akemann2011} that have found applications, e.g., in the context of Euclidean QCD at finite baryonic chemical potential. 
However, according to the full classification of symmetries in non-Hermitian random matrices~\cite{bernard2002,kawabata2019}, there are $38$ non-Hermitian universality classes. It turns out that the Ginibre ensemble is only one of the three universality classes~\cite{hamazaki2020,jaiswal2019} for local level statistics in non-Hermitian random matrix theory. The other two, called AI$^\dagger$ and AII$^\dagger$, are defined in terms of transposition symmetry instead of complex conjugation. For instance, they describe the spectral correlations of the QCD Dirac operator in two-color QCD coupled to a chiral $U(1)$
gauge field~\cite{Halasz:1997fc,kanazawa2021}.

The progress in the development of a full classification scheme has not yet fully translated into a systematic spectral characterization of non-Hermitian quantum chaotic systems, see Ref.~\cite{altland2021} for a recent study focused on open fermionic quantum matter.
First, there is no equivalent of the
BGS conjecture, so the relation between dynamics and level statistics is unclear.
There are also additional technical problems: Correlations of complex eigenvalues are weakened, and the necessary unfolding of eigenvalues
may be problematic~\cite{akemann2019}, in particular, when the
eigenvalue distribution is not radially symmetric. However, these problems have been ameliorated in the last years with the introduction of spectral observables that do not require unfolding for short-range correlators, such as the ratio of spacings between nearest-neighbor eigenvalues~\cite{sa2020}, 
which have found applications in the study of collective-spin Liouvillians~\cite{rubio2021}, non-Hermitian Anderson transitions~\cite{huang2020a,luo2021,luo2021a}, directed random graphs~\cite{peron2020}, nonunitary open quantum circuits~\cite{sa2021,prosen2021}, two-color QCD~\cite{kanazawa2021}, and the classical-quantum transition~\cite{tarnowski2021}. The study of long-range correlators such as the number variance~\cite{huang2020,lacroix2019,lebowitz1999,jancovici1981} or spectral form factor~\cite{fyodorov1997,chan2021} (which requires unfolding) further suggests that some weakened form of spectral rigidity is still present in non-Hermitian systems and will be the subject of a separate publication~\cite{usnext}.
 
To address the issues mentioned in the previous paragraph, one needs a many-body system where it is possible to test the results of random matrix theory, including the existence of many more universality classes; at the same time this system should be simple enough to be amenable to an analytical and numerical treatment so that it is possible to independently probe the nature of the non-Hermitian quantum dynamics. In this paper, we propose that this model is the non-Hermitian SYK (nHSYK) model. We focus our study on the local level statistics of this model, which probe the quantum dynamics for long timescales of the order of the Heisenberg time. For $q=2$, we find that the spectrum is largely uncorrelated and well described by the equivalent of Poisson statistics. For $q>2$, there is excellent agreement with the predictions of non-Hermitian random matrix theory in short-range correlators like the complex spacing ratio~\cite{sa2020} and microscopic correlators near the hard edge. The latter, in particular, have remained unaddressed for all but the Ginibre universality classes.

Depending on both $q$ with values $3$, $4$, or $6$, and $N$, we have explicitly identified $19$ of the $38$ classes for non-Hermitian systems and, in particular, $14$ of the $15$ corresponding to non-pseudo-Hermitian Hamiltonians. This encompasses nHSYK models belonging to different complex Ginibre, complex symmetric, and quaternionic symmetric universality classes; non-Hermitian Wishart-Sachdev-Ye-Kitaev (nHWSYK) models, a chiral extension of the nHSYK model realizing further non-Hermitian chiral classes; and non-Hermitian complex-fermion SYK models.

Finally, we note that different non-Hermitian variants of the SYK model have recently been investigated~\cite{garcia2021,garcia2021b,pengfei2021,zhang2021,zhang2021a} in the literature. However, both the employed models and the focus of these studies are quite different from ours. In Ref.~\cite{garcia2021}, the gravity dual of two complex conjugate copies of a $q=4$ nHSYK model was identified as an Euclidean wormhole by an analysis of the free energy. The role of replica-symmetry-breaking solutions in the low-temperature limit of the free energy of two copies of an nHSYK has been investigated in Ref.~\cite{garcia2021b}. The entanglement entropy and an emergent replica conformal symmetry were recently studied in chains of $q = 2$ nHSYK models~\cite{zhang2021,zhang2021a}. The Page curve~\cite{page1993}, describing the process of black hole evaporation~\cite{pengfei2021}, was computed by an effectively open SYK model, where the role of the environment is played by a $q=2$ SYK.  

\begin{figure}[t!]
	% 	 \subfigure[]{
	% 	\subfigure[]{ 
	\includegraphics[width=8cm]{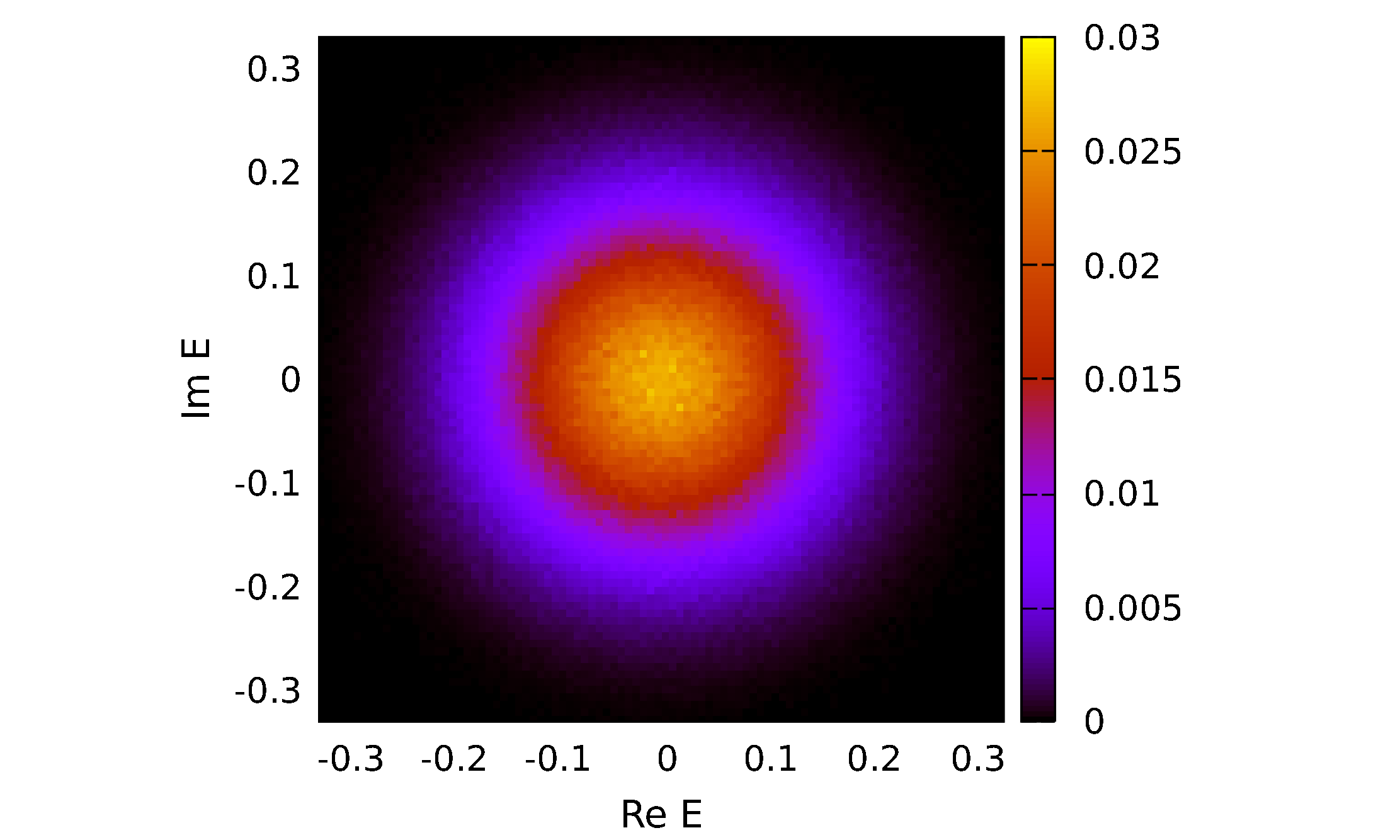}
	\includegraphics[width=8cm]{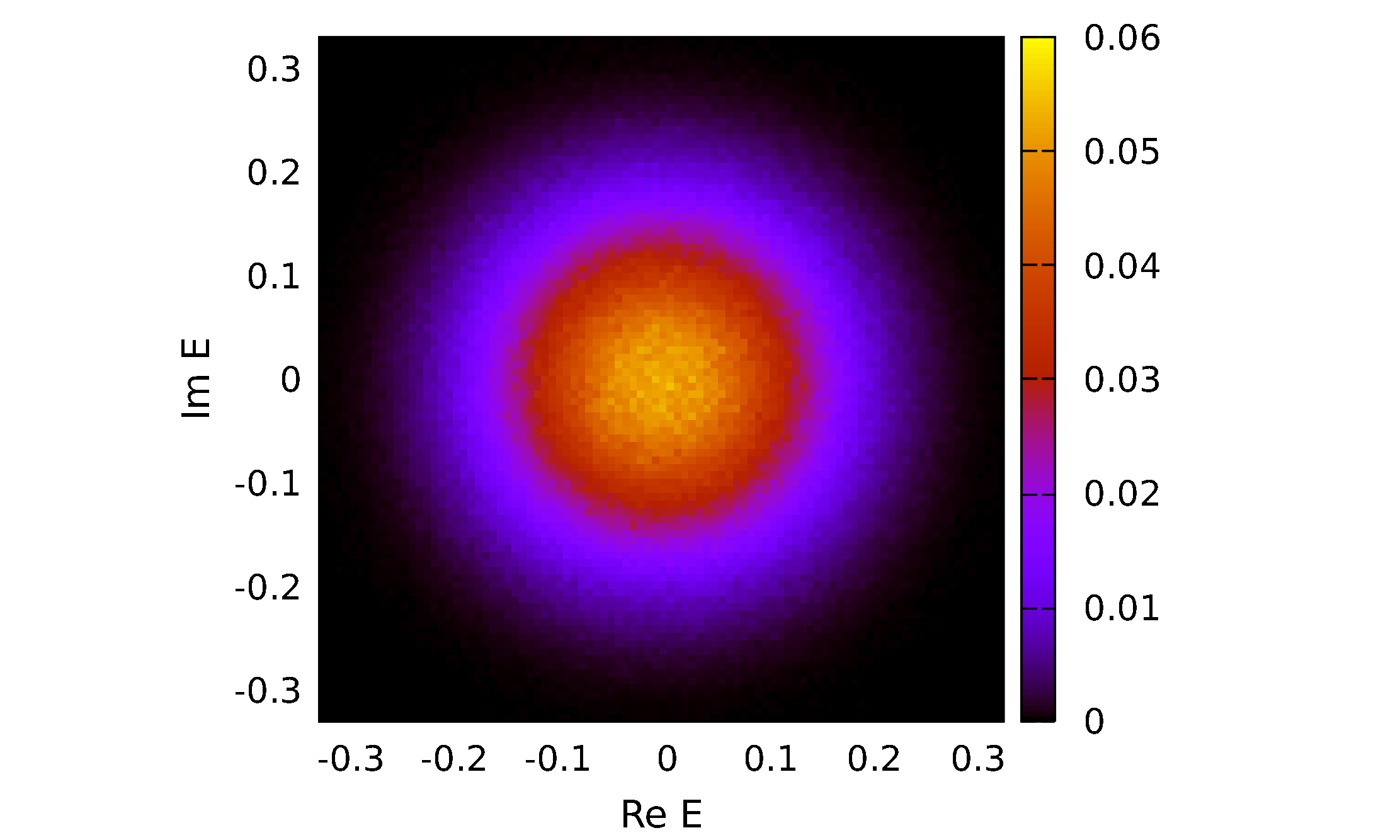}
	\caption{Spectral density associated with the complex eigenvalues of the nHSYK Hamiltonian of Eq.~(\ref{hami}) for $q=2$ and $N=18$ (left panel) and $20$ (right panel).
	  The spectral density is radially symmetric, and it has 
  a maximum in the $|E| \sim 0$ region and then decreases monotonically
  with no sign of a sharp spectral edge. There is no qualitative dependence on $N$. We see that the level correlations are not quantum chaotic. 
	}
	\label{fig:denzq2}
\end{figure}

\vfill

\section{The Sachdev-Ye-Kitaev model with complex couplings}
 
We study a single non-Hermitian SYK model of $N$ Majorana fermions in $(0 + 1)$ dimensions with $q$-body interactions in Fock space, with complex instead of real couplings:
\begin{align}\label{hami}
 	H \, &= \, \sum_{i_1<i_2<\cdots<i_q}^{N} (J_{i_1i_2\cdots i_q}+i M_{i_1i_2\cdots i_q}) \, \psi_{i_1} \, \psi_{i_2} \, \cdots \, \psi_{i_q},
\end{align}
where $J_{i_1\cdots i_q}$ and $ M_{i_1\cdots i_q}$ are Gaussian-distributed random variables with zero average and variance~\cite{kitaev2015,maldacena2016} 
\be
\sigma^2 =\frac 1{6(2N)^{q-1}},
\ee
and $\{ \psi_{i}, \psi_{j} \} = 2\delta_{ij}$. The Majorana fermions can be represented by $2^{N/2}$-dimensional Hermitian Dirac $\gamma$ matrices, and below, we
denote them by $\gamma_i$.
For our analytical considerations, we work in a basis where the odd- and even-numbered $\gamma$ matrices are represented by real symmetric and purely imaginary antisymmetric matrices, respectively.
We shall consider the cases $q=2,\;3,\;4$, and $6$, and the number of fermions, $N$, is always even.
 
The Hamiltonian has, in general, a complex spectrum. However, depending on $q$, it could have some additional symmetry. For instance, for $q=3$, we see that it has a chiral symmetry $E_i \to -E_i$, where $E_i$ is, in general, complex.

\begin{figure}[t]
	% 	 \subfigure[]{denzgt9k1q3hm20000f.pdf
	\includegraphics[width=8cm]{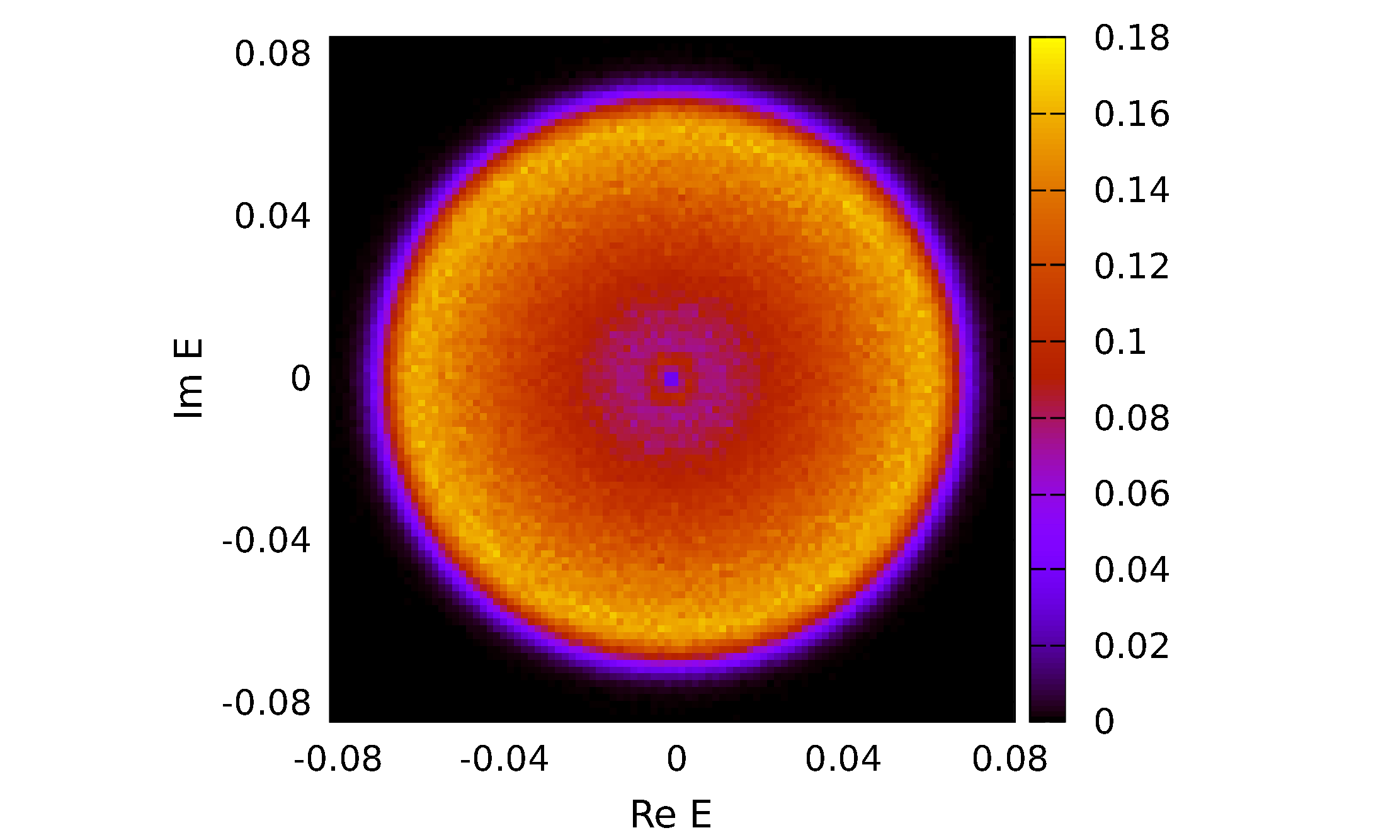}
	% 	\subfigure[]{ 
	\includegraphics[width=8cm]{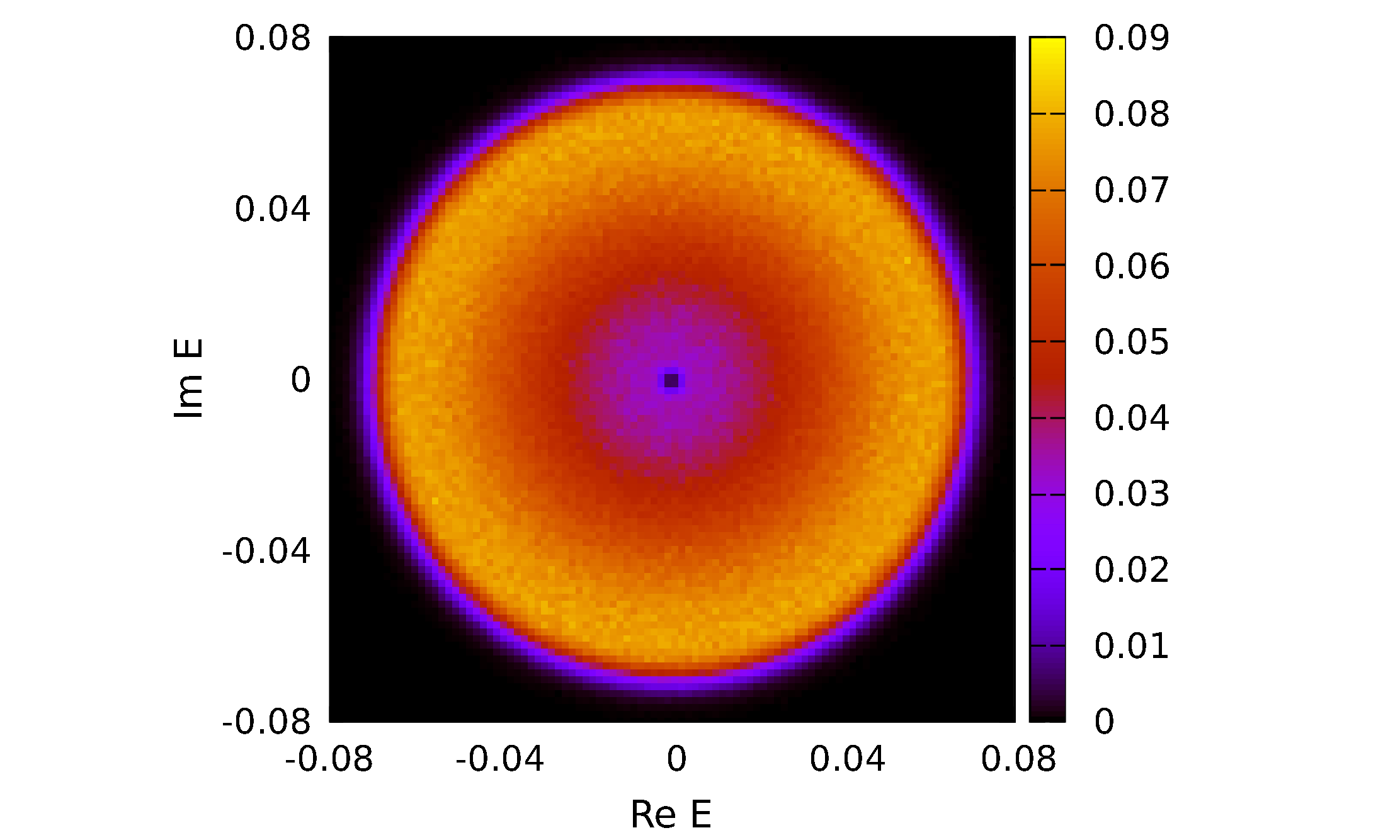}
	\includegraphics[width=8cm]{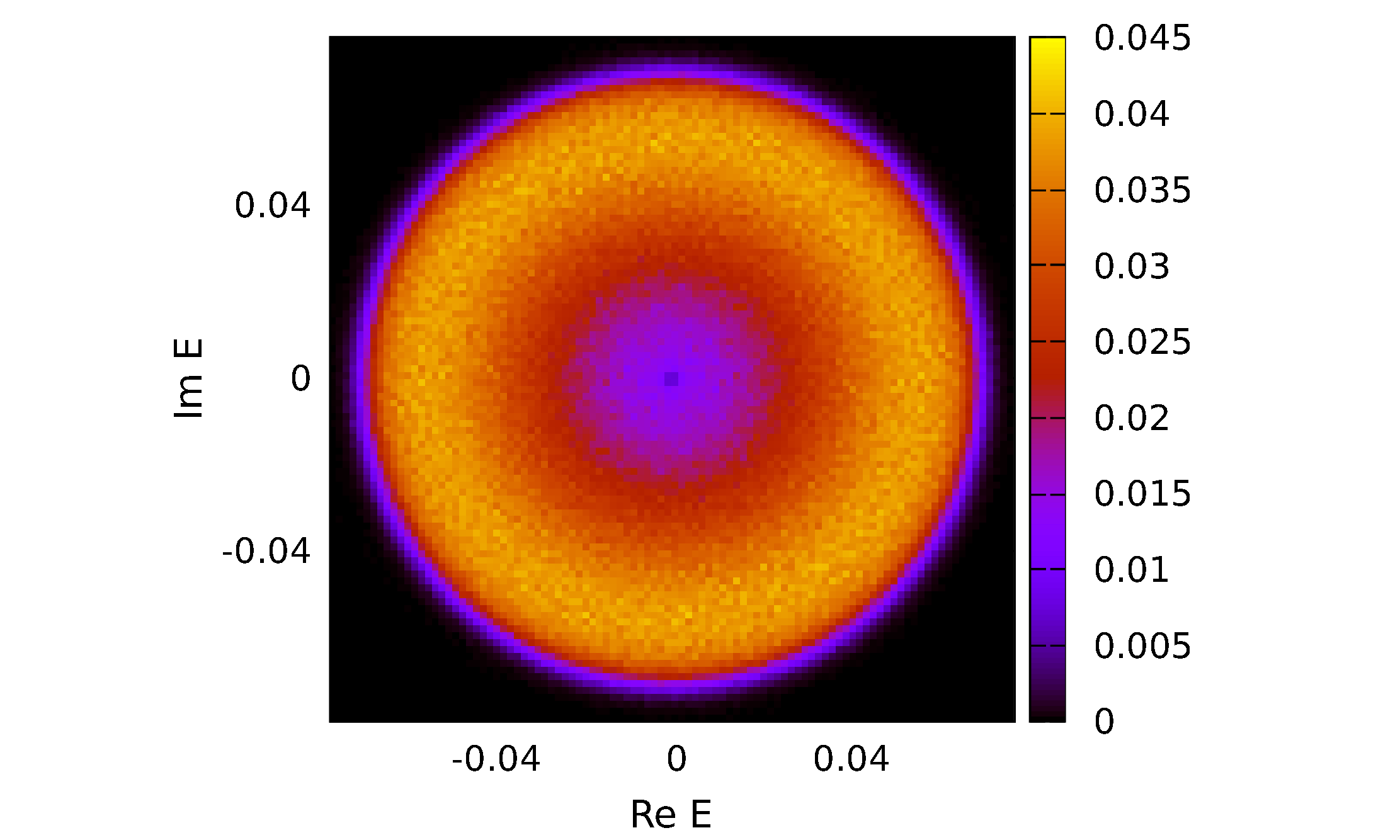}
	\includegraphics[width=8cm]{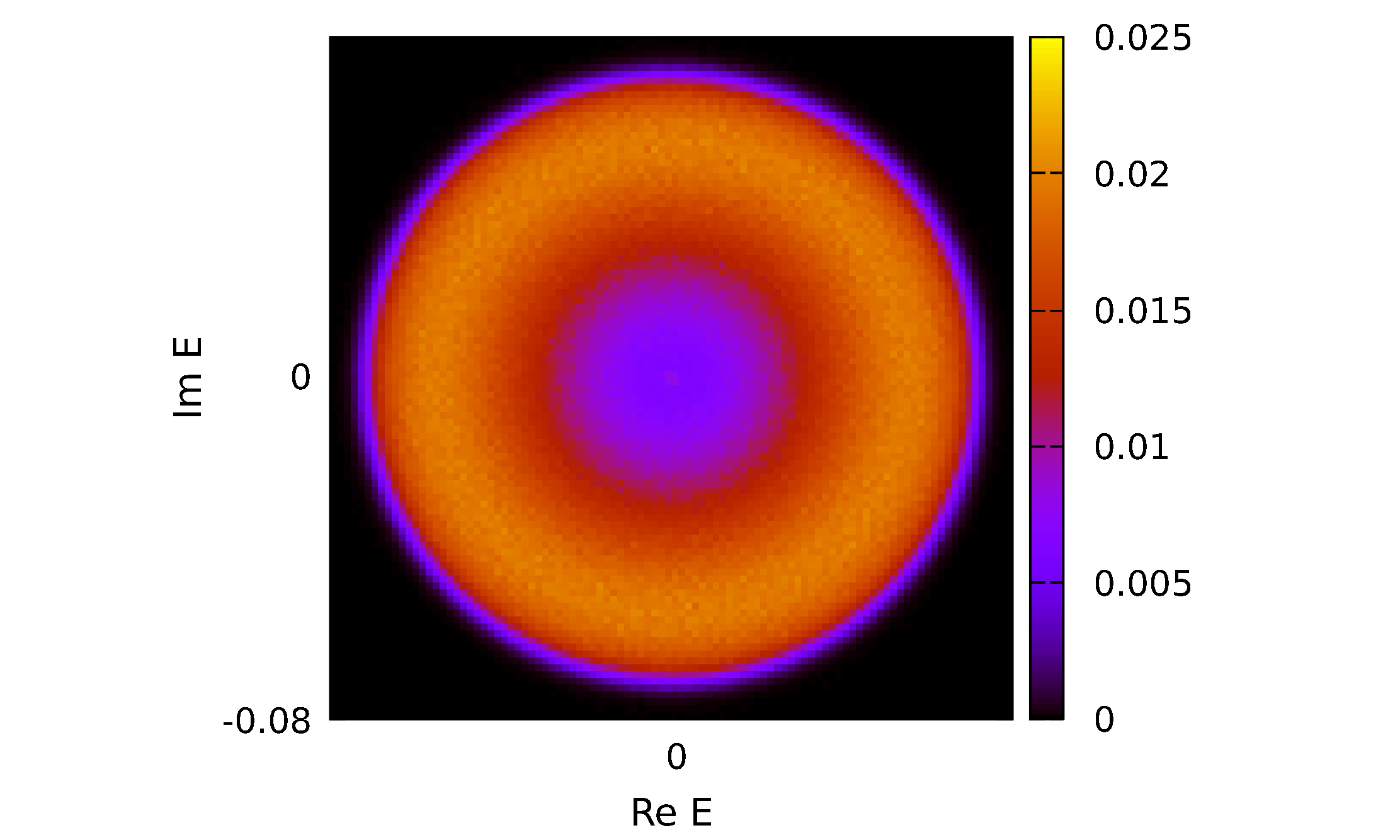} 
	\caption{
	  Spectral density associated with the eigenvalues of the nHSYK Hamiltonian of Eq.~(\ref{hami}) for $q=3$ and $N = 18$ (top-left panel), $20$ (top-right panel), $22$ (bottom-left panel), and $24$ (bottom-right panel).
          The spectral density is radially symmetric but qualitatively different from the $q=2$ case. In all cases, the spectrum has a sharp edge that becomes
          discontinuous in the thermodynamical limit, and
the maximum is not at the center but rather in a ring not far from the edge. The chiral symmetry of the spectrum $E \to -E$ has a rather profound effect on the density in the region $|E|\sim0$. We can observe the characteristic oscillations for $N=18$ near $|E|=0$. The spectral density, is in general, suppressed in this region, though the suppression strength and pattern are dependent on $N$.
   These features are related to different non-Hermitian universality classes.
	}
	\label{fig:denzq3}
\end{figure}

 \begin{figure}[t]
	% 	 \subfigure[]{denzgt9k1q3hm20000f.pdf
	\includegraphics[width=8cm]{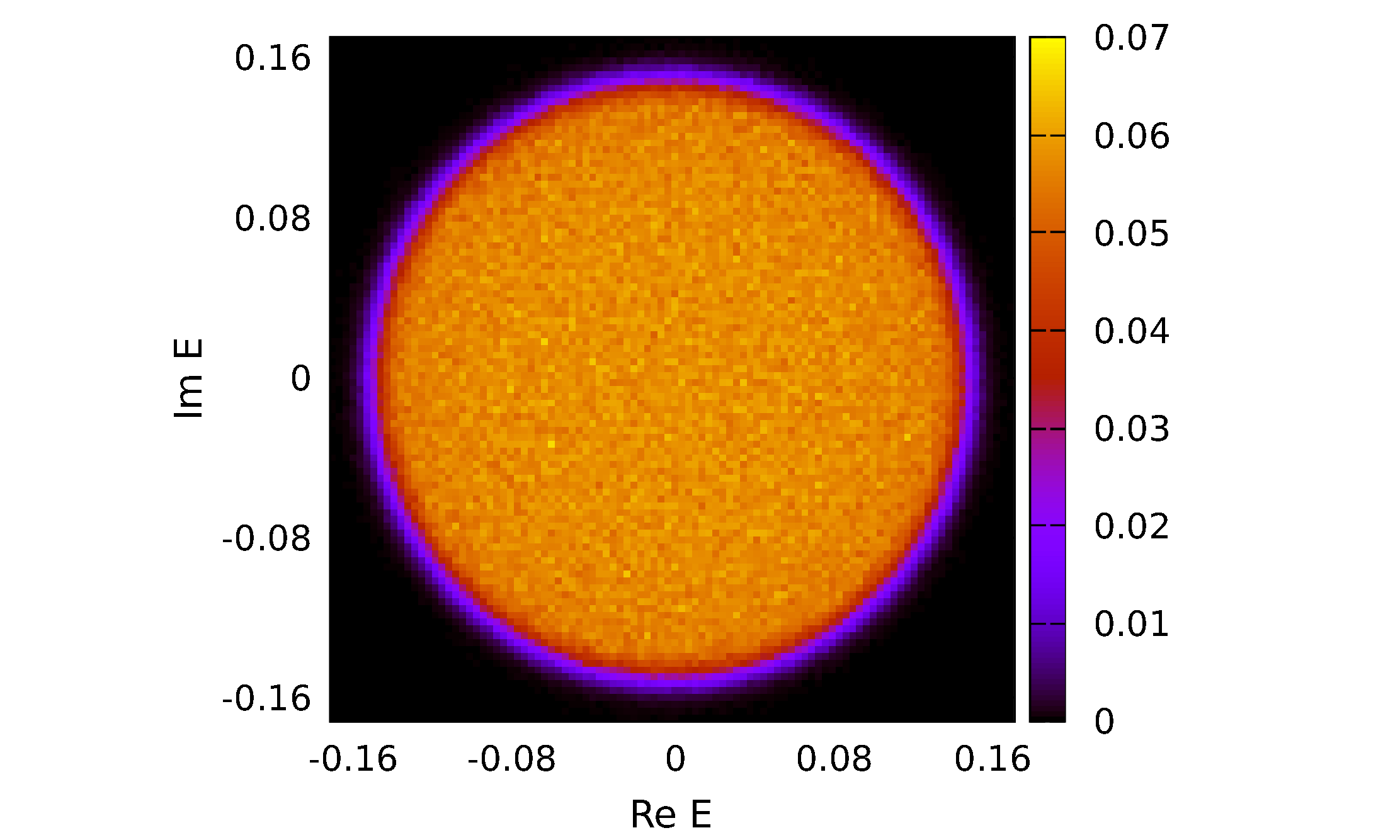}
	% 	\subfigure[]{ 
	\includegraphics[width=8cm]{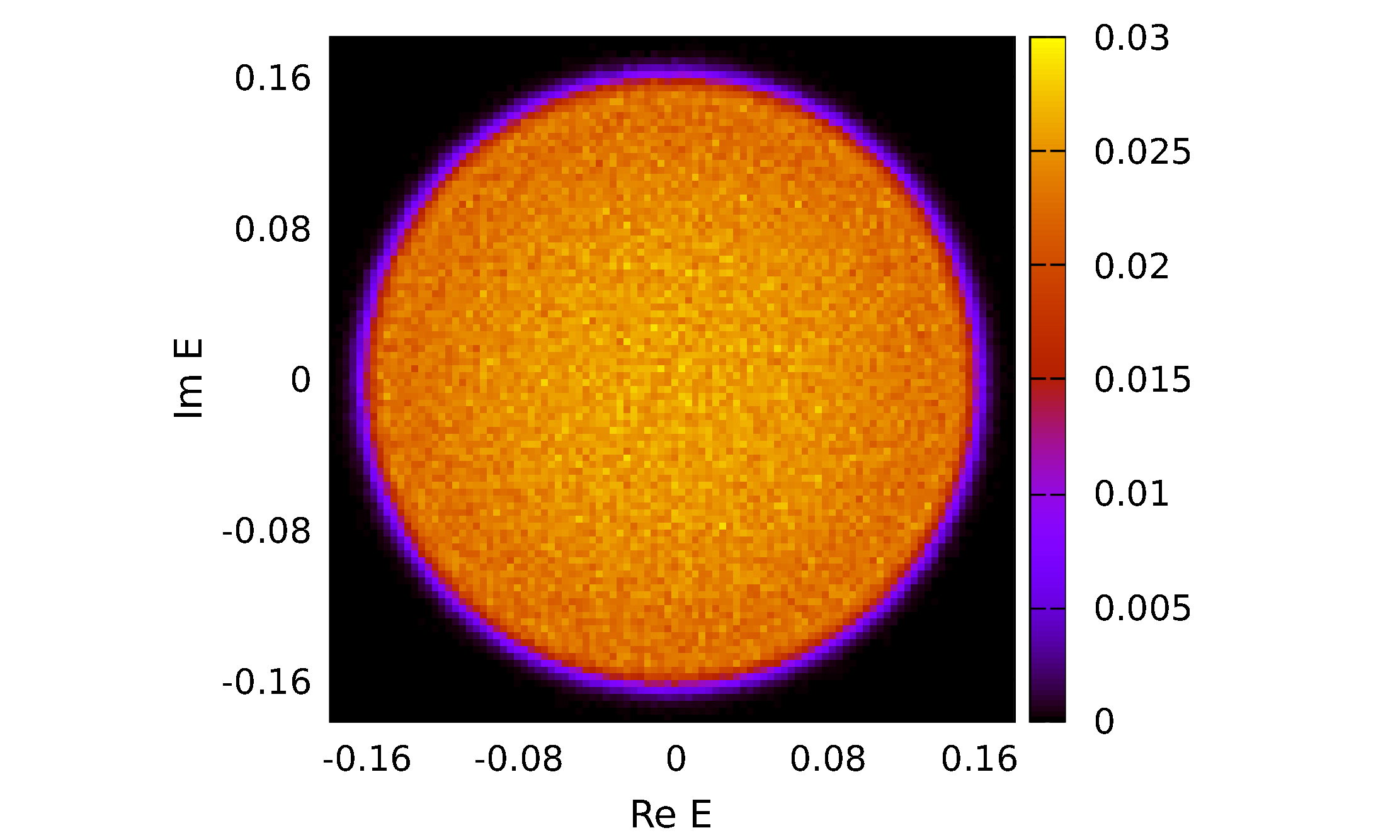}
	\includegraphics[width=8cm]{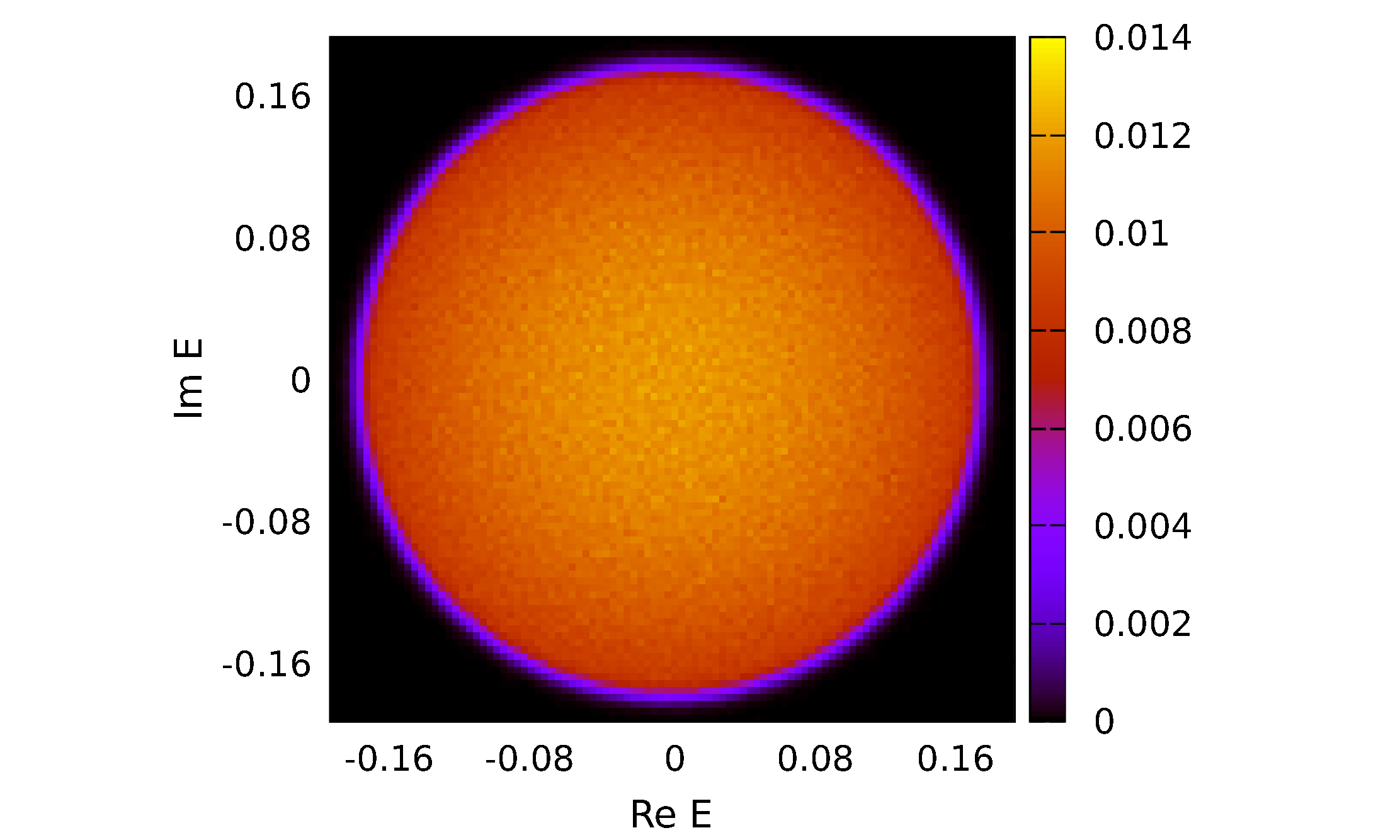}
	\includegraphics[width=8cm]{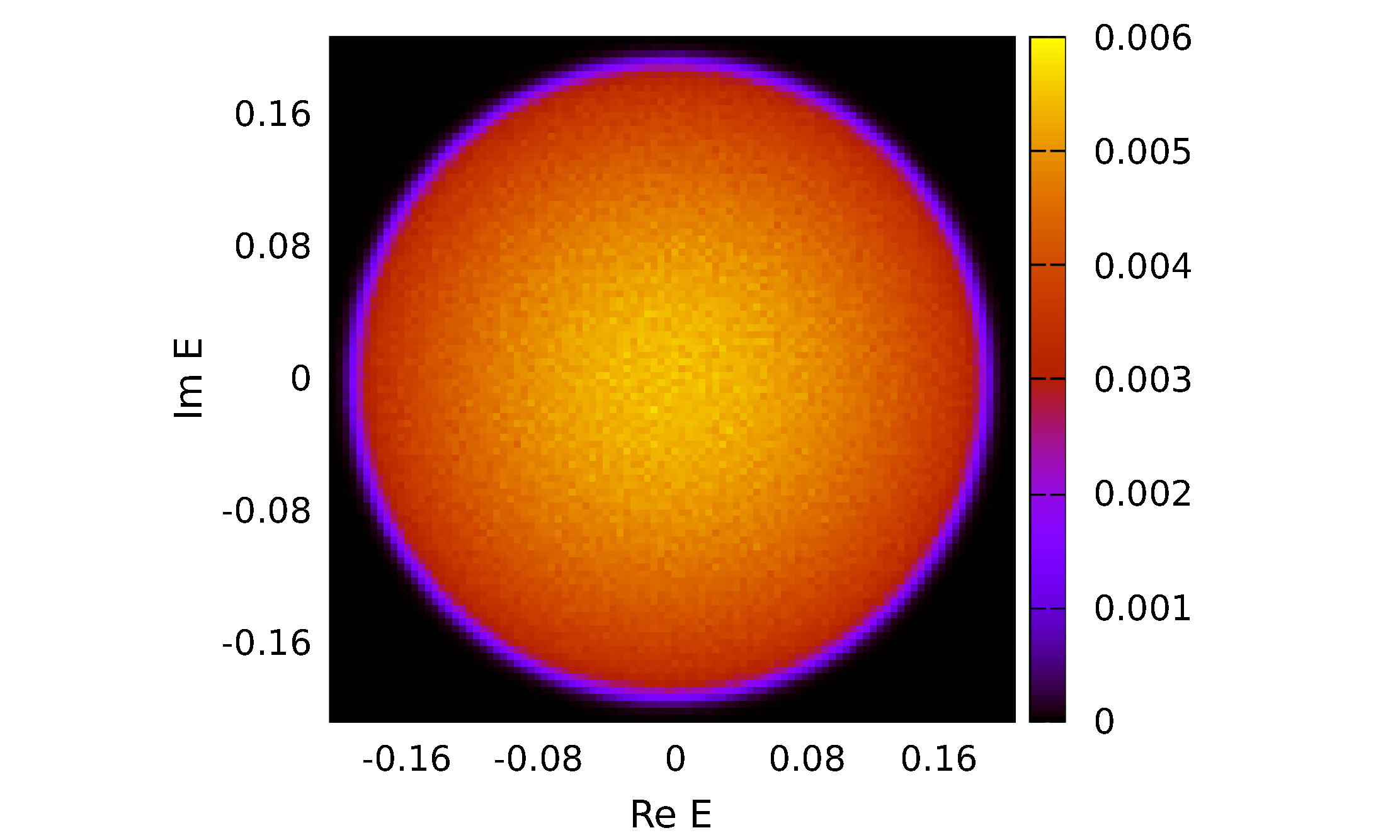}
	\caption{Spectral density associated with the complex eigenvalues of the nHSYK Hamiltonian of Eq.~(\ref{hami}) for $q=4$ and $N = 18$ (top-left panel), $20$ (top-right panel), $22$ (bottom-left panel), and $24$ (bottom-right panel).
          As in the previous cases, the average spectral density is radially symmetric, but, unlike the $q=3$ case, it is rather unstructured, with a broad maximum located in the central part, followed by a slow decay for larger $|E|$ energies. Finally, as is the case for $q=3$, a rather sharp edge is observed where the density vanishes abruptly. Because of the absence of inversion symmetry, the spectral density is not suppressed
          or enhanced in the $|E|\sim0 $ region, as was the case for
 $q = 3$.}
	\label{fig:denzq4}
\end{figure}

\begin{figure}[t]
	% 	 \subfigure[]{denzgt9k1q3hm20000f.pdf
	%	\includegraphics[width=8cm]{./denzgt8k1q6nhSYKhm10000s1.pdf}
	% 	\subfigure[]{ 
	\includegraphics[width=8cm]{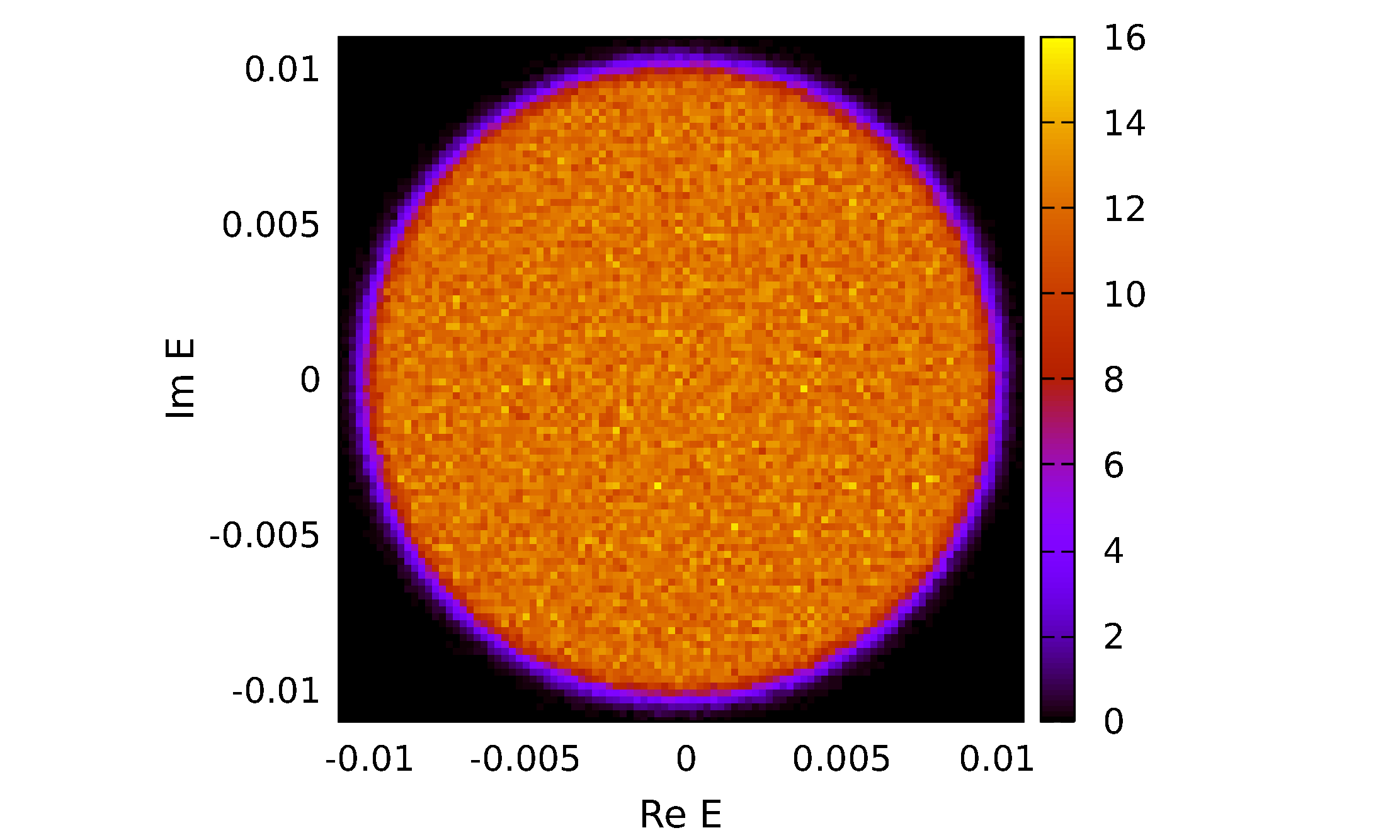}
	\includegraphics[width=8cm]{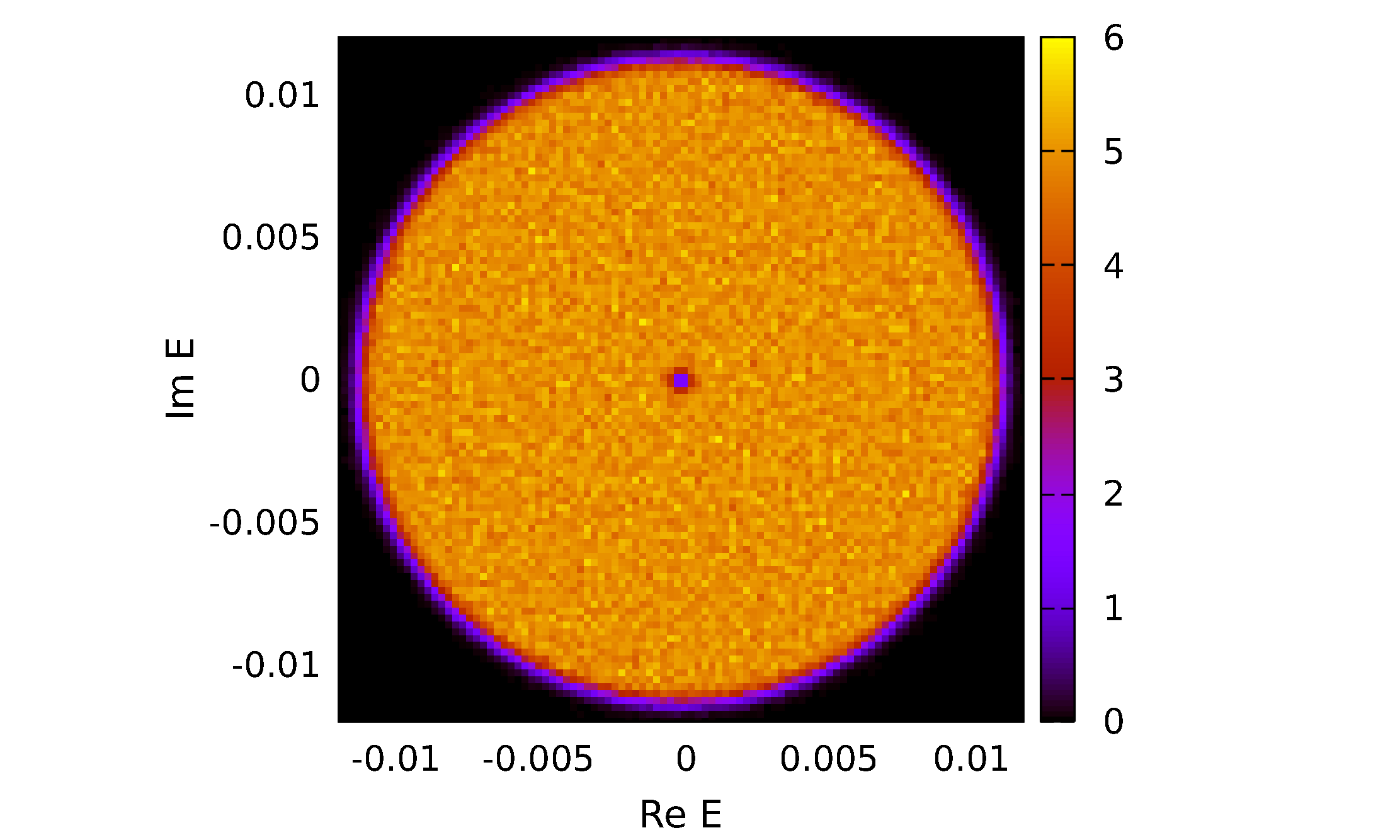}
	\includegraphics[width=8cm]{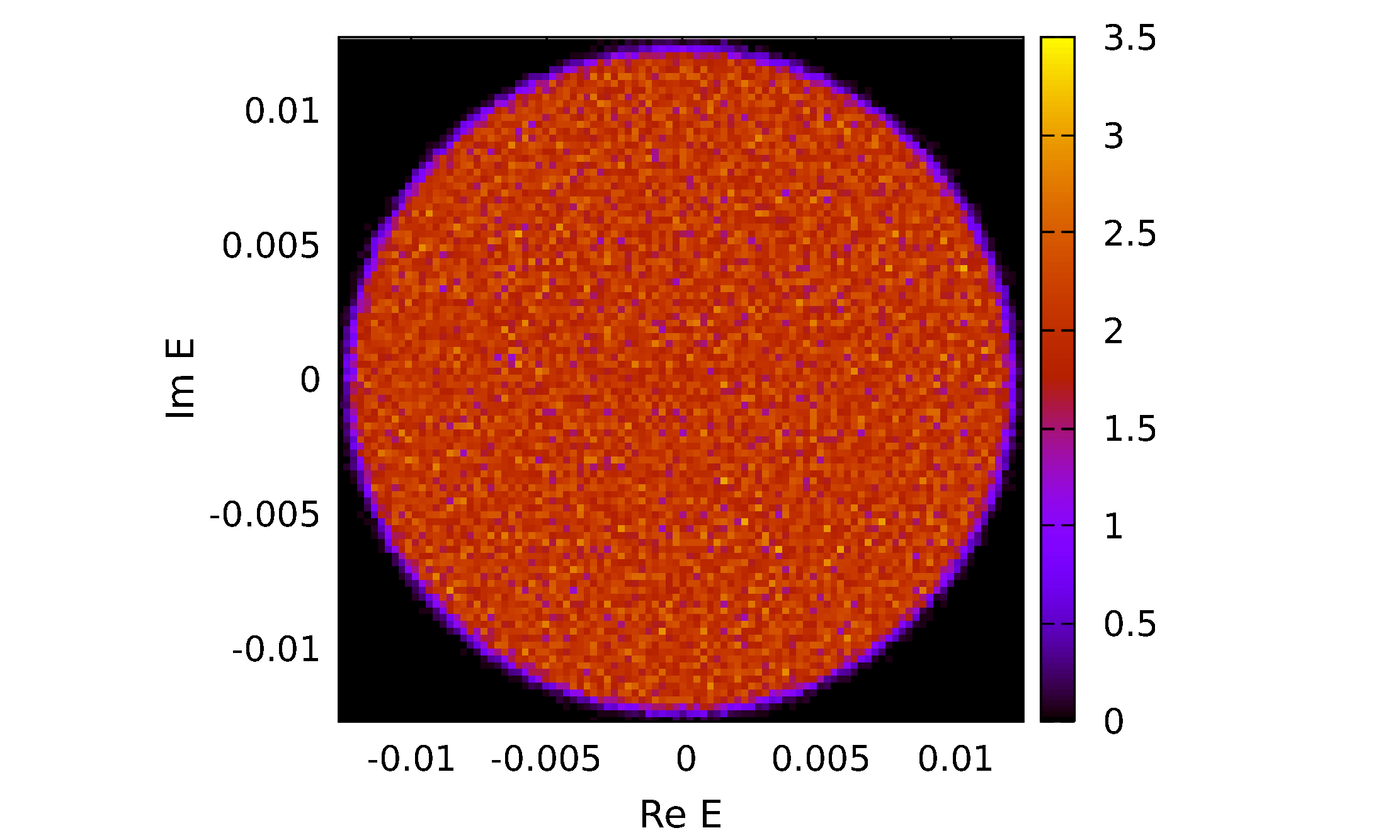}
	\includegraphics[width=8cm]{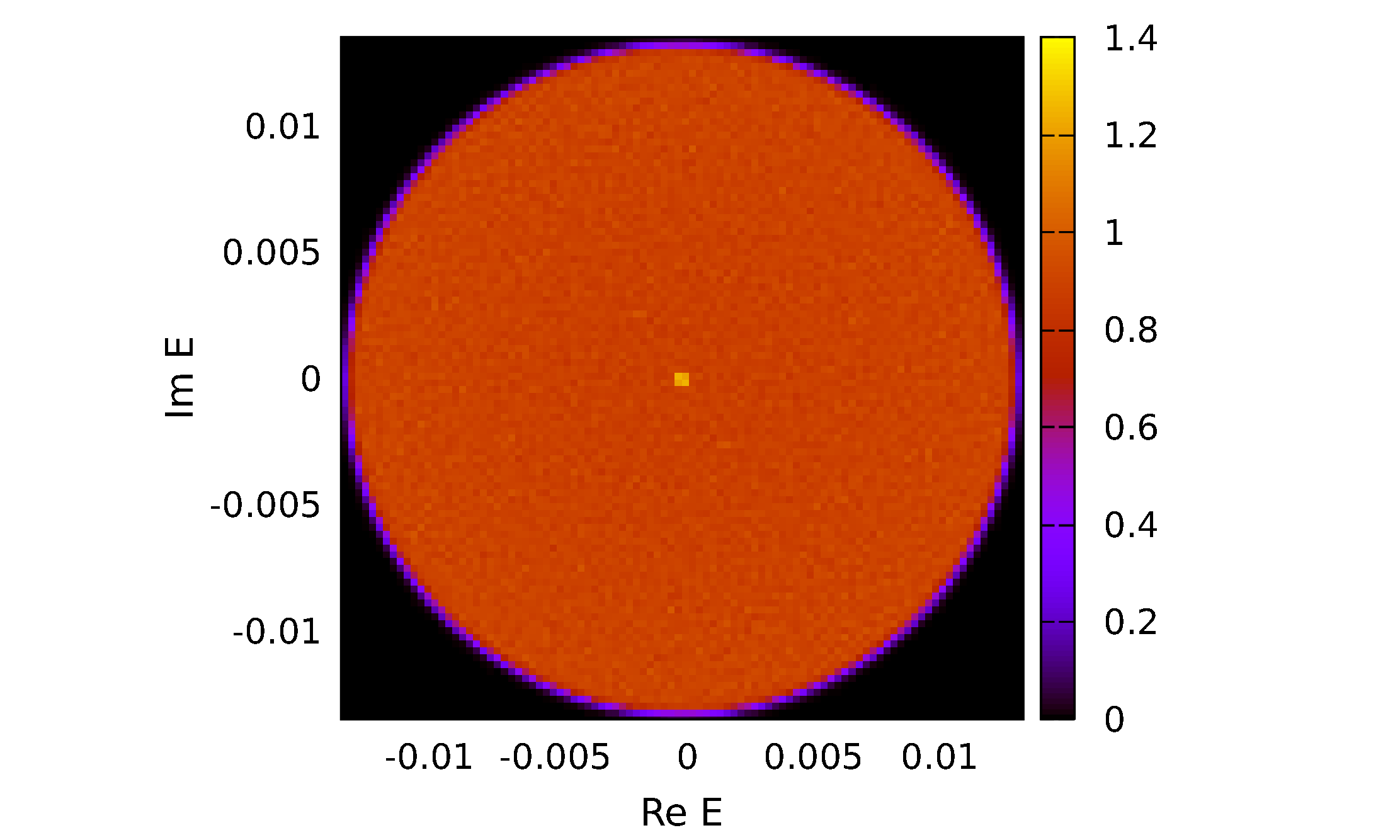}
	\caption{Spectral density of the complex eigenvalues of the nHSYK Hamiltonian of Eq.~(\ref{hami}) for $q=6$ and $N = 18$ (top-left panel), $20$ (top-right panel), $22$ (bottom-left panel), and $24$ (bottom-right panel). As in the previous cases, the average spectral density is radially symmetric. However, we observe distinctive features as well. The spectrum has inversion symmetry for $N=20,\; 24$, but, unlike for $q=3$ and to a lesser extent for $q=4$, the density is rather unstructured except in the $|E| \to 0$ region where, depending on $N$, we observe either a sharp suppression ($N = 20$) or a strong enhancement ($N = 24$). Also in this case, we
	find a sharp edge.
	}
	\label{fig:denzq6}
\end{figure}

\section{Spectral density}
We start our analysis with the study of the spectral density of the nHSYK model. We compute its spectrum by exact diagonalization techniques for values of
$N \leq 28$. 
For each set of parameters, we obtain at least $10^6$ eigenvalues. We recall that for real couplings and a fixed finite $N\gg 1,q>2$, or $N,q\to\infty$ with $q/\sqrt{N}$ finite~\cite{erdos2014}, the spectral density close to the ground state grows exponentially, a feature also typical of quantum black holes.
For odd $q$, when the ``Hamiltonian'' is the supercharge of a supersymmetric
theory, the system (i.e., the supercharge) has chiral symmetry.
The spectral density close to zero energy, controlled by the chiral symmetry, is well described~\cite{garcia2018a,garcia2021d} by the universal predictions of random matrix theory~\cite{verbaarschot1993a} belonging to a universality class that depends on the number $N$ of Majoranas. Numerical results for the complex spectral density, depicted in Figs.~\ref{fig:denzq2}--\ref{fig:denzq6}, indicate a rotationally invariant spectrum for all $q$ and $N$.
The reason for this is that the probability density can be
written as $\exp(-c \Tr H H^\dagger)$, which is invariant under multiplication of the spectrum by a constant phase.
For $q > 2$, the spectral support ends in a rather sharp edge similar to that observed in random matrix ensembles. 

For $q=2$ and $q=4$---see Figs.~\ref{fig:denzq2} and \ref{fig:denzq4}, respectively---the spectral density has a maximum in the central region of small $|E|$ with a monotonous decay towards a rather sharp edge for $q =4$.
In contrast, for $q=3$, the spectral density depicted in Fig.~\ref{fig:denzq3} has a richer structure.
Depending on $N$, we observe different degrees of suppression, or enhancement, of the spectral density, followed by a maximum at intermediate distances. In some cases, we also observe
universal oscillations close to
zero on the scale of the level spacing due to a spectral inversion symmetry $E \to -E$. For $q = 6$, see Fig.~\ref{fig:denzq6}, we also observe the effect of inversion symmetry for
$N=20$ and $N=24$, but otherwise, the spectral density is almost constant.
In the next section, we study in more detail the relation between these special spectral features and
additional global symmetries of the nHSYK Hamiltonian.
This will eventually lead to a full match between different universality classes and specific nHSYK Hamiltonians, depending on $N$ and $q$, that implement them.

\section{Symmetry classification}
\label{sec:symm_class}

As is the case with Hermitian Hamiltonians, non-Hermitian Hamiltonians are classified by symmetries of the irreducible blocks of the unitary symmetries.
What remains are anti-unitary symmetries and involutive symmetries. There are two possibilities for anti-unitary symmetries of $H$, $[\sT_+, H]=0$ with $\sT_+^2 =\pm 1$. [Here, the subscript refers to the sign in the commutation relation in the first
column of Eqs.~(\ref{eq:nHsym_TRS})--(\ref{eq:nHsym_pH}) below].
Regarding involutive symmetries, we have the chiral symmetry $\{\Pi,H \}=0$ and
the anti-unitary symmetry $\{\sT_-,H \}= 0$, again with $\sT_-^2=\pm1$.
For non-Hermitian matrices, we have the additional possibility to
map $H \to H^\dagger$, in combination with the anti-unitary symmetries.
\newpage
We thus arrive at the following symmetries~\cite{Halasz:1997fc,bernard2002,magnea2008,lieu2018,kawabata2019,zhou2019,ashida2020,hamazaki2020,lieu2020,altland2021}:
\begin{alignat}{99}
\label{eq:nHsym_TRS}
&\sT_+ H \sT_+^{-1} = H,\qquad 
&&\sT_+^2=\pm 1, \qquad
&& \sT_+ \;\text{anti-unitary} \qquad
&&\text{(Time-Reversal Symmetry)},
\\
\label{eq:nHsym_PHSd}
&\sT_- H \sT_-^{-1} = -H,\qquad 
&&\sT_-^2=\pm 1, \qquad
&& \sT_- \;\text{anti-unitary}\qquad 
&&\text{(Particle-Hole Symmetry)},
\\
\label{eq:nHsym_TRSd}
&\sC_+ H^\dagger \sC_+^{-1} = H,\qquad 
&&\sC_+^2=\pm 1, \qquad 
&& \sC_+ \;\text{anti-unitary} 
&&\text{(Time-Reversal Symmetry)},
\\
\label{eq:nHsym_PHS}
&\sC_- H^\dagger \sC_-^{-1} = -H,\qquad 
&&\sC_-^2=\pm 1, \qquad
&& \sC_-\; \text{anti-unitary} \qquad 
&&\text{(Particle-Hole Symmetry)},
\\
\label{eq:nHsym_CSd}
&\Pi H \Pi^{-1} = -H,\qquad 
&&\Pi^2=1, \qquad 
&&\Pi\; \text{unitary}
&&\text{(Chiral Symmetry)},
\\
\label{eq:nHsym_pH}
&\eta H^\dagger \eta^{-1} = H,\qquad 
&&\eta^2=1, \qquad
&&\eta\;\text{unitary}
&&\text{(Pseudo-Hermiticity)}.
\end{alignat}

In the Hermitian case, the classification simplifies to the symmetries of Eqs.~(\ref{eq:nHsym_TRS}), (\ref{eq:nHsym_PHSd}), and (\ref{eq:nHsym_CSd}).
In Ref.~\cite{kawabata2019}, there is one more
symmetry given by the substitution $H \to \i H$ in Eq.~(\ref{eq:nHsym_pH}) (pseudo-anti-Hermiticity), but this does not give any new classes for non-Hermitian Hamiltonians. In the case of pseudo-Hermiticity, Eq.~(\ref{eq:nHsym_pH}), one would normally consider the modified Hamiltonian $\eta H$, which is Hermitian, as $(\eta H)^\dagger =\eta H$. A well-known example of this is the Wilson Dirac operator, where the symmetry is known as
$\Gamma_5$-Hermiticity \cite{Kieburg:2011uf}.
In the Bernard-LeClair (BL) classification scheme~\cite{bernard2002,zhou2019},
Eqs.~(\ref{eq:nHsym_TRS}) and (\ref{eq:nHsym_PHSd}) involve complex conjugation of the Hamiltonian matrix $H$ and are referred to as K symmetries; Eqs.~(\ref{eq:nHsym_TRSd}) and \eref{eq:nHsym_PHS} describe a transposition symmetry of $H$ and are dubbed C symmetries; chiral symmetry, Eq.~(\ref{eq:nHsym_CSd}), is called P symmetry; and pseudo-Hermiticity, Eq.~(\ref{eq:nHsym_pH}), is referred to as Q symmetry in Ref.~\cite{bernard2002}.
As in the Hermitian case, chiral symmetry and pseudo-Hermiticity can be written as a composition of time-reversal and particle-hole symmetries, and they are only non-trivial in the absence of the latter. Furthermore, they can either commute or anti-commute with time-reversal and particle-hole symmetries. 
Carefully accounting for all inequivalent combinations of independent anti-unitary symmetries and their commutation or anti-commutation relations with chiral symmetry and pseudo-Hermiticity gives $38$ non-Hermitian symmetry classes,
of which 23 are pseudo-Hermitian~\cite{bernard2002,magnea2008,zhou2019,kawabata2019}.

For the nHSYK model, we consider the charge-conjugation operators
\cite{you2016,li2017,kanazawa2017,sun2020}
\begin{align}\label{eq:antiunitary_conj}
\sP=K\prod_{i=1}^{N/2}\gamma_{2i-1}
\quad \text{and} \quad
\sR=K\prod_{i=1}^{N/2}\i\gamma_{2i},
\end{align}
where $K$ denotes the complex-conjugation operator, which square to
\begin{align}\label{eq:PR_square}
\sP^2=(-1)^{\frac 12 N/2(N/2-1)}
\quad \text{and} \quad 
\sR^2=(-1)^{\frac 12 N/2(N/2+1)}.
\end{align}
The combination of these two operators yields the Hermitian operator,
\begin{equation}\label{eq:chiral_op}
\sS=\sP\sR=\i^{N^2/4}\prod_{i=1}^{N}\gamma_{i},
\end{equation}
which squares to the identity. This operator is also known as $\Gamma_5$ or the chiral-symmetry operator.  

The operators $\sP$ and $\sR$ act on Majorana fermions as $\sP \gamma_i\sP^{-1}=-(-1)^{N/2}\gamma_i$ and $\sR \gamma_i \sR^{-1}=(-1)^{N/2}\gamma_i$. The complex couplings $J_{i_1\cdots i_q}+\i M_{i_1\cdots i_q}$ are invariant under transposition, and, hence, the nHSYK Hamiltonian has the involutive symmetries
\begin{align}
\label{eq:P_transform_H}
\sP H^\dagger \sP^{-1}&=(-1)^{q(q+1)/2}(-1)^{qN/2}H,
\\
\label{eq:R_transform_H}
\sR H^\dagger \sR^{-1}&=(-1)^{q(q-1)/2}(-1)^{qN/2}H,
\\
\label{eq:S_transform_H}
\sS H \sS^{-1}&=(-1)^q H.
\end{align}
Comparing with Eqs.~(\ref{eq:nHsym_TRS})--(\ref{eq:nHsym_pH}), we see that $\sP$ and $\sR$ play the role of either $\sC_+$ or $\sC_-$, while $\sS$ either commutes or anti-commutes with the Hamiltonian.
For the nHSYK model, the many-body matrix elements are manifestly complex, and no anti-unitary symmetries that map $H$ back to itself exist. Then, only the involutive symmetries~(\ref{eq:nHsym_TRSd})--(\ref{eq:nHsym_CSd}) can occur.\footnote{In the Hermitian SYK model, the couplings are real and thus invariant under complex conjugation. However, in that case, $\sT_\pm$ and $\sC_\pm$ are equivalent to each other.}
The nHSYK model thus belongs to one of the ten BL symmetry classes without reality conditions~\cite{bernard2002}, which are in one-to-one correspondence with the Hermitian Altland-Zirnbauer (AZ) classes~\cite{altland1997} and are summarized in Table~\ref{tab:sym_class}. The difference between the AZ classes and the BL classes without reality conditions is that complex conjugation is replaced by transposition and the Hermiticity constraint is lifted. For example, a Hermitian real Hamiltonian is to be replaced by a non-Hermitian complex symmetric Hamiltonian.
Following the Kawabata-Shiozaki-Ueda-Sato nomenclature~\cite{kawabata2019}, they are dubbed A, AIII$^\dagger$, AI$^\dagger$, AII$^\dagger$, D, C, AI$^\dagger_+$, AII$^\dagger_+$, AI$^\dagger_-$, and AII$^\dagger_-$.\footnote{Although they share the same name under the nomenclature of Ref.~\cite{kawabata2019}, the non-Hermitian symmetry classes A, C, and D are \emph{not} the same as the Hermitian classes A, C, and D.}
Here, we have adopted a shorthand notation, where a subscript in the name of a class indicates that the chiral symmetry is commuting (subscript $+$) or anti-commuting
(subscript $-$) with the time-reversal or particle-hole symmetries. For example, AI$^\dagger_+$ denotes class AI$^\dagger$ with chiral symmetry that commutes with the time-reversal symmetry ($\Pi \sT_+=+\sT_+\Pi$).

{
	\setlength\cellspacetoplimit{0.5ex}
	\setlength\cellspacebottomlimit{0.5ex}
	\begin{table*}[]
	  \caption{Non-Hermitian symmetry classes without reality conditions, nine of which are realized in the non-Hermitian SYK model (i.e., all except class AIII$^\dagger$). For each class, we list its anti-unitary and chiral symmetries, an explicit matrix realization~\cite{magnea2008},
            its name under the Kawabata-Shiozaki-Ueda-Sato classification~\cite{kawabata2019}, and the corresponding
            Hermitian ensemble. In the matrix realizations, $A$, $B$, $C$, and $D$ are arbitrary non-Hermitian matrices unless specified otherwise, and empty entries correspond to zeros. In the last column, we list the AZ classes~\cite{altland1997} that are in one-to-one correspondence with the non-Hermitian classes without reality conditions.
		}
		\label{tab:sym_class}
		\begin{tabular}{ Sc Sc Sc Sc Sl Sl@{}}
			\toprule
			$\sC_+^2$     & $\sC_-^2$  & $\sS^2$ & Matrix realization & Class   & Hermitian corresp.     \\ \midrule
			$0$           & $0$        & $0$     & $A$                & A               & GUE (A)         \\
			$0$           & $0$        & $1$     & $\matAIIId$        & AIII$^\dagger$  & chGUE (AIII)    \\
			$+1$          & $0$        & $0$     & $A=A^\top$         & AI$^\dagger$    & GOE (AI)        \\
			$-1$          & $0$        & $0$     & $\matAIId$         & AII$^\dagger$   & GSE (AII)       \\
			$0$           & $+1$       & $0$     & $A=-A^\top$        & D               & BdG-S (D)       \\
			$0$           & $-1$       & $0$     & $\matC$            & C               & BdG-A (C)       \\
			$+1$          & $+1$       & $1$     & $\matAIdp$         & AI$^\dagger_+$  & chGOE (BDI)     \\
			$-1$          & $-1$       & $1$     & $\matAIIdp$        & AII$^\dagger_+$ & chGSE (CII)     \\
			$+1$          & $-1$       & $1$     & $\matAIdm$         & AI$^\dagger_-$  & chBdG-S (CI)    \\
			$-1$          & $+1$       & $1$     & $\matAIIdm$        & AII$^\dagger_-$ & chBdG-A (DIII)  \\ \bottomrule
		\end{tabular}
	\end{table*}
}

The nHSYK symmetry classification can be performed systematically by evaluating Eqs.~(\ref{eq:PR_square}) and (\ref{eq:P_transform_H})--(\ref{eq:S_transform_H}) for different values of $q\mod4$ and $N\mod8$. Note that while the physical interpretation of the operators $\sP$, $\sR$, and $\sS$ is different in the SYK and nHSYK models, the defining relations in Eqs.~(\ref{eq:PR_square}) and (\ref{eq:P_transform_H})--(\ref{eq:S_transform_H}) are formally the same. It follows that the symmetry classification of the former~\cite{you2016,garcia2016,cotler2016,li2017,kanazawa2017,garcia2018a,behrends2019,sun2020} also holds for the latter, provided that one replaces any reality condition by a transposition one. We now investigate in more detail the dependence of these symmetries of the odd or even nature of $q$ in the nHSYK Hamiltonian Eq.~(\ref{hami}).

\paragraph*{Even $q$.}
According to Eq.~(\ref{eq:S_transform_H}), $H$ commutes with $\sS$ (which is proportional to the fermion parity operator), the Hilbert space is split into sectors of conserved even and odd parity, and the Hamiltonian is block-diagonal. There is no chiral symmetry. From Eqs.~(\ref{eq:P_transform_H}) and (\ref{eq:R_transform_H}), we see that $H$ transforms similarly under both $\sP$ and $\sR$ (when they act within the same block) and it suffices to consider the action of one, say $\sP$. We have the
commutation relation
\be
\sS\sP =(-1)^{\frac N2} \sP\sS.
\ee
\begin{itemize}
\item When $N\mod8=2,6$, $\sP$ is a fermionic operator that anti-commutes with $\sS$. Note that $\sP$ is not an involutive symmetry of the Hamiltonian in a diagonal block representation, as it maps blocks of different parity into each other. The two blocks are the transpose of each other and have no further constraints (class A or complex Ginibre).

\item When $N\mod8=0,4$ and $q\mod4=0$, $\sP$ is a bosonic operator that commutes with $\sS$. Each block of the Hamiltonian has the involutive symmetry, $\sP H^\dagger \sP^{-1}=+H$. If $N\mod8=0$, $\sP^2=1$, and we
can find a basis in which the Hamiltonian is symmetric. This is the universality class of complex symmetric matrices, also known as AI$^\dagger$. 
If $N\mod8=4$, $\sP^2=-1$, we can find a basis in which $H^\top = I H I^{-1} $, with $I$ the symplectic unit matrix. This class is AII$^\dagger$.

\item When $N\mod8=0,4$ and $q\mod4=2$, $\sP$ is again a bosonic
operator that commutes with $\sS$. Within each block, we have the involutive symmetry, $\sP H^\dagger \sP^{-1}=-H$. If $N\mod8=0$, $\sP^2=1$, we can find a basis in which the Hamiltonian becomes anti-symmetric and the universality class is given by that of complex anti-symmetric matrices (non-Hermitian class D); if $N\mod8=4$, $\sP^2=-1$, and we can find a basis where $H^\top =-I H I^{-1}$. Complex matrices satisfying this constraint belong to non-Hermitian class C.
\end{itemize}

\paragraph*{Odd $q$.}

In this case, $\sS$ is a chiral symmetry operator that anti-commutes with $H$,
so that $H$
acquires an off-diagonal block structure in a chiral basis.
The operators $\sP$ and $\sR$ now act differently on $H$ (one
satisfies $X H^\dagger X^{-1} = H $ and the other $X H^\dagger X^{-1} = -H$ with
$X$ either $\sP$ or $\sR$). 
Hence both must be considered if we use
the square of $\sP$, $\sR$, and $\sS$ to classify the matrices. However, for the derivation of the block structure, as given in Table~\ref{tab:sym_class}, we of course only need either
$\sP$ or $\sR$, and $\sS$.
\begin{itemize}
\item When $N\mod8=0$, both $\sP$ and $\sR$ are bosonic operators squaring to $+1$. Since they commute with $\sS$, and one of them satisfies $X H^\dagger X^{-1} = H$ ($\sP$ if $q\mod4=3$, $\sR$ if $q\mod4=1$), the Hamiltonian in a suitable basis is a complex matrix with vanishing diagonal blocks and off-diagonal blocks that are the transpose of each other (class AI$^\dagger_+$), irrespective of whether $q\mod4=1$ or $3$.

\item When $N\mod8=4$, both $\sP$ and $\sR$ are bosonic operators commuting with $\sS$ and squaring to $-1$. Hence the off-diagonal blocks $A$ and $B$ of $H$ are related by
$B^\top = I A I^{-1}$. This is the universality class AII$^\dagger_+$, irrespective of whether $q\mod4=1$ or $3$.

\item When $N\mod8=2,6$ we have that $\{\sS,\sP\}=0$ and  $\{\sS,\sR\}=0$. So $\sP$ and $\sR$ have vanishing diagonal blocks, and depending on $N$, $\sP^2=1$ and $\sR^2=-1$ ($N\mod8=2$), or $\sP^2=-1$ and $\sR^2=1$ ($N\mod8=6$). We choose the operator that squares to $1$. Since it anti-commutes with $\sS$, it has the block structure
  \be
  X=\bmat 0 & x^{-1} \\ x & 0 \emat.
  \ee
If $X H^\dagger X^{-1} =H$ the blocks of $H$ satisfy $ x A^\top x^{-1} = A$ and
$ x B^\top x^{-1} = B$, so we can find a basis in which the blocks are symmetric.
This is the case for $ N \mod 8=2$ and $q\mod4 =1$ or $ N \mod 8=6$ and $q\mod4 =3$
(class AI$^\dagger_-$).
If $X H^\dagger X^{-1} = -H$ the blocks of $H$ satisfy $ x A^\top x^{-1} = -A$ and
$ x B^\top x^{-1} =- B$, so we can find a basis in which the blocks are skew-symmetric. This is the case for $ N \mod 8=2$ and $q\mod4 =3$ or $N \mod 8=6$ and $q\mod4 =1$ (class AII$^\dagger_-$).
\end{itemize}

The complete symmetry classification of the nHSYK Hamiltonian
for all $q$ and even $N$ in terms of
BL classes
is summarized in Table~\ref{tab:nHSYK_class}. Note that the tenth symmetry class AIII$^\dagger$ is not realized by the nHSYK Hamiltonian Eq.~(\ref{hami}).\footnote{It is, however, realized in the chiral nHSYK model; see Sec.~\ref{sec:nHWSYK}.}
We note that one could also view the
non-Hermitian structures as complexified real structures relating them
to the classification in terms of symmetric spaces; see Ref.~\cite{Zirnbauer:2010gg} for related remarks.

It is important to stress that, since the symmetry classification is algebraic, it is not necessarily related 
to specific features of spectral correlations that probe the quantum dynamics for long timescales of the order of the Heisenberg time. 
However, we see in the next two sections that this is the case. 
By studying local bulk and hard-edge spectral correlations for different values of $q$ and $N$, we find not only excellent agreement with the predictions of non-Hermitian RMT for $q > 2$ but also that the different universality classes resulting from the symmetry classification can be characterized by the analysis of level statistics. Assuming that the relation between RMT level statistics and quantum chaos persists for non-Hermitian Hamiltonians, our results provide direct evidence that the nHSYK model is versatile enough to describe all possible quantum ergodic states to which quantum chaotic systems with a complex spectrum relax after a sufficiently long time.   

 \begin{table}[]
	\caption{Complete symmetry classification of the nHSYK Hamiltonian into BL classes without reality conditions for all $q$ and even $N$.}
	\label{tab:nHSYK_class}
	\begin{tabular}{@{}l cccc@{}}
		\toprule
		$N\,\mathrm{mod}\,8$   & 0              & 2               & 4               & 6               \\ \midrule
		$q\,\mathrm{mod}\,4=0$ & AI$^\dagger$   & A               & AII$^\dagger$   & A               \\
		$q\,\mathrm{mod}\,4=1$ & AI$^\dagger_+$ & AI$^\dagger_-$  & AII$^\dagger_+$ & AII$^\dagger_-$ \\
		$q\,\mathrm{mod}\,4=2$ & D              & A               & C               & A               \\
		$q\,\mathrm{mod}\,4=3$ & AI$^\dagger_+$ & AII$^\dagger_-$ & AII$^\dagger_+$ & AI$^\dagger_-$  \\ \bottomrule
	\end{tabular}
\end{table}

\section{Level statistics: complex spacing ratios}
We initiate our analysis of spectral correlations by studying spacing ratios, also called adjacent gap ratios,which are a spectral observable that has the advantage of not requiring unfolding. It was introduced to describe short-range spectral correlations of real spectra~\cite{oganesyan2007,atas2016,brody1981} but it has recently been generalized~\cite{sa2020} to complex spectra. Because of their short-range nature, these ratios probe the quantum dynamics for late timescales of the order of the Heisenberg time. However, as was mentioned previously, a full dynamical characterization of level statistics, equivalent to the BGS conjecture~\cite{bohigas1984}, for non-Hermitian systems is still missing. Therefore, strictly speaking, we perform a comparison with the predictions of non-Hermitian random matrix theory, implicitly assuming that good agreement still implies quantum chaotic motion.

\begin{figure}
	% 	 \subfigure[]{denzgt9k1q3hm20000f.pdf
	\includegraphics[width=8cm]{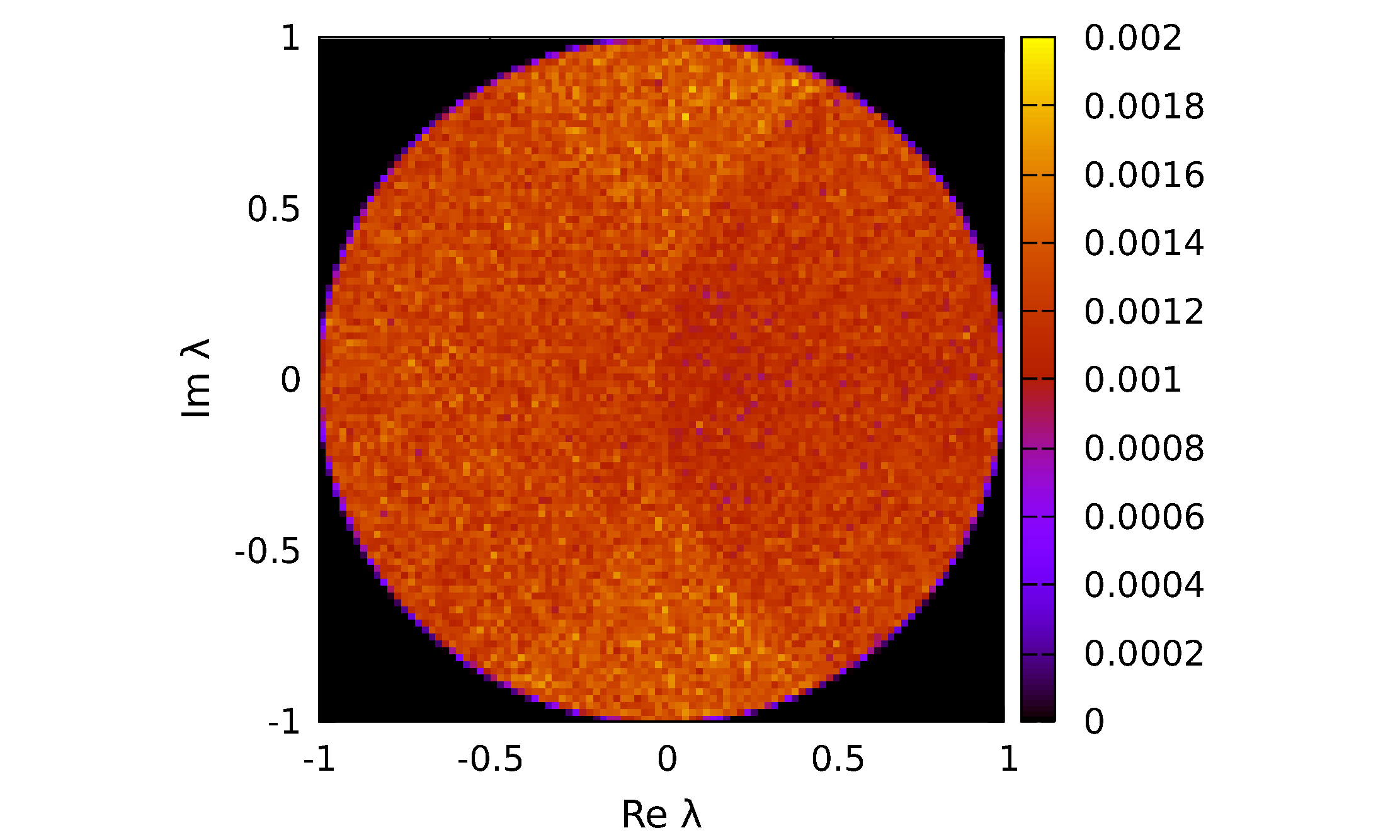}
	% 	\subfigure[]{ 
	\includegraphics[width=8cm]{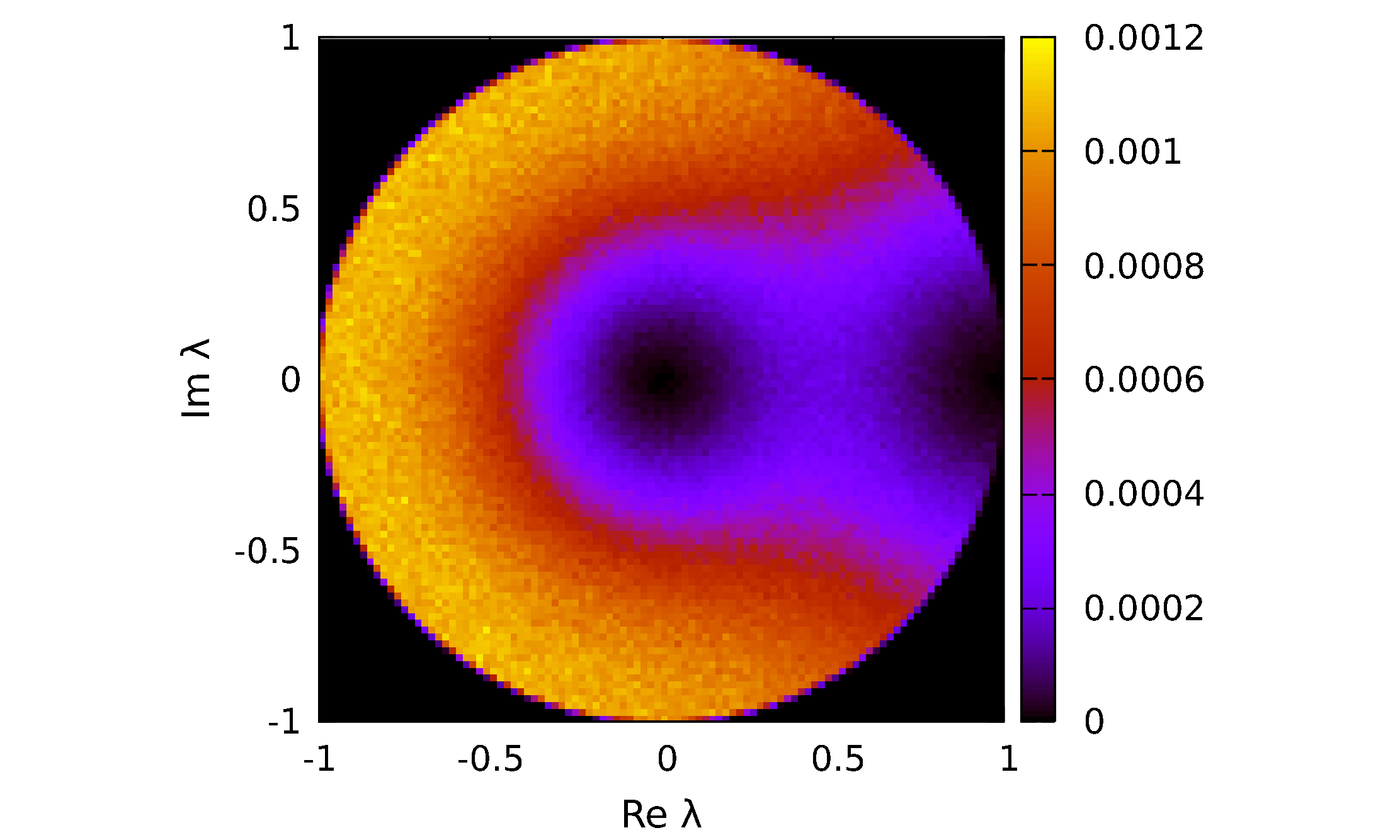}
	\includegraphics[width=8cm]{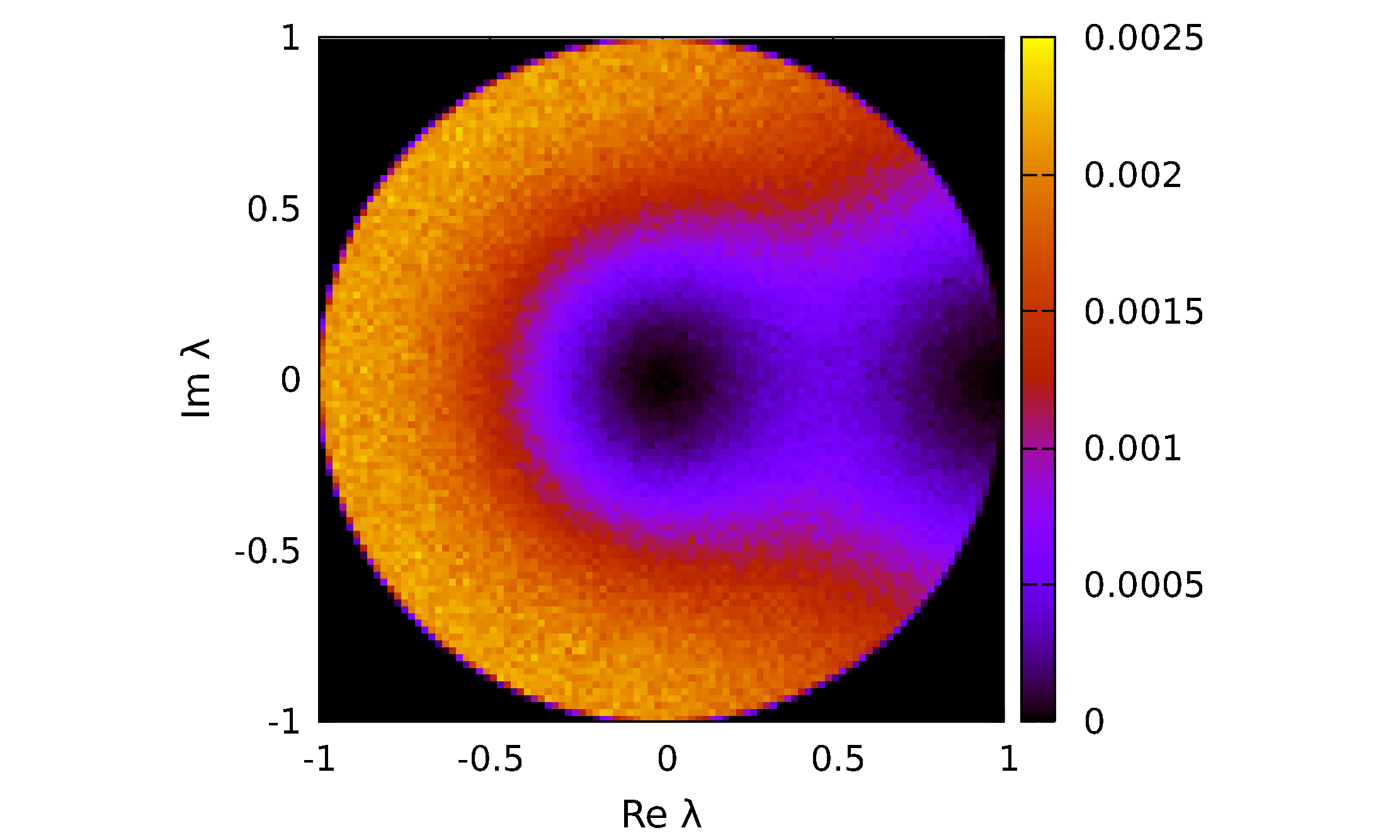}
	\includegraphics[width=8cm]{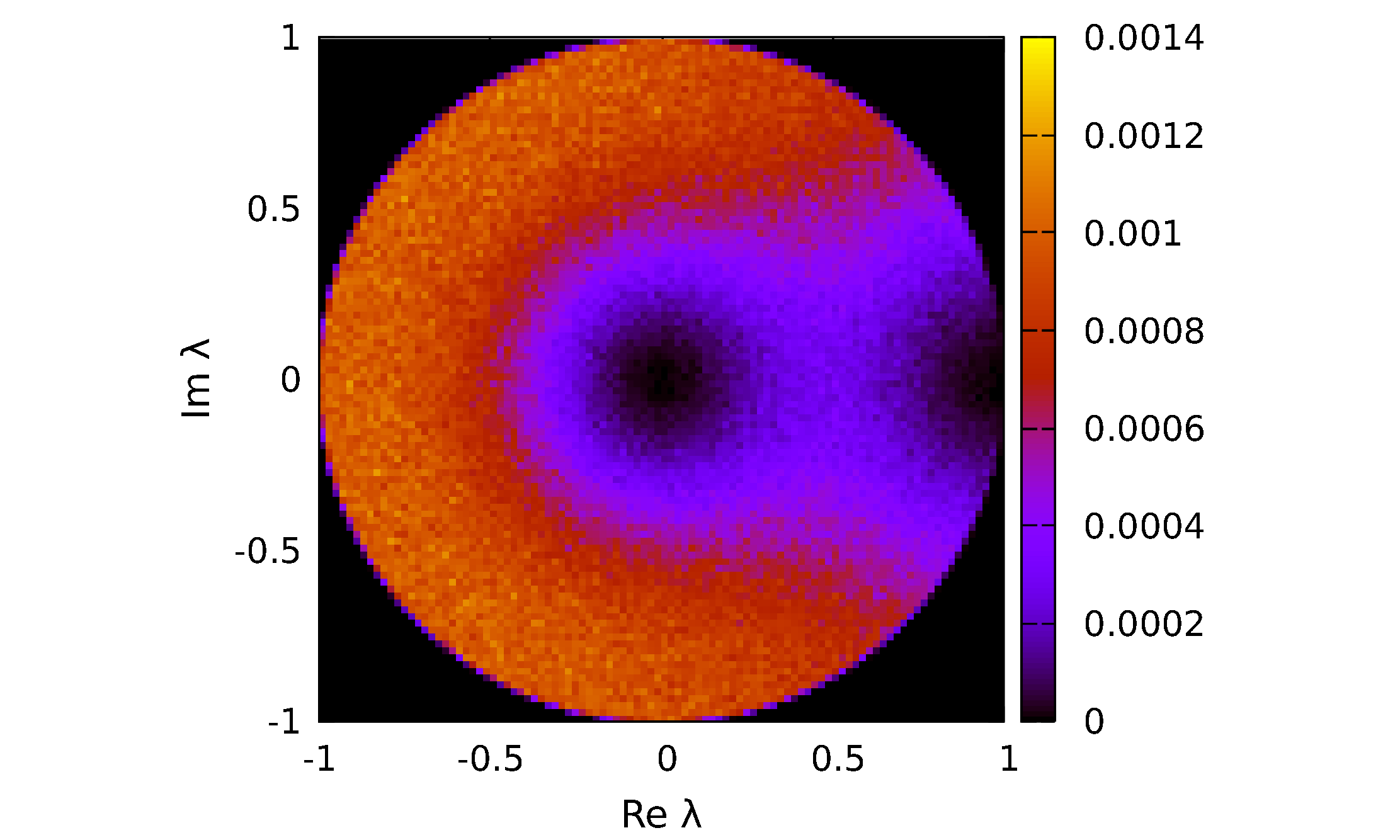}
	\caption{Distribution of complex spacing ratios, Eq.~(\ref{eq:cspacing}), for $N=20$ and $q=2$ (top-left panel), $3$ (top-right panel), $4$ (bottom-left panel), and $6$ (bottom-right panel). Suppression of the complex spacing ratio density around the origin and along the real positive semi-axis is observed for $q>2$. This is also the random matrix theory prediction~\cite{sa2020}, which should indicate quantum chaotic dynamics.
	}
	\label{fig:denzadjN20}
\end{figure}

\begin{figure}
	% 	 \subfigure[]{denzgt9k1q3hm20000f.pdf
	\includegraphics[width=8cm]{./denzadjgt12k1q2nhSYKhm3500s1.pdf}
	% 	\subfigure[]{ 
	\includegraphics[width=8cm]{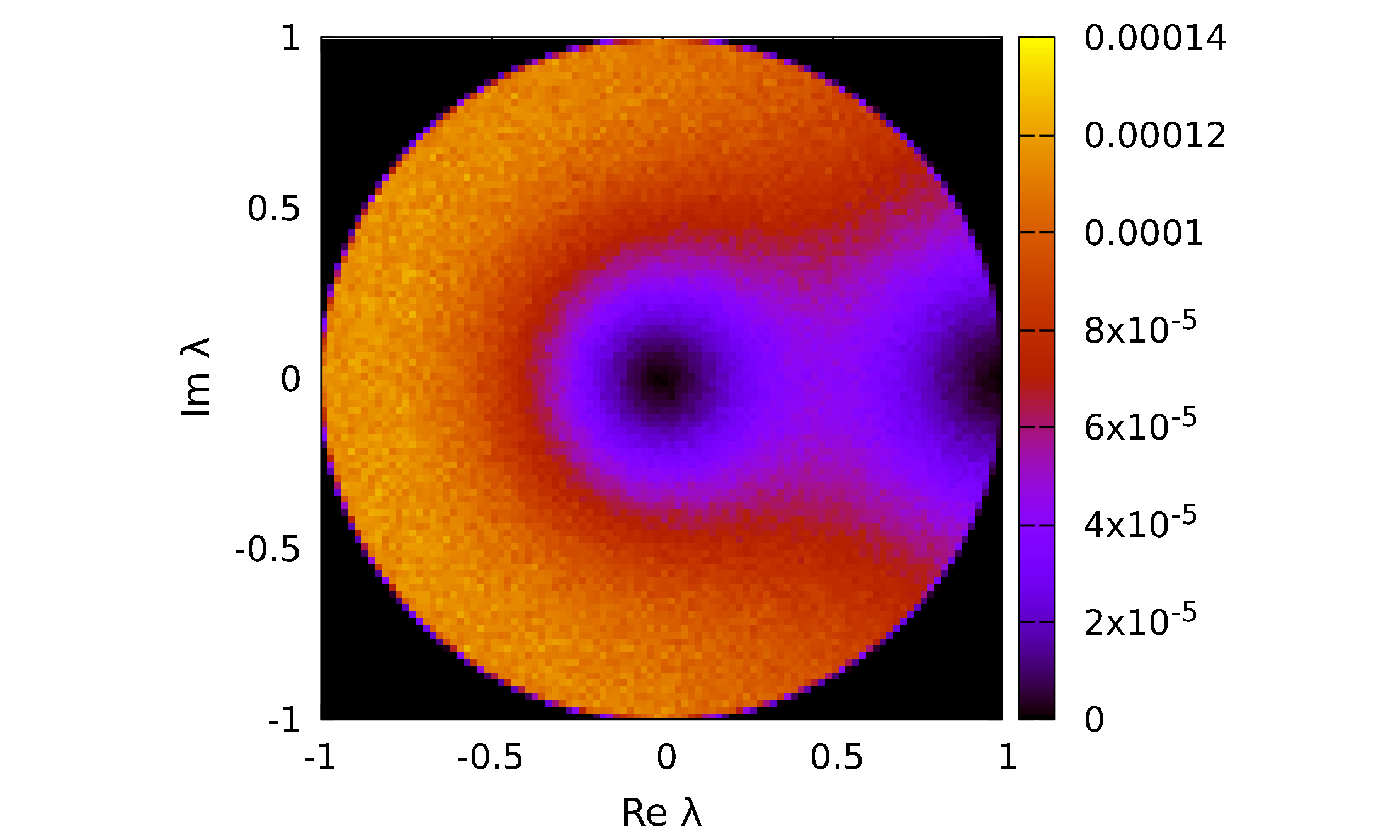}
	\includegraphics[width=8cm]{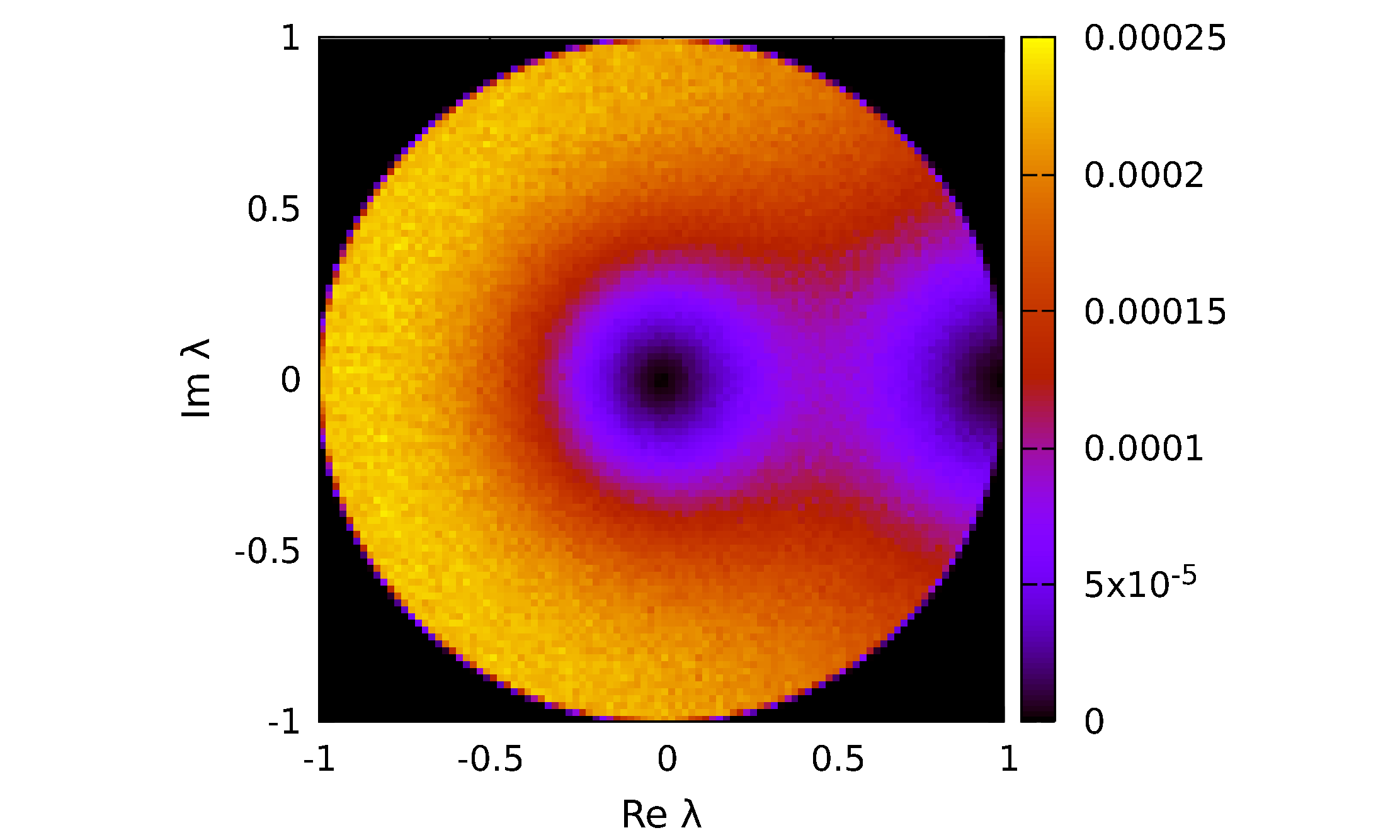}
	\includegraphics[width=8cm]{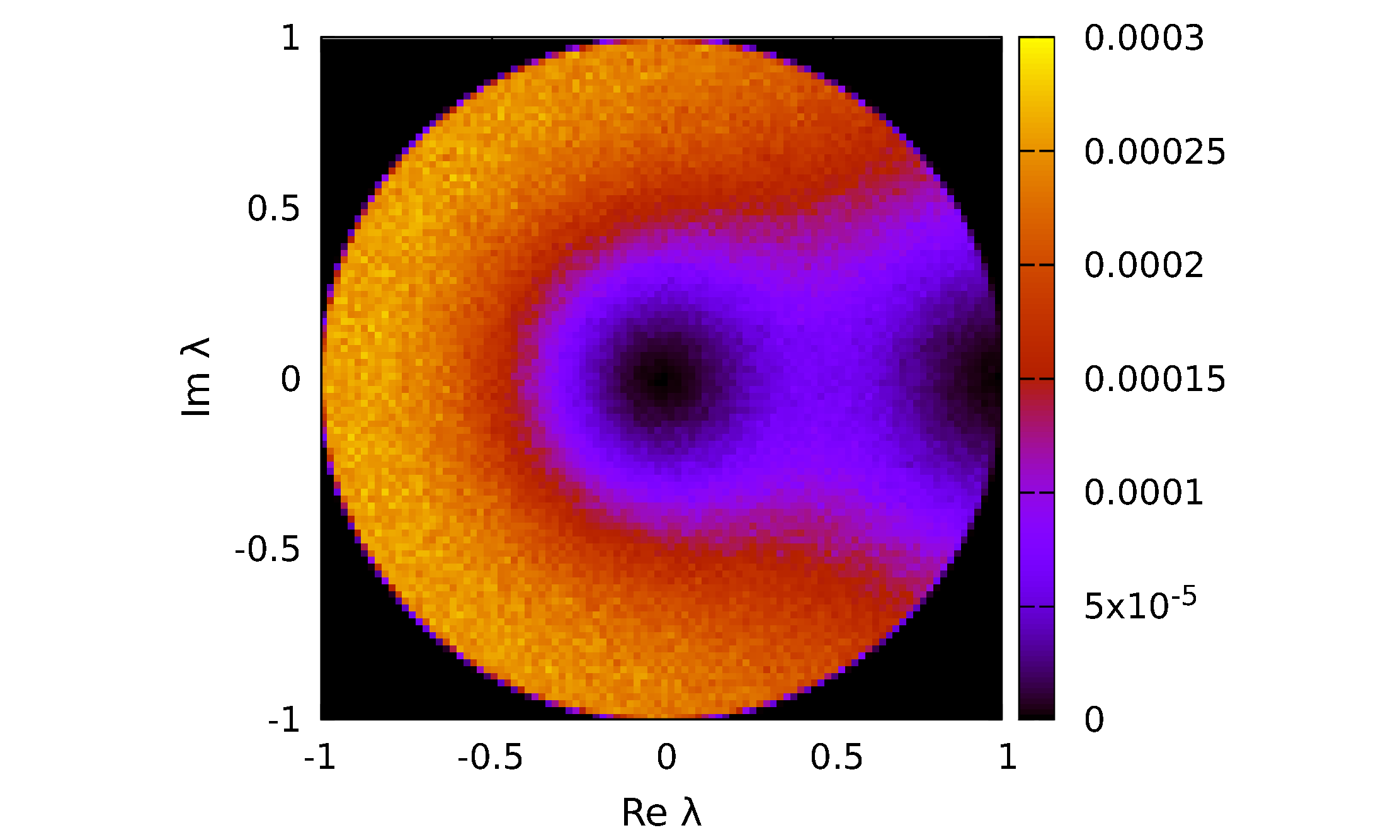}
	\caption{Distribution of complex spacing ratios, Eq.~(\ref{eq:cspacing}), for $N=24$ and $q=2$ (top-left panel), $3$ (top-right panel), $4$ (bottom-left panel), and $6$ (bottom-right panel). The results are qualitatively similar to those for $N =20$. 
	}
	\label{fig:denzadjN24}
\end{figure}

\begin{figure}
	% 	 \subfigure[]{denzgt9k1q3hm20000f.pdf
	\includegraphics[width=8cm]{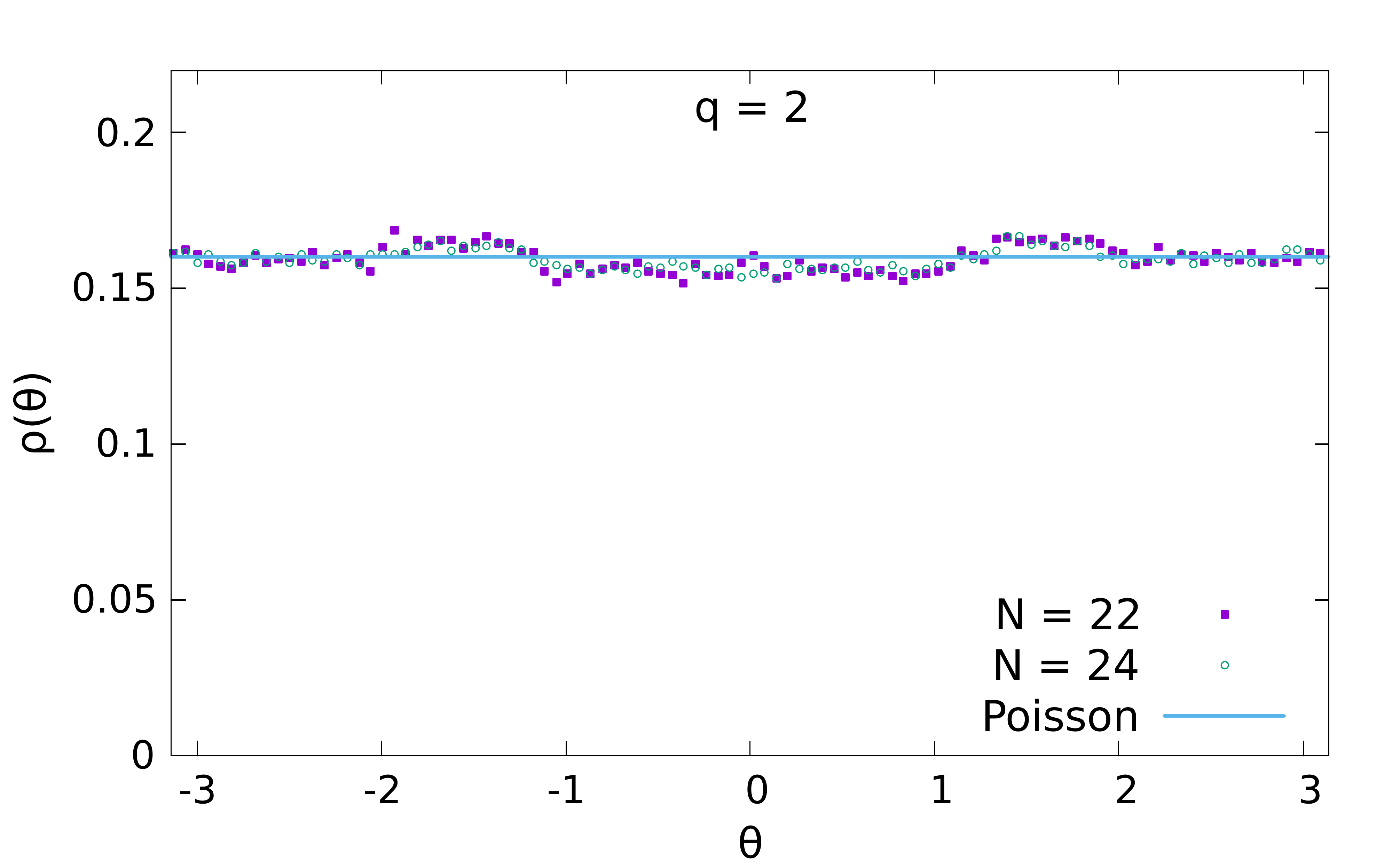}
	% 	\subfigure[]{ 
	\includegraphics[width=8cm]{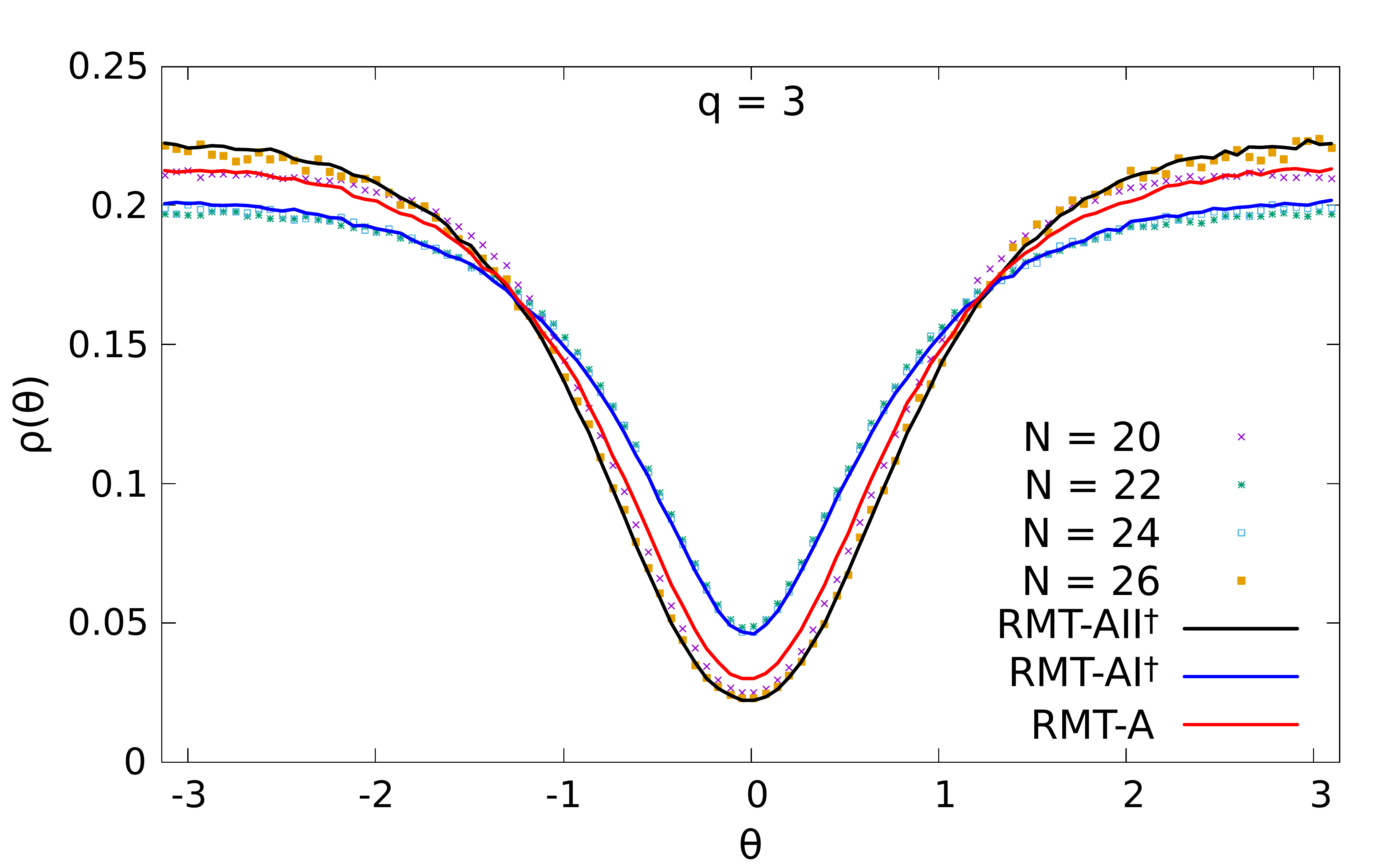}
	\includegraphics[width=8cm]{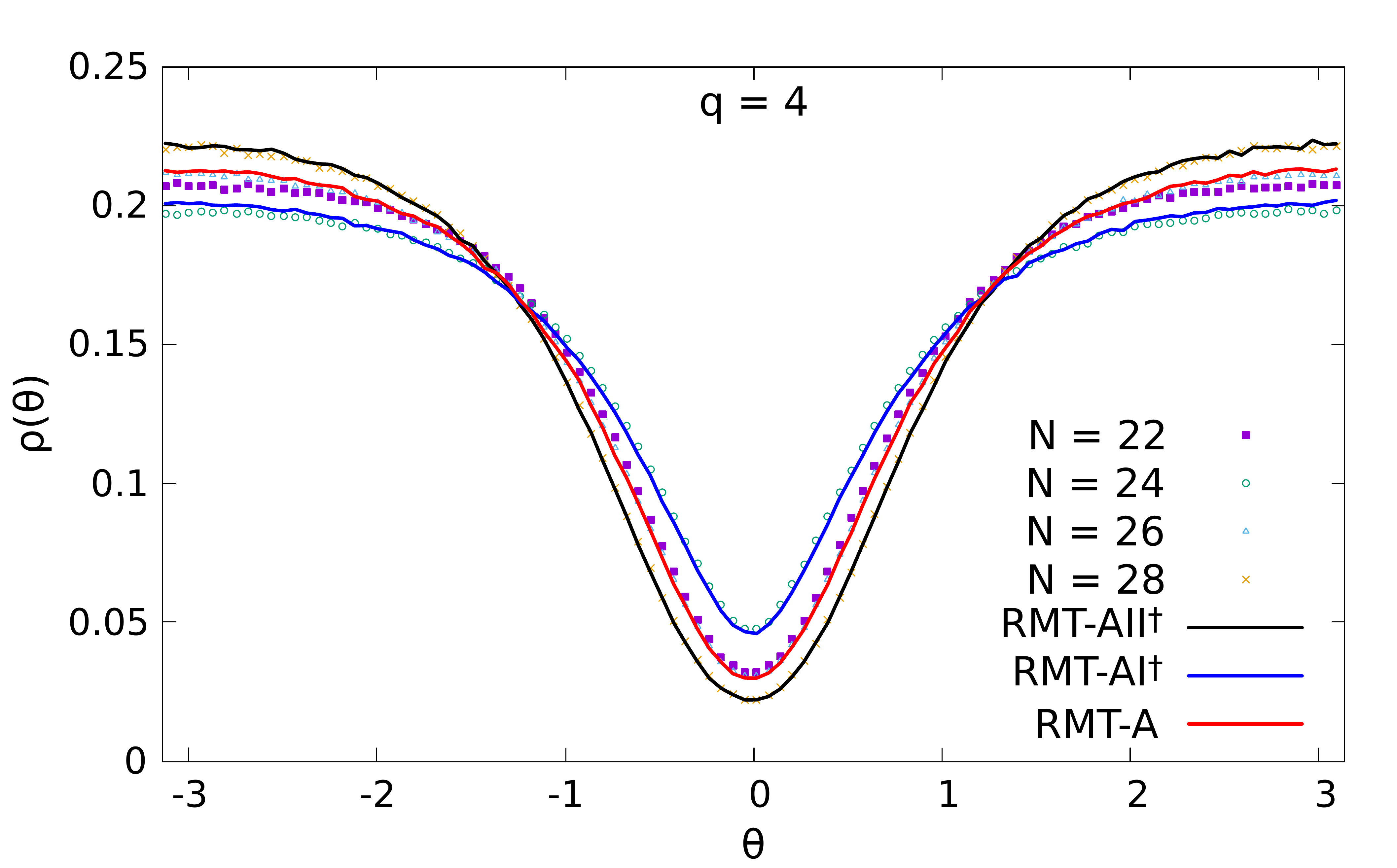}
	\includegraphics[width=8cm]{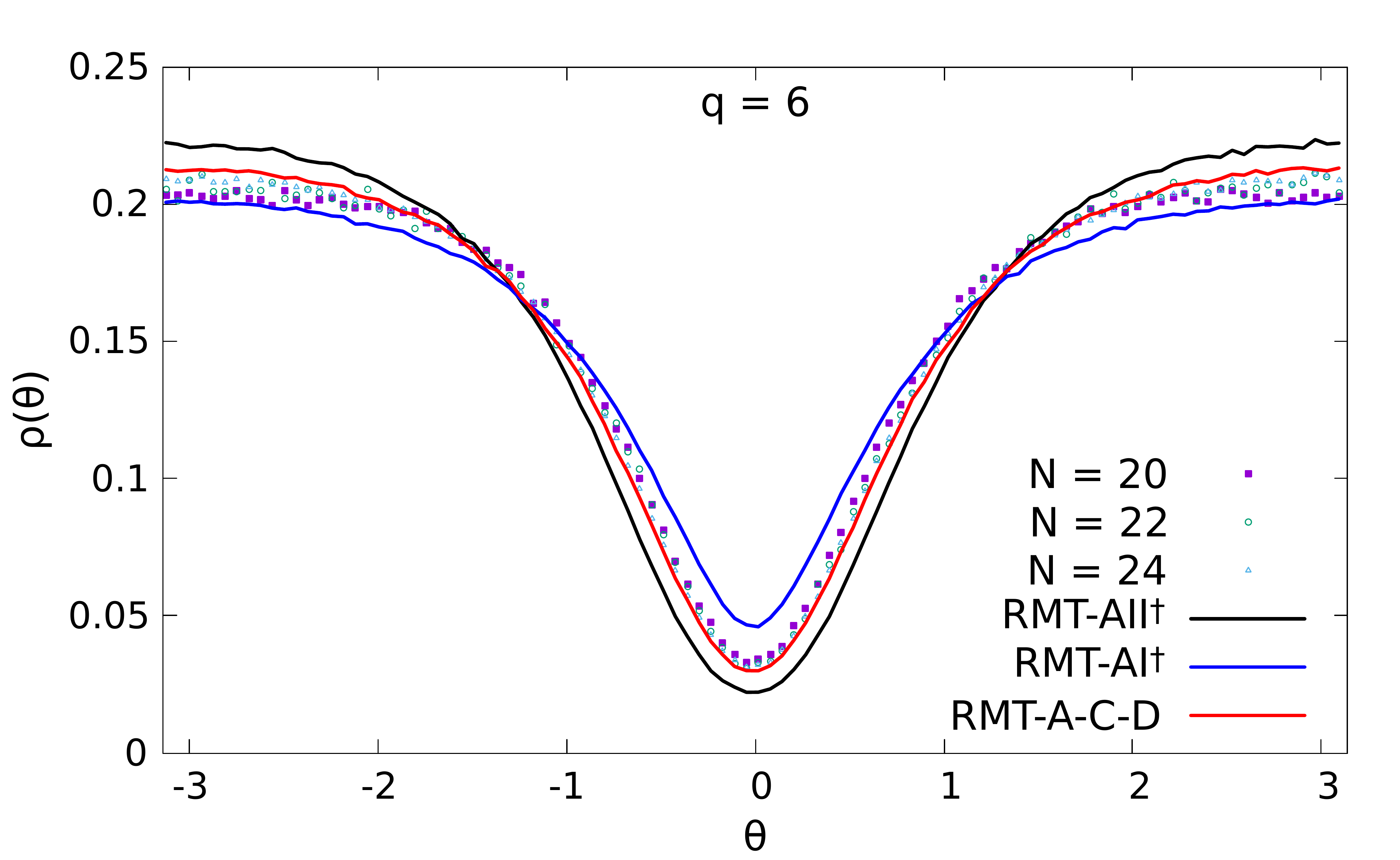}
	\caption{Angular density of the complex spacing ratio related to the eigenvalues of the nHSYK Hamiltonian Eq.~(\ref{eq:cspacing}) for different values of $N$ and $q$. We see agreement with the predictions of Table~\ref{tab:nHSYK_class} in all cases.
   	\label{fig:denadjtheta}}
\end{figure}

\begin{figure}
	% 	 \subfigure[]{denzgt9k1q3hm20000f.pdf
	\includegraphics[width=8cm]{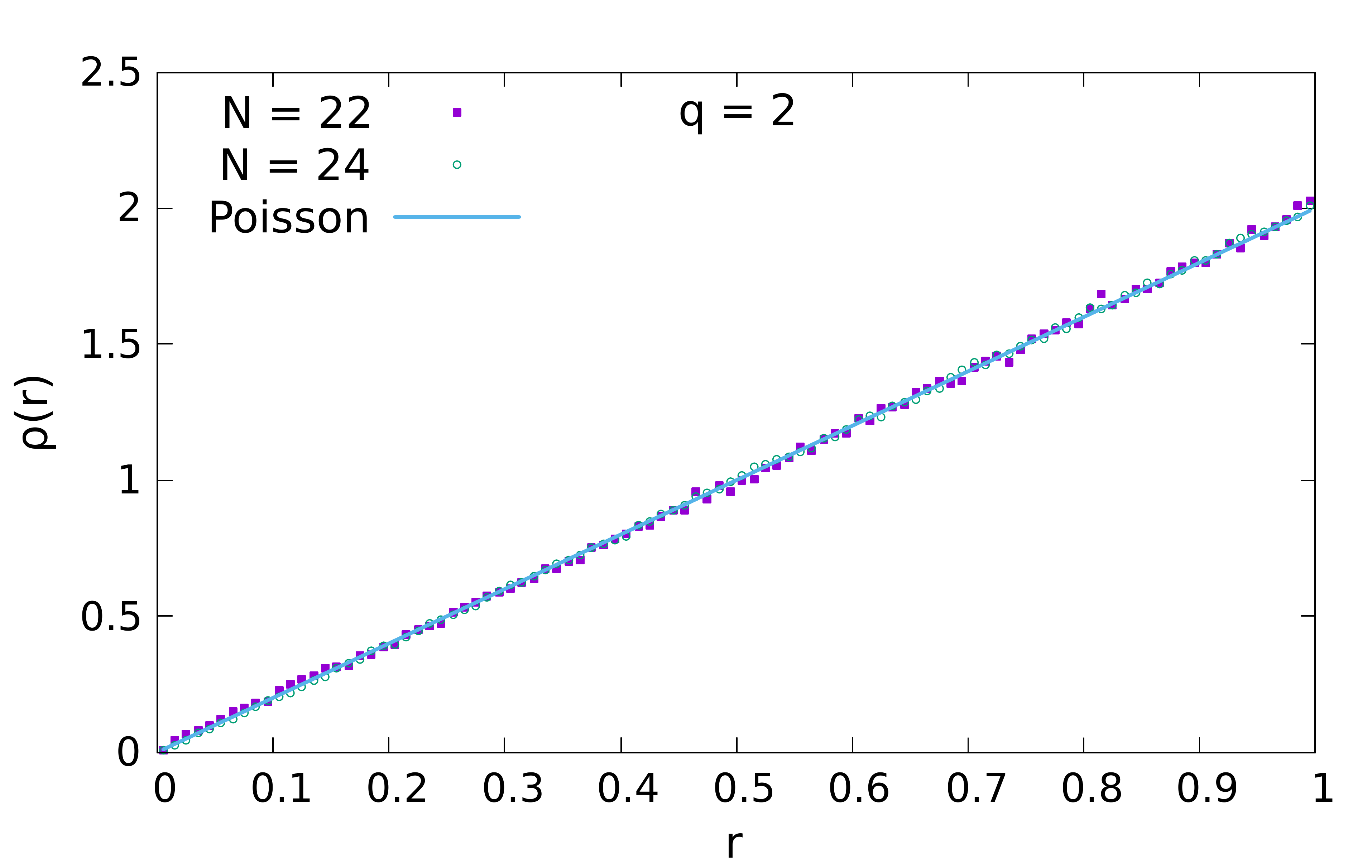}
	% 	\subfigure[]{ 
	\includegraphics[width=8cm]{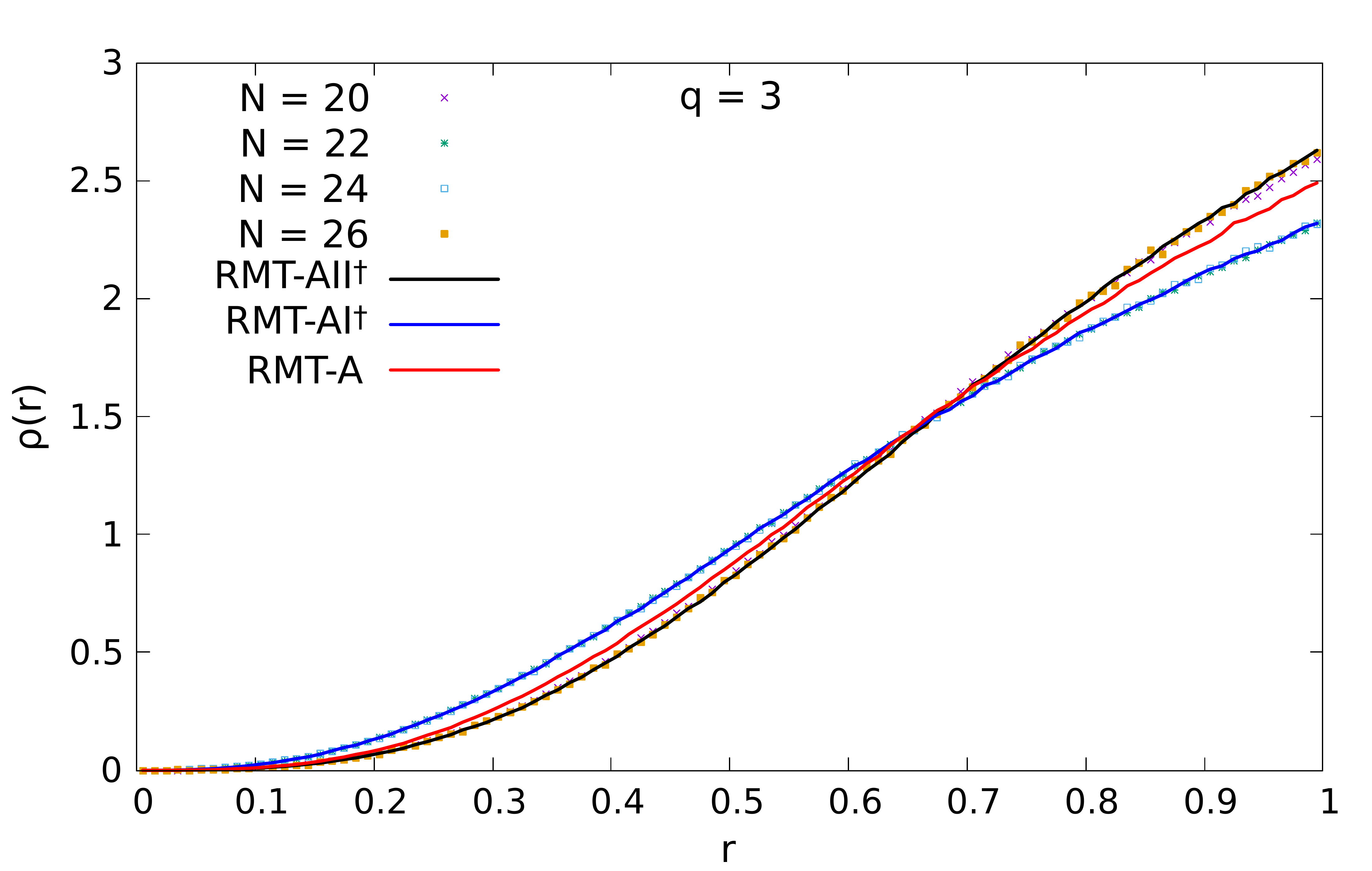}
	\includegraphics[width=8cm]{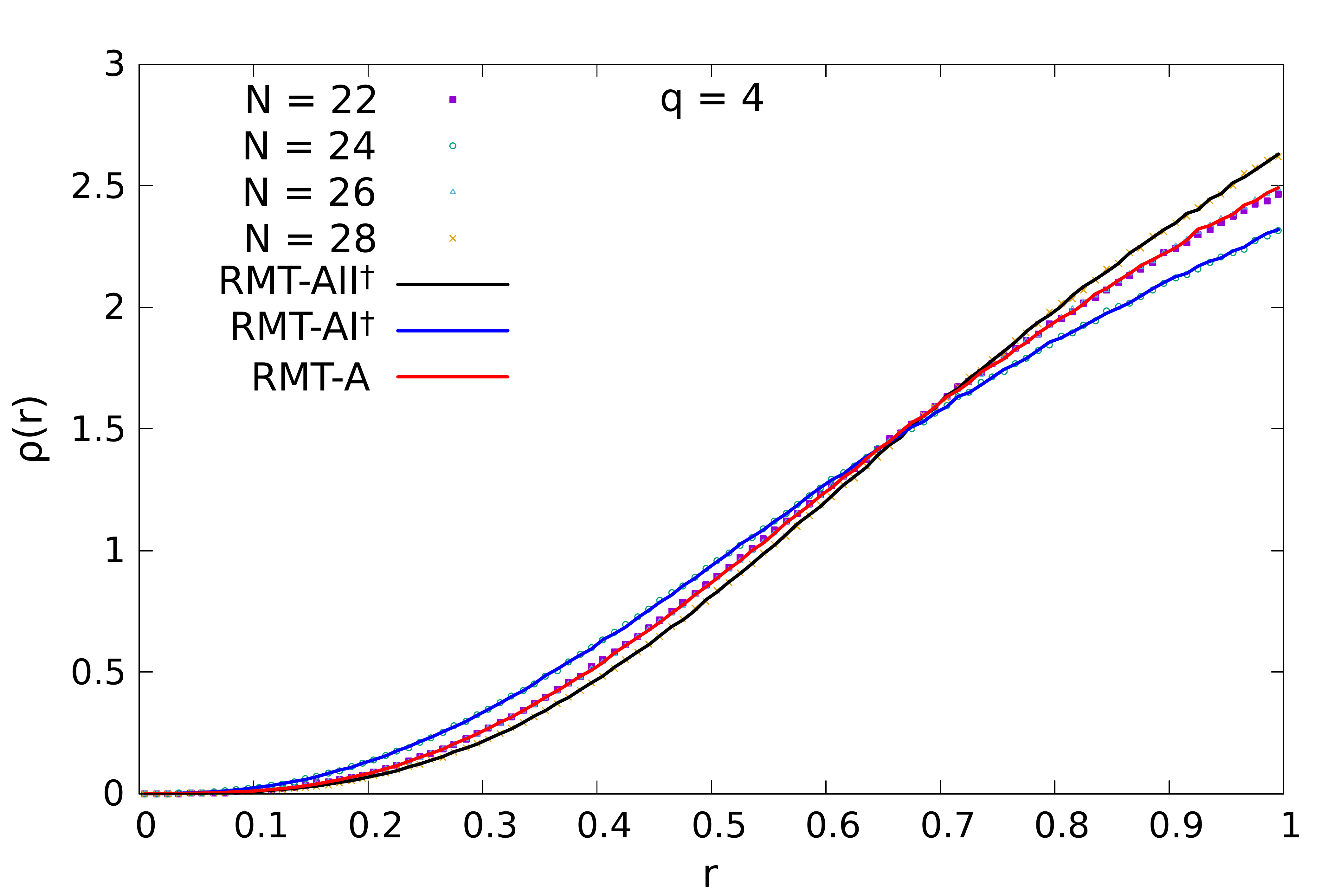}
	\includegraphics[width=8cm]{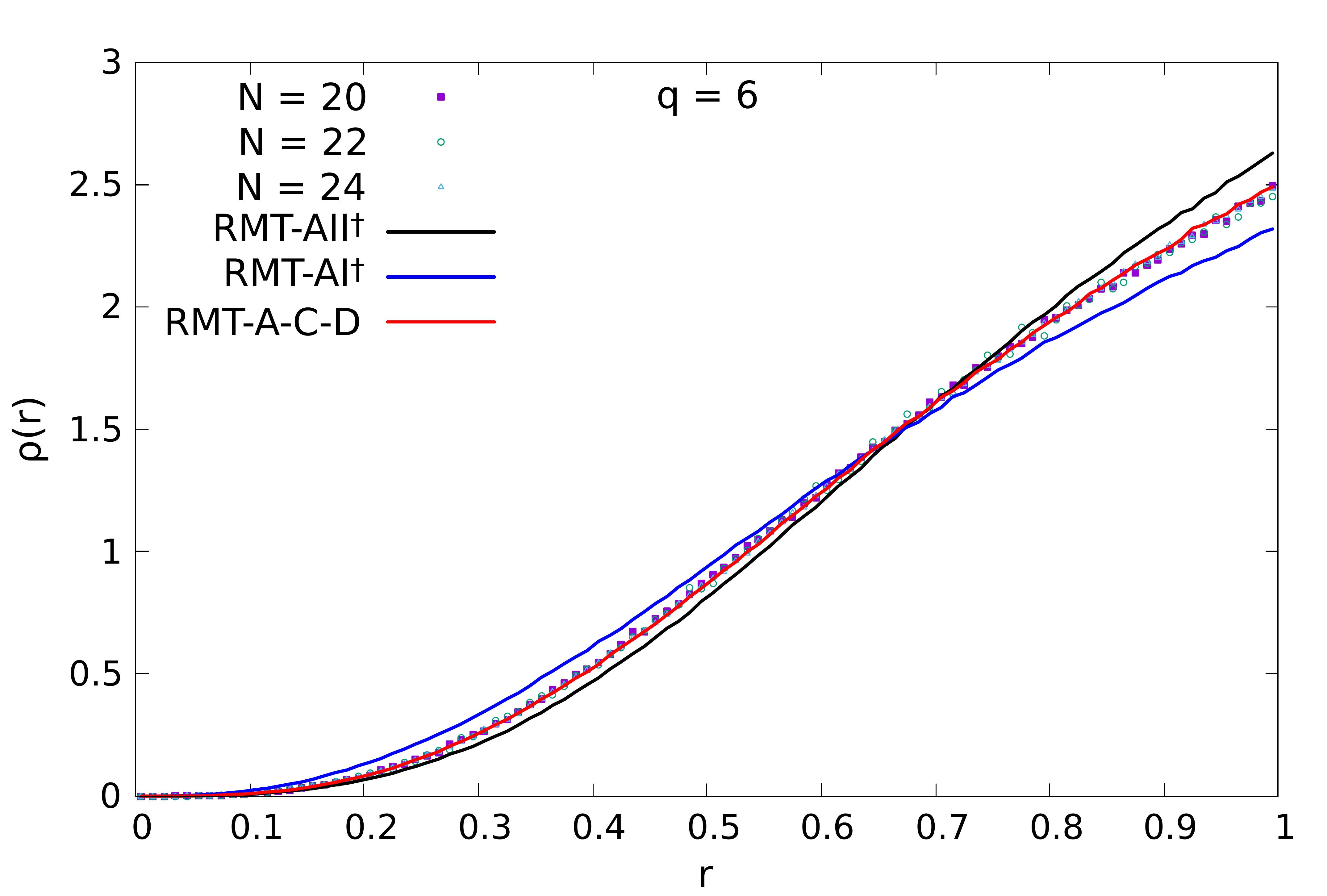}
	\caption{Radial density of the complex spacing ratio related to the eigenvalues of the nHSYK Hamiltonian Eq.~(\ref{eq:cspacing}) for different values of $N$ and $q$. We see agreement with the predictions of Table~\ref{tab:nHSYK_class} in all cases.
         \label{fig:denadjrho}}
\end{figure}

\begin{figure}[tbp]
	\centering
	\includegraphics[width=0.45\textwidth]{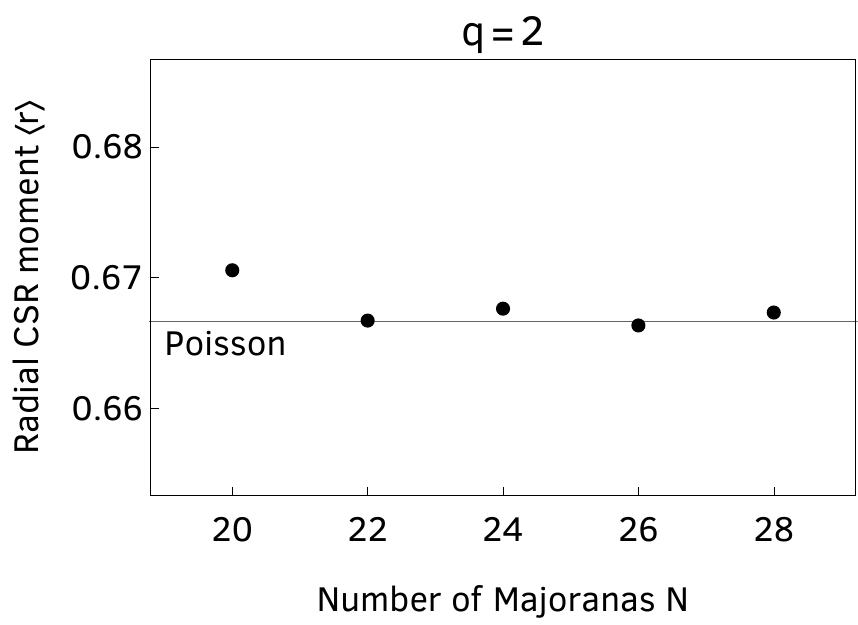}
	\includegraphics[width=0.45\textwidth]{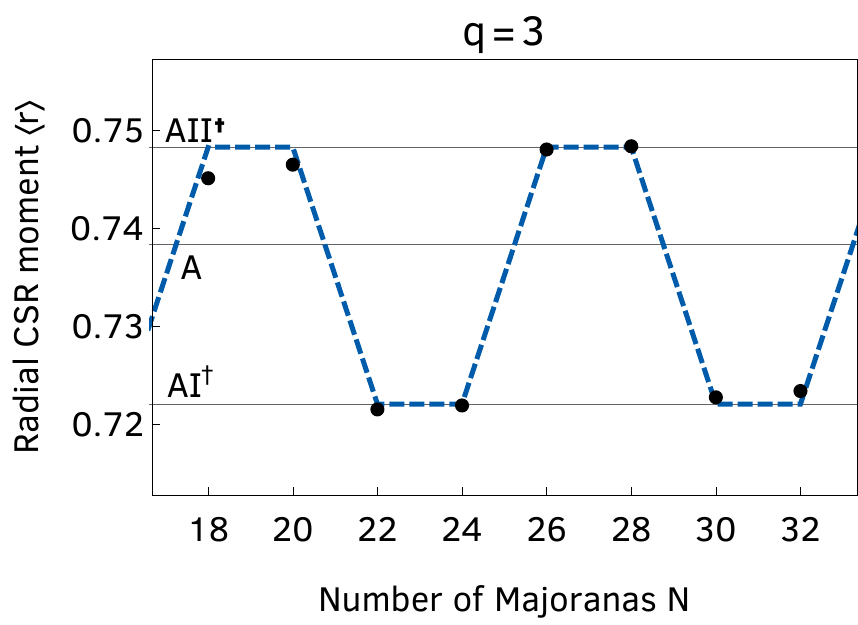}
	\includegraphics[width=0.45\textwidth]{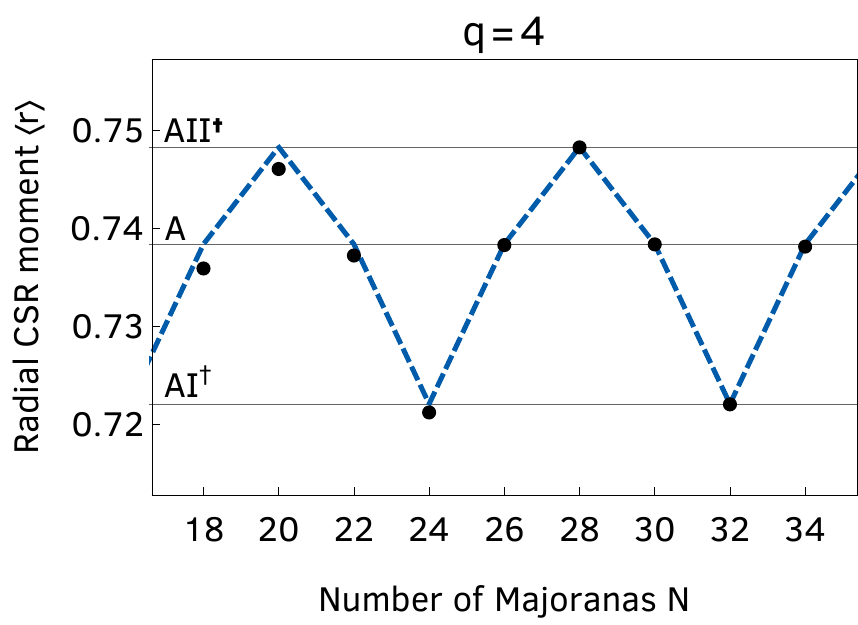}
	\includegraphics[width=0.45\textwidth]{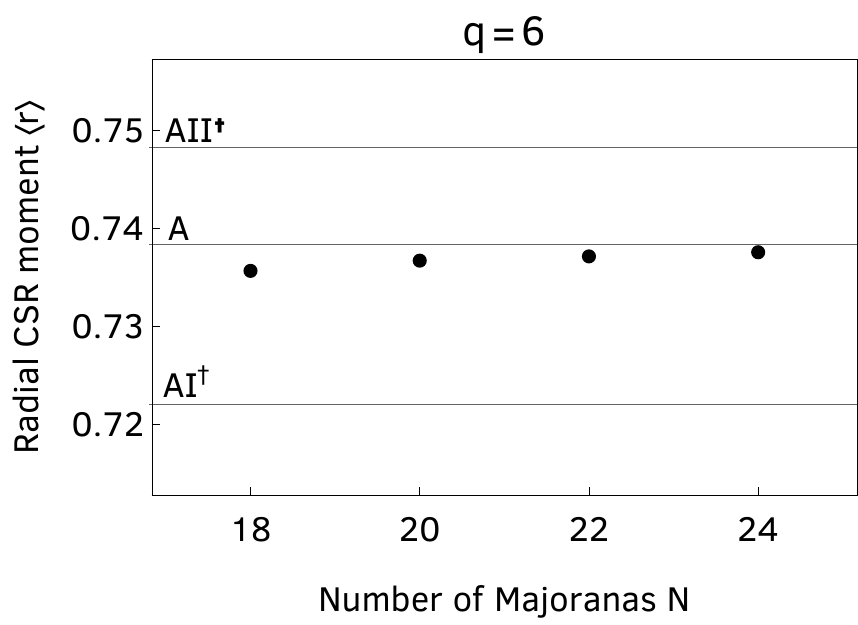}
	\caption{First radial moment $\av{r}$ of the complex spacing ratio (CSR)
          distribution as a function of the number of Majoranas, $N$, for the $q=2,3,4,6$ nHSYK model. The dots are the results of exact numerical diagonalization, the horizontal solid lines are the values of $\av{r}$ for the three universal bulk statistics (A, AI$^\dagger$, and AII$^\dagger$), and the dashed curve follows the classification scheme of Table~\ref{tab:nHSYK_class}. For $q=2$, $\av{r}$ goes to the Poisson value $2/3$ as $N$ increases, showing that no RMT correlations exist around the Heisenberg time. For $q=6$, the three classes
          realized for different values of $N$ all have the same bulk correlations (those of class A). We see excellent agreement between the nHSYK results and the RMT predictions, except for the smaller values of $N$, where finite-size effects are more pronounced.}
	\label{fig:CSRavr}
\end{figure}

We define the complex spacing ratio as
\begin{equation}\label{eq:cspacing}
	\lambda_k=\frac{E_k^\mathrm{NN}-E_k}{E_k^\mathrm{NNN}-E_k}.
\end{equation}
where $E_k$ with $k = 1,2,\ldots, 2^{N/2}$ 
is the complex spectrum for a given disorder realization, $E_k^\mathrm{NN}$ is the closest eigenvalue to $E_k$ using the standard distance in the complex plane, and $E_k^\mathrm{NNN}$ is the second closest eigenvalue. By construction, $|\lambda_k| \leq 1$, so it is restricted to the unit disk.
This is the most natural generalization of the real spacing ratio to the complex case. We perform ensemble averaging, such that we have a minimum of $10^6$ eigenvalues for each set of parameters $N,q$. 
The distribution of the resulting averaged complex spacing ratio $\lambda_k$ is depicted in Figs.~\ref{fig:denzadjN20}~and~\ref{fig:denzadjN24}. We observe qualitative differences between $q=2$ and $q>2$. 
For the former, it is rather unstructured (i.e., flat) with no clear signature of level repulsion for small spacing. In the real case, this is a signature of the absence of quantum chaos. By contrast, for $q>2$, the complex spacing ratio is heavily suppressed for small spacings, especially at small angles, which is a signature of level repulsion. Indeed, a very similar pattern is observed for non-Hermitian random matrices~\cite{sa2020}. We now show that the three universality classes of local bulk correlations---AI$^\dagger$,
A (GinUE), and AII$^\dagger$~\cite{hamazaki2020}---with increasing level repulsion, can be clearly distinguished by the complex spacing ratio distribution.

In order to gain a more quantitative understanding of the spectral correlations, we compute the marginal angular, $\rho(\theta)$, and radial, $\rho(r)$, complex spacing ratio distributions, where $\lambda_k=r_k e^{\i \theta_k}$. The results, presented in Figs.~\ref{fig:denadjtheta}~and~\ref{fig:denadjrho}, confirm the existence, depending on $N$ and $q$, of the three universality classes mentioned above. However, only for $q >2$, do we observe agreement with the random matrix prediction, which indicates that, as in the real case, this is a requirement for non-Hermitian many-body quantum chaos. We note that the full spectrum was employed in the evaluation of these marginal distributions. Therefore, $\rho(r)$ and $\rho(\theta)$ cannot distinguish between, for instance, class A and classes C and D, as the last two only differ from class A in the region $|E| \sim 0$ of small eigenvalues. In summary, the complex spacing ratios of the nHSYK Hamiltonian~(\ref{hami}) distinguish universality classes A, AI$^\dagger$, and AII$^\dagger$. 

To get a more visual confirmation of the symmetry classification, we can characterize the complex spacing ratio distribution by a single number, say its first radial moment, $\av{r}=\int \d r\, r \rho(r)$, as a function of $N$ and $q$. The values of $\av{r}$ for the three universal bulk statistics (A, AI$^\dagger$, and AII$^\dagger$) are given in Table~\ref{tab:numerics}. The results presented in Fig.~\ref{fig:CSRavr} show that $\av{r}$ computed numerically for the nHSYK model closely follows the predicted RMT pattern for $q=3$, $q=4$, and $6$, while it goes to the Poisson value for $q=2$.

These results confirm the predictions of Table~\ref{tab:nHSYK_class} for the local bulk correlations (level repulsion). In the next section, we study the distribution of the eigenvalue with the lowest absolute value. The shape of this observable is expected to have a universal form for each universality class that should agree with the random matrix prediction provided the spectrum has one of the inversion symmetries studied previously. This enables us to identify
 additional universality classes depending on the type of inversion symmetry of the nHSYK Hamiltonian by the study of level statistics.  

\section{Level statistics: hard-edge universality}

\begin{figure}[tbp]
	\centering
	\includegraphics[width=0.9\textwidth]{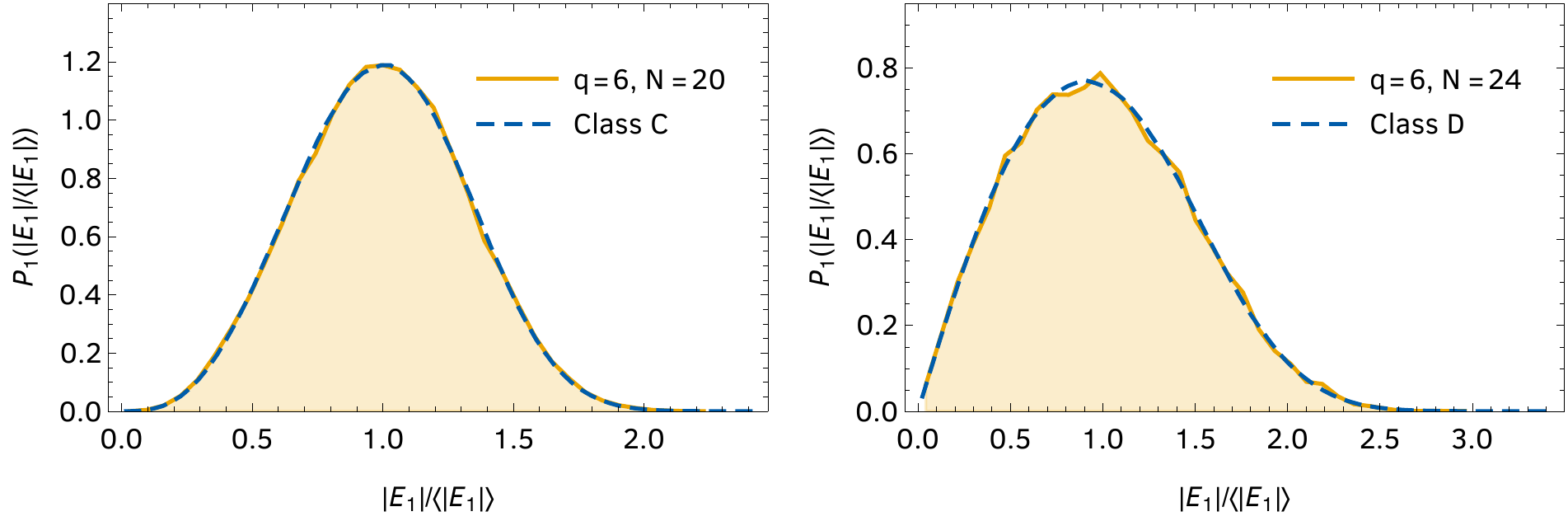}
	\caption{Distribution of the eigenvalues of $H$ with the smallest absolute value for the $q=6$ nHSYK model with $N=20$ and $N=24$ (filled histogram), compared with the random matrix theory predictions for the non-Hermitian classes C and D, respectively (dashed curves). The RMT predictions are obtained by exactly diagonalizing $10^7$ random matrices of dimension $100$ structured according to  the fourth column of Table~\ref{tab:sym_class}. The nHSYK results (solid curves) are obtained from an ensemble of approximately $4.5\times10^{4}$ and $3\times10^{4}$ realizations for $N=20$ and $N=24$, respectively, resulting in much larger finite-size effects. The comparison is parameter-free and does not involve any fitting.}
	\label{fig:EMINdistq6}
\end{figure}

Through the use of complex spacing ratios, we can only distinguish three universality classes
of bulk correlations. In addition, the classes with spectral inversion
symmetry (chiral or particle-hole) show universal repulsion from the spectral origin, the so-called hard edge for real spectra. This universal behavior can be captured by zooming in on the eigenvalues closest to the origin, on a scale of up to a few level spacings, the so-called microscopic limit. In particular, the distribution of the eigenvalue with the smallest modulus, $P_1(\abs{E_1})$, gives, when combined with the bulk complex spacing ratio distribution, a measure to uniquely distinguish the ten non-Hermitian symmetry classes without reality conditions. As an example, in Fig.~\ref{fig:EMINdistq6}, we show the distribution of $\abs{E_1}$ for the nHSYK model with $q=6$ and $N=20$ and $24$, and compare it with the prediction of non-Hermitian random matrix theory for classes C and D, respectively.
In order to carry out a parameter-free comparison with the nHSYK model, we normalize the distribution to unity and rescale $|E_1|$ by its average.
We thus see that, while the $q = 6$ nHSYK Hamiltonian has the same bulk statistics for all $N$, we can still resolve the Bott periodicity, which enables us to distinguish universality classes, through the statistics of $|E_1|$.

\begin{figure}[tbp]
	\centering
	\includegraphics[width=0.45\textwidth]{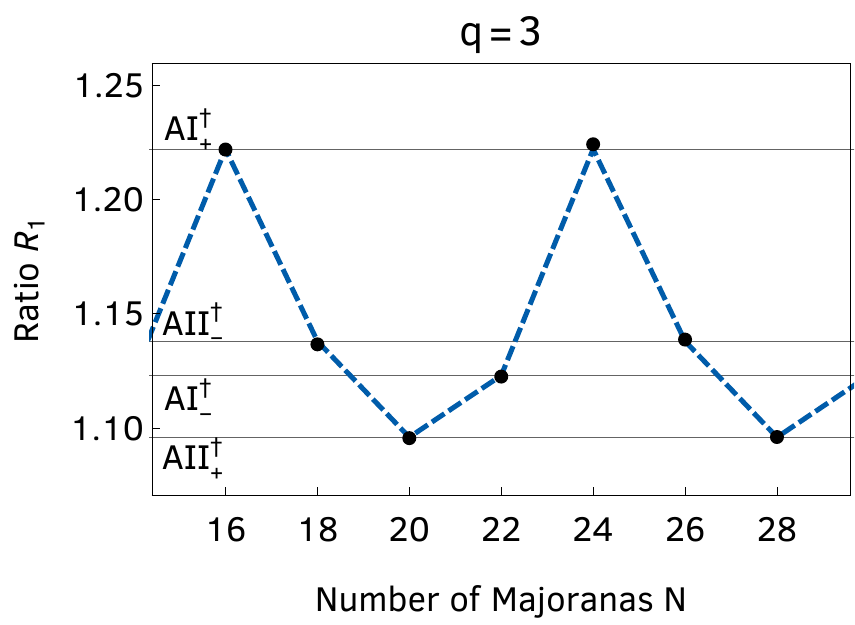}
	\includegraphics[width=0.45\textwidth]{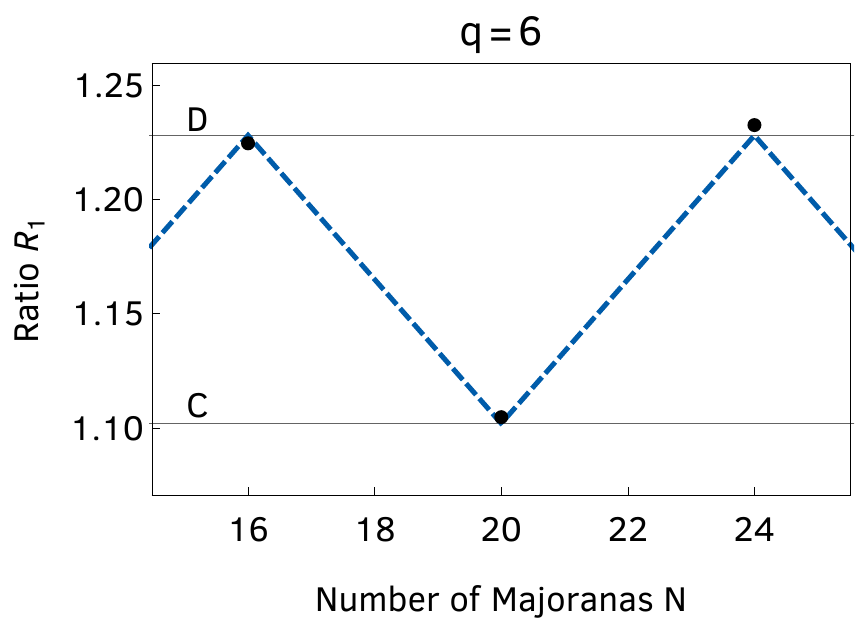}
	\caption{Ratio $R_1$ in Eq.~(\ref{eq:r1}) as a function of the number of Majoranas, $N$, for the $q = 3$ and the $q = 6$ nHSYK model. The dots are the results of exact numerical diagonalization of the corresponding nHSYK Hamiltonian, the horizontal solid lines are the values of the ratio for the six classes with spectral inversion symmetry realized in the  nHSYK model, and the dashed curves follow the classification scheme of Table~\ref{tab:nHSYK_class}. For $q = 3$, the Hamiltonian has a chiral symmetry for all even $N$, while for $q = 6$, there is a particle-hole symmetry only if $N$ is a multiple of $4$. We see excellent agreement between the nHSYK results and the RMT predictions for all available system sizes.}
	\label{fig:EMINrat}
\end{figure}

A convenient way to capture the hard-edge universality by a single number is to consider the ratio (normalized variance)
\begin{equation}\label{eq:r1}
	R_1=\frac{\av{\abs{E_1}^2}}{\av{\abs{E_1}}^2}=\frac{\int \d |E| \, |E|^2 P_1(|E|)}{\(\int \d |E| \, |E|\, P_1(|E|)\)^2},
\end{equation}
following the proposal by Sun and Ye for Hermitian random matrices~\cite{sun2020}. The values of $R_1$ for the seven non-Hermitian classes with inversion symmetry listed in Table~\ref{tab:sym_class} are tabulated in Table~\ref{tab:numerics}. To further confirm our symmetry classification, in Fig.~\ref{fig:EMINrat}, we show the value of $R_1$ as a function of $N$ for the $q=3$ and $q=6$ nHSYK models. We again see excellent agreement with the random matrix predictions, thus fully confirming the symmetry classification of Sec.~\ref{sec:symm_class}.

\begin{table}[tbp]
  \caption{Universal single-number signatures of the non-Hermitian universality classes without reality conditions. The first radial moment~$\av{r}$ of the complex spacing distribution measures the bulk level repulsion of the three universal bulk classes A, AI$^\dagger$, and AII$^\dagger$, while the ratio $R_1$ in Eq.~(\ref{eq:r1}) gives the repulsion between the hard edge and the eigenvalue with smallest absolute value for the seven classes with spectral inversion symmetry. The values of $\av{r}$ are obtained by numerical exact diagonalization of $2^{15}\times2^{15}$ random matrices of the corresponding class
averaging over an ensemble of $2^8$ realizations. Meanwhile, the value of $\av{r}$ has been obtained analytically for class A~\cite{dusa2022} in the thermodynamical limit. Its value is equal to $\av{r}=0.73866$, which is in agreement with our numerical value. In order to compute the ratio $R_1$ in Eq.~(\ref{eq:r1}), we numerically diagonalize $10^7$ $100\times100$ matrices of the corresponding universality class. Note that for the complex spacing ratio distribution (and its moments), it was shown in Ref.~\cite{sa2020} that they have large finite-size corrections for Gaussian-distributed random matrices; hence,
very large matrices have to be considered to converge to the 
universal result of the thermodynamical limit. In contrast, we have verified numerically that the smallest eigenvalue distribution (and thus $R_1$) converges to a universal distribution very rapidly with the system size; therefore, using relatively small matrices is justified.}
\label{tab:numerics}
	\begin{tabular}{@{}l ccc@{}}
		\toprule
		Class    & A      & AI$^\dagger$  & AII$^\dagger$ \\ \midrule
		$\av{r}$ & 0.7384  & 0.7222         & 0.7486         \\ \bottomrule
	\end{tabular}
	\quad
	\begin{tabular}{@{}l ccccccc@{}}
		\toprule
		Class & AIII$^\dagger$ & D      & C      & AI$^\dagger_+$ & AII$^\dagger_+$ & AI$^\dagger_-$ & AII$^\dagger_-$ \\ \midrule
		$R_1$ & 1.129          & 1.228  & 1.102  & 1.222          & 1.096          & 1.123          & 1.138            \\ \bottomrule
	\end{tabular}
\end{table}

\section{Additional universality classes in generalized nHSYK models}
\label{sec:nHWSYK}

We have already identified nine universality classes by different choices of $N$ and $q$ of
the nHSYK Hamiltonian (\ref{hami}). However, of the 38 universality classes,
23 are pseudo-Hermitian, which brings us in the ball park of
realizing the 15 universality classes of non-Hermitian or non-pseudo-Hermitian Hamiltonians in the nHSYK model.
Indeed, there is only one universality class, AIII$^{\dagger}$ (also known as chGinUE), of the original tenfold way still to be identified in nHSYK. In this section, we consider generalizations of the nHSYK that belong to AIII$^{\dagger}$ and to several other universality classes. 

\subsection{Symmetry classification of chiral nHSYK models}

\begin{table*}[t]
	\caption{Five non-Hermitian symmetry classes with reality conditions realized in the nHWSYK model. For each class, we list its anti-unitary and chiral symmetries and their commutation relations, an explicit matrix realization~\cite{magnea2008}, and its name under the Kawabata-Shiozaki-Ueda-Sato classification~\cite{kawabata2019}. The symbols $\epsilon_{XY}$ indicate whether the two operators $X$ and $Y$ commute, e.g., $\sT \cH \sT^{-1}=\epsilon_{\sT \cH} \cH$, while the symbols $\eta_X$ denote the square of operator $X$, e.g., $\sT^2=\eta_\sT$. In the matrix realizations, $A$, $B$, $C$, and $D$ are arbitrary non-Hermitian matrices unless specified otherwise, and empty entries correspond to zeros.}
	\label{tab:sym_class_2}
	\begin{tabular}{@{}Sc Sc Sc Sc Sc Sc Sc Sc Sl@{}}
		\toprule
		$\epsilon_{\sT \cH}$ & $\eta_{\sT}$ & $\epsilon_{\sC \cH^\dagger}$ & $\eta_\sC$ & $\epsilon_{\Pi\sT}$ & $\epsilon_{\Pi\sC}$ & $\epsilon_{\Pi\eta}$ & Matrix  realization & Class \\ \midrule
		---          & ---      & ---          & ---      & ---             & ---             & $-1$            & $\matAIIIm$         & AIII$_-$ \\
		$+1$         & $+1$     & $+1$         & $+1$     & $+1$            & $-1$            & $-1$            & $\matBDIdmp$        & BDI$^\dagger_{-+}$\\
		$+1$         & $-1$     & $+1$         & $-1$     & $+1$            & $-1$            & $-1$            & $\matCIIdmp$        & CII$^\dagger_{-+}$\\
		$-1$         & $+1$     & $-1$         & $+1$     & $+1$            & $-1$            & $-1$            & $\matBDIpm$        & 	BDI$_{+-}$ \\
		$-1$         & $-1$     & $-1$         & $-1$     & $+1$            & $-1$            & $-1$            & $\matCIIpm$         & CII$_{+-}$\\
		\bottomrule
	\end{tabular}
\end{table*}

We start by investigating models with a built-in chiral symmetry, where the Hamiltonian, termed non-Hermitian chiral or Wishart SYK (nHWSYK)~\cite{garcia2021d}, in an appropriate basis is represented by two off-diagonal blocks, where each block is either an Hermitian or non-Hermitian SYK. 
By tuning $N$ and $q$ in this model, we describe six new universality classes.  
More specifically, we consider the Hamiltonian,
\begin{equation}\label{haminh}
\cH=\adiag(H,H'):=\begin{pmatrix}
& H \\
H' & 
\end{pmatrix},
\end{equation}
equivalent to the two-matrix model $\cH'=HH'$, where $H$ and $H'$ are independent Hermitian or non-Hermitian SYK Hamiltonians (both with the same $N$ and $q$). Throughout, $\diag$ and $\adiag$ represent diagonal and anti-diagonal block matrices, respectively. The model $\cH$ gives the non-Hermitian generalization of the Wishart-SYK model~\cite{garcia2021d}, where, in the latter, $H'=H^\dagger$ to ensure Hermiticity. This model has built-in chiral symmetry represented by the operator $\Pi=\diag(\id,-\id)$ that commutes with the Hamiltonian. Since the (nH)SYK model with odd $q$ already has a chiral symmetry and two symmetries of the same type only generate an additional commuting unitary symmetry, we consider only even $q$.

For all even $q$, if $N\mod8=2,6$ and the couplings are complex, $H$ and $H'$ have no symmetries, and the chiral symmetry is the only symmetry of $\cH$, which thus belongs to class AIII$^\dagger$. This is the only non-Hermitian class without reality conditions (see Table~\ref{tab:sym_class}) not realized by the original nHSYK model~(\ref{hami}), thus completing the corresponding tenfold way. Contrary to the Hermitian case, there are still additional classes beyond those (i.e., classes with reality conditions), some of which are realized in nHWSYK for other values of $q$ and $N$. In Table~\ref{tab:sym_class_2}, we list these extra classes, together with their defining symmetries and a matrix representation.

For all even $q$, if $N\mod8=2,6$ and the couplings are real, $H$ and $H'$ are Hermitian and thus also have trivial pseudo-Hermiticity. However, when the trivial pseudo-Hermiticity of $H$, $H'$ passes down to $\cH$, it becomes nontrivial and is implemented by $\eta=\adiag(\id,\id)$. Since $\Pi$ and $\eta$ anti-commute, $\cH$ belongs to class AIII$_-$.

If $q\mod 4=0$, $N\mod8=0$, and the couplings are complex, then $H$ and $H'$ belong to class AI$^\dagger$ and have a transposition symmetry with anti-unitary $\sP$, $\sP H^\dagger \sP^{-1}=+H$, see Eq.\eref{eq:P_transform_H}, that squares to $\sP^2=+1$. It then follows that $\cH$ has a time-reversal symmetry $\sC_+ \cH^\dagger \sC_+^{-1}=+\cH$
with anti-unitary $\sC_+=\adiag(\sP,\sP)$ that squares to $\sC_+^2=+1$ and anti-commutes with $\Pi$. We thus see that $\cH$ belongs to class AI$^\dagger_-$ (see also Table~\ref{tab:sym_class_2}). Proceeding similarly, we find that if $q\mod 4=2$ and $N\mod8=4$, $\cH$ also belongs to class AI$^\dagger_-$, while for $q\mod 4=0$, $N\mod8=4$ and $q\mod 4=2$, $N\mod8=0$, it belongs to class AII$^\dagger_-$. These two classes were already implemented in the original nHSYK model.

If $q\mod 4=0$, $N\mod8=0$, and the couplings are real, then $H$ and $H'$ are Hermitian SYK models belonging to the AZ class AI. As in the preceding paragraph, $\cH$ has transposition anti-unitary symmetry $\sC_+ \cH^\dagger \sC_+^{-1}=+\cH$
with $\sC_+^2=+1$ that anti-commutes with the chiral symmetry operator $\Pi$, but it also has additional pseudo-Hermiticity implemented by $\eta=\adiag(\id,\id)$ that anti-commutes with $\Pi$ and commutes with $\sC_+$. The combined transposition symmetry and pseudo-Hermiticity induce a complex-conjugation symmetry with unitary $\sT_+=\eta \sC_+=\diag(\sP,\sP)$, $\sT_+\sH \sT_+^{-1}=+\sH$, that squares to $\sT_+^2=+1$ and commutes with $\Pi$. We thus conclude that $\cH$ belongs to class BDI$^\dagger_{-+}$. Proceeding similarly, we find that if $q\mod 4=0$, $N\mod8=4$, it belongs to class CII$^\dagger_{-+}$; if $q\mod4=2$, $N\mod8=0$, to class BDI$_{+-}$; and if $q\mod4=2$, $N\mod8=4$, to class CII$_{+-}$.

The symmetry classification of the chiral two-matrix model $\cH$ with Hermitian and non-Hermitian SYK blocks is summarized in Tables.~\ref{tab:two_matrix_class_nHSYK} and \ref{tab:two_matrix_class_SYK}.

\begin{table}[tb]
	\caption{Symmetry classes realized in the chiral two-matrix model $\cH$ with nHSYK blocks $H,$ $H'$.}
	\label{tab:two_matrix_class_nHSYK}
	\begin{tabular}{@{}lcccc@{}}
		\toprule
		$N\,\mathrm{mod}\,8$   & 0               & 2              & 4               & 6              \\ \midrule
		$q\,\mathrm{mod}\,4=0$ & AI$^\dagger_-$  & AIII$^\dagger$ & AII$^\dagger_-$ & AIII$^\dagger$ \\
		$q\,\mathrm{mod}\,4=2$ & AII$^\dagger_-$ & AIII$^\dagger$ & AI$^\dagger_-$  & AIII$^\dagger$ \\ \bottomrule
	\end{tabular}
\end{table}

\begin{table}[tb]
	\caption{Symmetry classes realized in the chiral two-matrix model $\cH$ with Hermitian SYK blocks $H,$ $H'$.}
	\label{tab:two_matrix_class_SYK}
	\begin{tabular}{@{}lcccc@{}}
		\toprule
		$N\,\mathrm{mod}\,8$   & 0                   & 2        & 4                  & 6        \\ \midrule
		$q\,\mathrm{mod}\,4=0$ & BDI$^\dagger_{-+}$  & AIII$_-$ & CII$^\dagger_{-+}$ & AIII$_-$ \\
		$q\,\mathrm{mod}\,4=2$ & BDI$_{+-}$ & AIII$_-$ & CII$_{+-}$  & AIII$_-$ \\ \bottomrule
	\end{tabular}
\end{table}

\subsection{Level statistics in the chiral class \texorpdfstring{AIII$^\dagger$}{AIII}}

\begin{figure}
	% 	 \subfigure[]{denzgt9k1q3hm20000f.pdf
	\includegraphics[width=8cm]{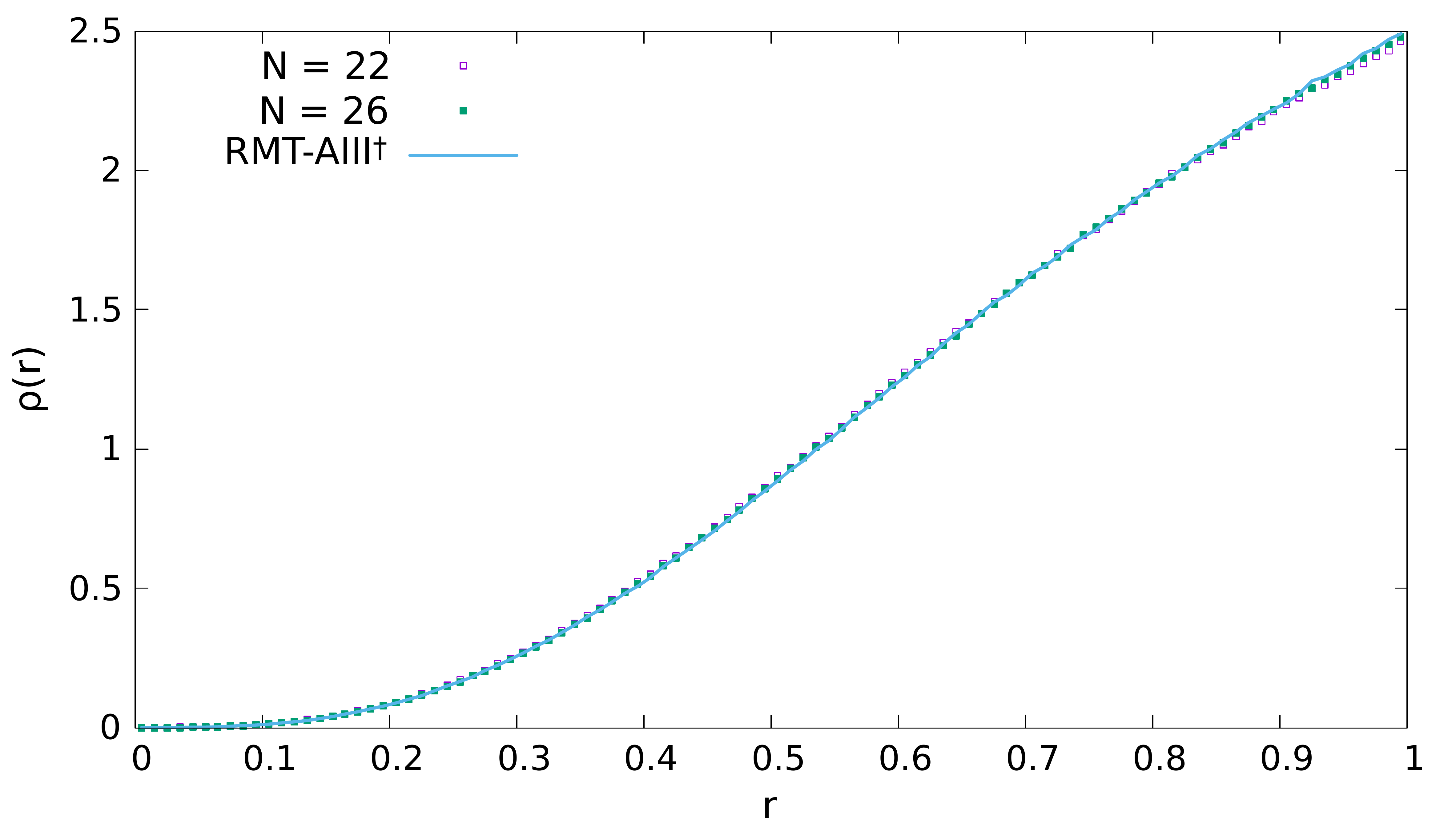}
	% 	\subfigure[]{ 
	\includegraphics[width=8cm]{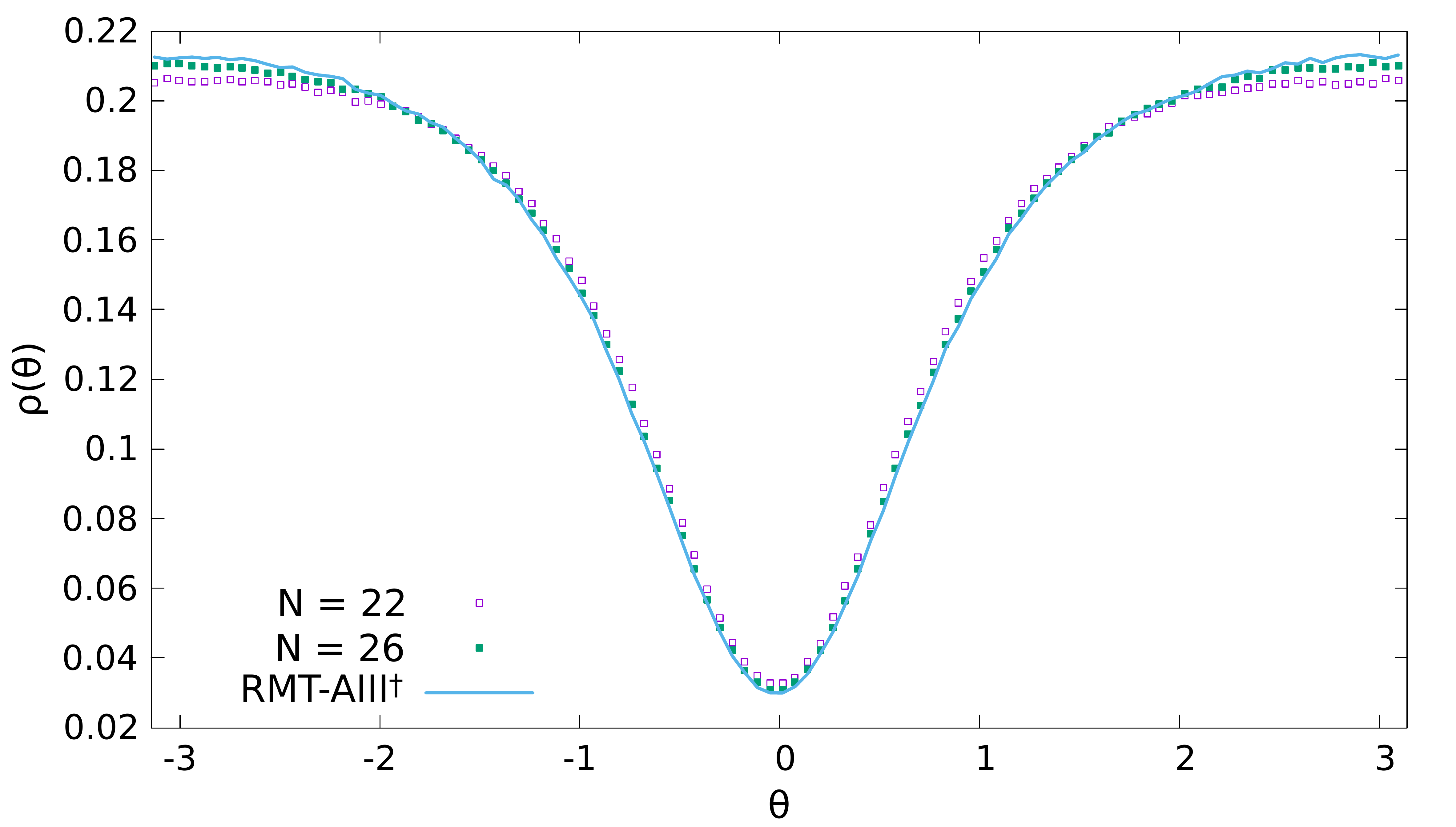}
	\caption{Radial (left panel) and angular (right panel)  density of the complex spacing ratio defined by Eq.~(\ref{eq:cspacing}) for the nHWSYK Hamiltonian Eq.~(\ref{haminh}) with $N=22$, $26$ and about $20000$ disorder realizations. We see agreement with the predictions of class AIII$^{\dagger}$ in all cases.
	}
	\label{fig:djtherhoAIII}
\end{figure}

Instead of an extensive numerical confirmation of these new universality classes in the SYK model,
we focus on the AIII$^\dagger$ universality class that completes the tenfold way. We note that only recently~\cite{garcia2021d} have the correlations of the Hermitian version of AIII$^\dagger$, the chGUE universality class, been identified in the SYK model. We study level statistics for the chiral nHSYK Hamiltonian~(\ref{haminh}) with $H$, $H'$ blocks given by independent $q=4$ nHSYK models. For $q  = 4$, AIII$^\dagger$ requires $N$mod$8 = 2,6$ so we stick to $N = 22, 26$. We obtained more than $10^7$ eigenvalues in each case by exact diagonalization techniques.
In analogy with the previous cases, we start with the analysis of short-range bulk spectral correlations represented by the radial $\rho(r)$ and angular $\rho(\theta)$ distributions of the complex spacing ratio density. Results depicted in Fig.~\ref{fig:djtherhoAIII} confirm an excellent agreement with the AIII$^\dagger$ random matrix prediction for both $N$.
However, this is a bulk observable, so the result for AIII$^\dagger$ is identical to that of the A universality class.

In order to differentiate these two universality classes, we study the distribution of the eigenvalue with the smallest modulus $|E_1|$. Because of the chiral symmetry, we expect the distribution to be universal and very well approximated by the random matrix prediction for AIII$^\dagger$. 

\begin{figure}[tbp]
	\includegraphics[width=10cm]{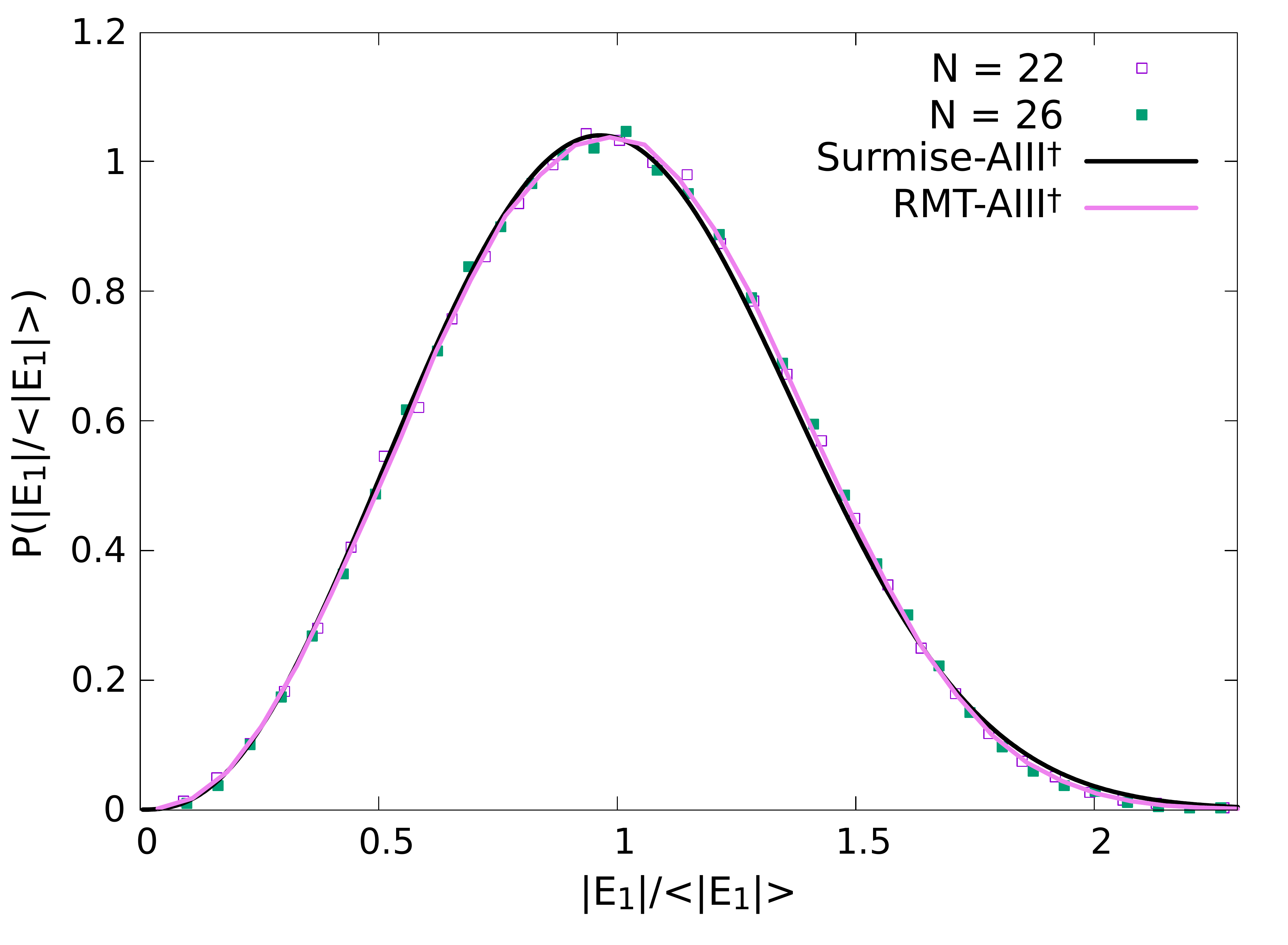}
	\caption{Distribution of the eigenvalue with the smallest absolute value $|E_1|$ for the chiral nHSYK model Eq.~(\ref{haminh}) with $q=4$ and $N=22,26$ and about $50000$ disorder realizations. We find excellent agreement with the random matrix prediction for the AIII$^\dagger$ universality class, obtained by diagonalization of $10^7$, $100\times 100$ matrices. The analytical surmise [Eq.~(\ref{eq:AIIId_P1_lambda_1}), black] is also very close to the numerical results. We note that the comparison is parameter-free.}
	\label{fig:dise1AIII}
\end{figure}

We first compute the normalized variance $R_1$, Eq.~(\ref{eq:r1}). For $N =22$, with around $20000$ disorder realizations, we obtain $R_1 = 1.131$, while for $N = 26$, with $12000$ disorder realizations, we get $R_1 = 1.128$. This is in excellent agreement with the random matrix result $R_1 = 1.129$.
The agreement is not restricted to this low-order moment. The full distribution of $|E_1|$, see Fig.~\ref{fig:dise1AIII}, is very close to the numerical random matrix result. The comparison does not involve any fitting procedure, only a rescaling by the average in each case. This is further confirmation that this chiral nHSYK model belongs to the AIII$^\dagger$ universality class.

\begin{figure}[tbp]
	\centering
	\includegraphics[width=\textwidth]{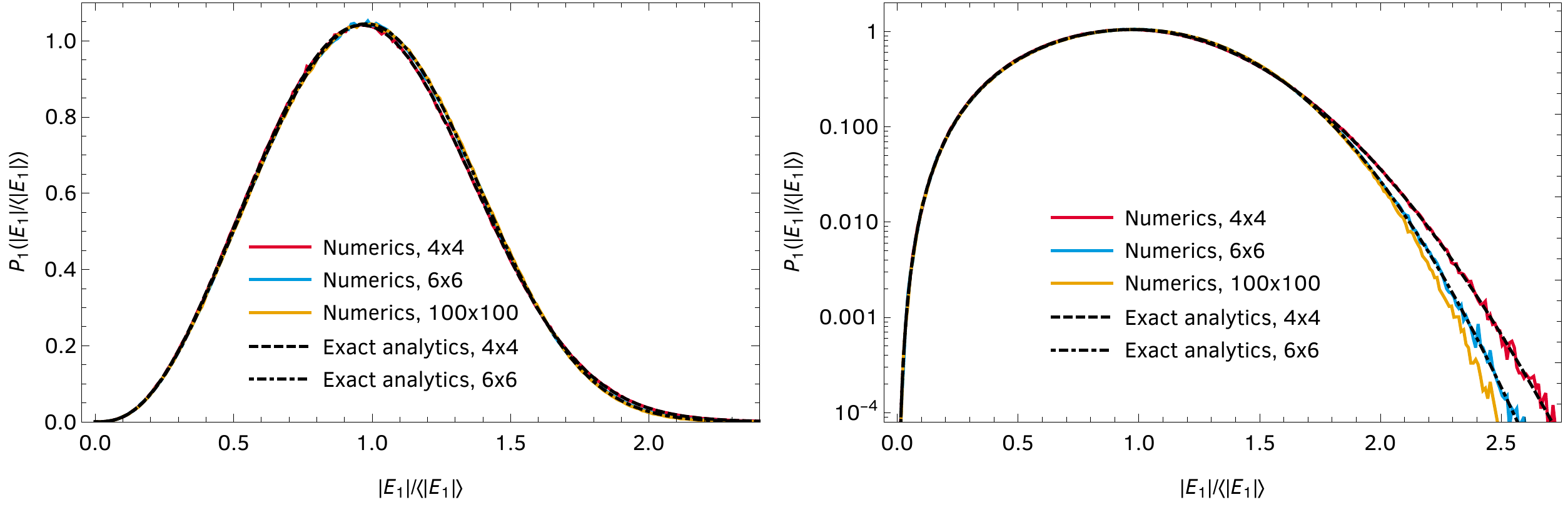}
	\caption{Distribution of the eigenvalues with smallest absolute values, $P_1(|E_1|/\av{|E_1|})$, for random matrices from class AIII$^\dagger$, on linear (left panel) and logarithmic (right panel) scales. The solid, colored curves are obtained from the exact diagonalization of $10^7$ random matrices of different sizes, while the dashed and dot-dashed black lines are the exact analytic results for $D=2$ [$4\times 4$ matrices, Eq.~(\ref{eq:AIIId_P1_lambda_1})] and $D=3$ ($6\times6$ matrices).}
	\label{fig:Emin_surmise_AIIId}
\end{figure}

We can also make an analytical prediction for the distribution of $\abs{E_1}$ in class AIII$^\dagger$. The radial spectral density is given
by~\cite{Splittorff:2003cu,Verbaarschot:2005rj}
\be
\rho(|E|) =\frac {2 u^2}{\pi} |E|^2 K_0(|E|^2 u) I_0(|E|^2u),
\label{micronh}
\ee
with
\be
u =\pi \lim_{|E|\to \infty} \rho(|E|).
\ee
For small $|E|$, the radial spectral density behaves as
\be
\rho(\abs{E}) \sim \abs{E}^3 \log \abs{E}.
\ee
We conclude that the radial distribution of the smallest eigenvalue must also behave as $\abs{E_1}^3\log \abs{E_1}$ for small~$\abs{E_1}$, which is in agreement with numerical results.
It is also possible to obtain an excellent analytical estimation of the random matrix result for the full distribution of $|E_1|$ (see also Refs.~\cite{akemann2009PRE,akemann2009JPhysA}). 
We consider a random matrix~$W$ from class AIII$^\dagger$ with square $D\times D$ off-diagonal blocks (recall Table~\ref{tab:sym_class}). The $2D$ eigenvalues of $W$ come in symmetric pairs and are denoted $E_1,-E_1,E_2,-E_2,\dots,E_D,-E_D$. We introduce $D$ new variables $z_j=E_j^2$ whose joint distribution is given by ~\cite{osborn2004}
\begin{equation}\label{eq:AIIId_jpdf}
P_\mathrm{joint}(z_1,\dots, z_D)\propto
\prod_{j=1}^D K_0(2D |z_j|) \prod_{1\leq j<k\leq D} \abs{z_j-z_k}^2.
\end{equation}
Because the distribution~(\ref{eq:AIIId_jpdf}) is invariant under permutations of eigenvalues $z_j$, we can choose $z_1$ as the eigenvalue with the smallest absolute value. Then, by construction, the distribution of $z_1$ is obtained by integrating out all the remaining eigenvalues outside the disk centered at the origin and with radius $\abs{z_1}$:
\begin{equation}\label{eq:AIIId_P1_z1}
P_1(|z_1|)\propto |z_1| K_0(2D|z_1|)
\int_0^{2\pi}\prod_{j=1}^D\d \phi_j \int_{\abs{z_1}}^{+\infty} 
\prod_{j=2}^D\d |z_j| |z_j|  K_0(2D|z_j|)
\prod_{1\leq j<k\leq D}
\abs{|z_j|e^{\i \phi_j}-|z_k|e^{\i \phi_k}}^2,
\end{equation}
where we denote $z_j=|z_j|e^{\i\phi_j}$ in polar coordinates. We proceed by evaluating the integrals in Eq.~(\ref{eq:AIIId_P1_z1}) for the smallest possible values of~$D$, i.e., $D=2,3,4,\dots$. In the spirit of the Wigner surmise for the spacing distribution, we expect the results to converge to the universal large-$D$ limit very quickly with $D$. Indeed, the $D=2$ result is already almost indistinguishable from large-$D$ numerical calculations on a linear scale, as we now show. Setting $D=2$ in Eq.~(\ref{eq:AIIId_P1_z1}), performing the angular integration of $\abs{|z_1|e^{\i\phi_1}-|z_2|e^{\i\phi_2}}^2$, and changing variables back to the chiral eigenvalue $|E_1|=\sqrt{|z_1|}$, we obtain
\begin{equation}\label{eq:AIIId_P1_lambda_1}
P_1(|E_1|)=
	\sN \(c|E_1|\)^3 K_0\(\(c|E_1|\)^2\)
\(
\(c|E_1|\)^4 
\int_{\(c|E_1|\)^2}^{+\infty} \d x\, x K_0(2 x) +
\int_{\(c|E_1|\)^2}^{+\infty} \d x\, x^3 K_0(2 x)
\),
\end{equation}
where the normalization constant is equal to $\sN=32 c$. The arbitrary energy scale can be chosen as $c=7/9$ such that $\av{|E_1|}=\int \d |E_1| |E_1| P_1(|E_1|)=1$. The remaining integrals over Bessel functions could be expressed in closed form in terms of Bessel and Lommel functions, but their precise form is very complicated. The two integrals can easily be evaluated numerically to high accuracy. Note that 
in agreement with Eq.~\eref{micronh}, the $D=2$ result for $P_1(|E_1|)$ for small $|E_1|$
also behaves as $\sim |E_1|^3\log |E_1|$.
In Fig.~\ref{fig:Emin_surmise_AIIId}, we show how the $D = 2$ surmise (corresponding to $4\times4$ random matrices) compares with exact diagonalization results ($10^7$ disorder realizations), both in a linear (left panel) and logarithmic scale (right panel). We see excellent agreement with the $4\times4$ numerical result, as expected since we are performing an exact calculation. On a linear scale, it is also hard to distinguish it from the numerical results for large $D=50$ ($100\times100$ matrices), while deviations in the right tail can be noted non a logarithmic scale. The agreement with the nHWSYK result, see Fig.~\ref{fig:dise1AIII}, is also excellent.

The procedure can be easily improved by considering $D = 3$ and $D = 4$. The resulting expressions are similar to Eq.~(\ref{eq:AIIId_P1_lambda_1}) but involve
additional  integrals over Bessel functions and quickly become very cumbersome. In Fig.~\ref{fig:Emin_surmise_AIIId}, we compare the distribution computed for $D = 3$ (not written out) with the corresponding numerics, again seeing perfect agreement. Moreover, we see a fast convergence towards the large-$D$ result (e.g., $D=50$). For most practical purposes (cf.\ Fig.~\ref{fig:dise1AIII} for the comparison with the nHWSYK model), the $D = 2$ result suffices.

In conclusion, the study of spectral correlations shows that the $q=4$ chiral nHSYK model, Eq.~(\ref{haminh}), belongs to class AIII$^\dagger$, and its dynamics is quantum chaotic for sufficiently long timescales. We expect that a similar agreement with the random matrix predictions will be obtained for the other chiral nHSYK
models introduced in the previous subsection.

\subsection{Classes with complex-conjugation symmetry in nHSYK models with complex fermions}

{
	\setlength\cellspacetoplimit{0.5ex}
	\setlength\cellspacebottomlimit{0.5ex}
	\begin{table*}[]
		\caption{Non-Hermitian symmetry classes with reality conditions but no pseudo-Hermiticity, four of which are realized in the complex-fermion nHSYK model (i.e., all except class AI$_-$). For each class, we list its anti-unitary and chiral symmetries, an explicit matrix realization~\cite{magnea2008}, its name under the Kawabata-Shiozaki-Ueda-Sato classification~\cite{kawabata2019}, and its name as a Ginibre ensemble (class AI$_-$ has no conventional Ginibre name). In the matrix realizations, $A$, $B$, $C$, and $D$ are arbitrary non-Hermitian matrices unless specified otherwise, and empty entries correspond to zeros.
		}
		\label{tab:sym_class_3}
		\begin{tabular}{@{}Sc Sc Sc Sc Sl Sl@{}}
			\toprule
			$\sT_+^2$     & $\sT_-^2$  & $\sS^2$ & Matrix realization & Class   & Ginibre ensemble   \\ \midrule
			$+1$          & $0$        & $0$     & $A=A^*$            & AI             & GinOE          \\
			$-1$          & $0$        & $0$     & $\matAII$          & AII            & GinSE          \\
			$+1$          & $+1$       & $1$     & $\matAIp$          & AI$_+$         & chGinOE        \\
			$-1$          & $-1$       & $1$     & $\matAIIp$         & AII$_+$        & chGinSE        \\
			$+1$          & $-1$       & $1$     & $\matAIm$          & AI$_-$         & ---            \\ \bottomrule
		\end{tabular}
	\end{table*}
}

In the previous subsections, we completed the tenfold BL classes without reality conditions. There exist only five remaining classes without pseudo-Hermiticity (two non-chiral and three chiral classes), which are listed in Table~\ref{tab:sym_class_3}.\footnote{Of the $23$ classes with pseudo-Hermiticity, we also presented five examples in the previous subsection.} These classes involve complex-conjugation symmetry but no transposition symmetry (because simultaneous complex-conjugation and transposition symmetries imply pseudo-Hermiticity). Four of these classes emerge naturally if we consider complex-fermion nHSYK models, i.e., a Hamiltonian
\begin{equation}
H_{\mathrm{c}}=\sum_{\substack{i<j\\k<\ell}}^{N_\mathrm{c}} w_{ij;k\ell} \,c^\dagger_i c^\dagger_j c_k c_\ell,
\end{equation}
where $w_{ij;k\ell}=-w_{ji;k\ell}=-w_{ij,\ell k}$ are independent random variables (real, complex, or quaternionic) and $c^\dagger_i$ and $c_i$ are the creation and annihilation operators of $N_\mathrm{c}=N/2$ complex fermions, respectively.
The complex fermions can be expressed in terms of Majorana fermions as $c_i=\psi_{2i-1}+\i \psi_{2i}$,
$c_i^\dagger =\psi_{2i-1}-\i \psi_{2i}$. In a basis where the odd $\gamma$-matrices are real and the even ones are purely imaginary (which is our convention
throughout this paper), the creation and annihilation operators are real,
\begin{equation}\label{eq:complex_fermions_c}
  c_i= c_i^* 
\qquad\text{and}\qquad
c^\dagger_i=c^\top_i.
\end{equation}
Since we have
\begin{equation}
H_{\mathrm{c}}^\dagger=\sum_{\substack{i<j\\k<\ell}}^{N_\mathrm{c}} w_{k\ell;ij} \,c^\dagger_i c^\dagger_j c_k c_\ell,
\end{equation}
if we set $w_{ij;k\ell}=w^*_{k\ell;ij}$, $H_{\mathrm{c}}$ is Hermitian and we recover the two-body random ensemble~\cite{mon1975}. However, if we relax this condition and let $w_{ij;k\ell}$ and $w_{k\ell;ij}$ become independent, $H_\mathrm{c}$ is non-Hermitian and has complex eigenvalues. The independence of $w_{ij;k\ell}$ and $w_{k\ell;ij}$, together with Eq.~(\ref{eq:complex_fermions_c}), further precludes the existence of both transposition symmetries, Eqs.~(\ref{eq:nHsym_TRSd}) and (\ref{eq:nHsym_PHS}), and complex-conjugation particle-hole symmetry, Eq.~(\ref{eq:nHsym_PHSd}).
The only remaining possible anti-unitary symmetry is complex-conjugation time-reversal symmetry, $\sT_+ H_\mathrm{c} \sT_+^{-1}= H_\mathrm{c}$, Eq.~(\ref{eq:nHsym_TRS}), depending on the reality of the couplings.

\begin{itemize}
	\item If $w_{ij;k\ell}$ is complex (i.e., arbitrary), there is no anti-unitary symmetry $\sT_+$, and the Hamiltonian belongs to the Ginibre Unitary Ensemble (GinUE) or non-Hermitian class A.
	\item If $w_{ij;k\ell}=w_{ij;k\ell}^*$ is real, then there exists a basis where $H_\mathrm{c}=H_\mathrm{c}^*$ is real and it belongs to the Ginibre Orthogonal Ensemble (GinOE) or non-Hermitian class AI.
	\item If $w_{ij;k\ell}$ is a quaternion, i.e., 
	\begin{equation}
	w_{ij;k\ell}=\begin{pmatrix}
	w^{(1)}_{ij;k\ell} & w^{(2)}_{ij;k\ell} \\
	-w^{(2)*}_{ij;k\ell} & w^{(1)*}_{ij;k\ell}
	\end{pmatrix}
	\end{equation} 
	with $w^{(1)}_{ij;k\ell}$ and $w^{(2)}_{ij;k\ell}$ complex, there exists a basis where $H_\mathrm{c}=I H_\mathrm{c}^* I^{-1}$, with $I$ the symplectic unit matrix, because the matrix elements of $c_i$ and $c_i^\dagger$ are real (and a real number times a self-dual quaternion remains a self-dual quaternion). We see that $H_\mathrm{c}$ belongs to the Ginibre Symplectic Ensemble (GinSE) or non-Hermitian class AII.
  \end{itemize}

Proceeding as before, we can now consider a chiral or Wishart complex-fermion nHSYK model, i.e., a Hamiltonian given by Eq.~(\ref{haminh}) with off-diagonal blocks given by $H_\mathrm{c}$. These models again have a chiral symmetry implemented by the unitary operator $\Pi=\diag(\id,-\id)$. As before, the symmetry classification depends solely on the reality conditions of the couplings.
\begin{itemize}
	\item If the couplings are complex, the chiral symmetry is the only symmetry of the chiral Hamiltonian and it immediately follows that it belongs to class AIII$^\dagger$.
	\item If the couplings are real, each block has an anti-unitary symmetry realized by $K$ (complex conjugation), which implies the chiral Hamiltonian satisfies $\sT_+\cH\sT_+^{-1} =\cH$, with $\sT_+=\diag(K,K)$ squaring to $\sT_+^2=+1$. Because $\Pi$ and $\sT_+$ commute, this defines the universality class AI$_+$.
	\item If the couplings are quaternionic, the anti-unitary
          symmetry of the blocks, realized by $I$ with $I^2=-1$, implies the existence of an anti-unitary
          symmetry $\sT_+$ of $\cH$, with $\sT_+=\diag(I,I)$, which squares to $\sT_+^2=-1$ and again commutes with the chiral unitary $\Pi$. This is the universality class AII$_+$.
\end{itemize}

The symmetry classification of the non-chiral and chiral nHSYK models with complex fermions is summarized in Table~\ref{tab:cfnHSYK_class}. The final non-Hermitian symmetry class, AI$_-$, consisting of chiral block matrices with zero diagonal blocks and off-diagonal blocks being the complex conjugates of each other (see last line in Table~\ref{tab:sym_class_3}), is realized neither in the nHSYK nor in its chiral or complex-fermion extensions. It would be interesting to find a variant of the model that implements this symmetry class, and, in that way, complete the full fifteenfold classification of non-pseudo-Hermitian matrices by the nHSYK model.

\begin{table}[tb]
	\caption{Symmetry classes realized in complex-fermion nHSYK and nHWSYK models.}
	\label{tab:cfnHSYK_class}
	\begin{tabular}{@{}lccc@{}}
		\toprule
		Couplings   & Real       & Complex        & Quaternionic  \\ \midrule
		Non-chiral  & AI         & A              & AII           \\
		Chiral      & AI$_+$     & AIII$^\dagger$ & AII$_+$        \\ \bottomrule
	\end{tabular}
\end{table}

\section{Outlook and conclusions} 
We have proposed a single-site SYK model with complex couplings as a toy model for many-body non-Hermitian quantum chaos. We have shown that 19 out of 38 universality classes in the symmetry classification of non-Hermitian quantum systems occur naturally in this SYK model and its chiral and complex-fermion extensions. In particular, we have reproduced all but one (AI$_-$) non-pseudo-Hermitian universality classes,
namely, A, AIII$^\dagger$, AI$^\dagger$, AII$^\dagger$, D, C, AI$^\dagger_+$, AII$^\dagger_+$, AI$^\dagger_-$, AII$^\dagger_-$, AI, AII, AI$_+$, and AII$_+$, see Tables.~\ref{tab:nHSYK_class}, \ref{tab:two_matrix_class_nHSYK}, \ref{tab:two_matrix_class_SYK}, and \ref{tab:cfnHSYK_class}. 
A detailed spectral analysis of the original nHSYK model, involving short-range correlators such as the complex spacing ratios in the bulk and the distribution of the smallest eigenvalue near the hard edge, has revealed an excellent agreement with the
RMT predictions for nine universality classes occurring for different choices of $N$ and $q$ in the Hamiltonian. Six additional universality classes were identified for the non-Hermitian version of the Wishart-SYK model. In particular, it realizes the class AIII$^\dagger$, the only non-Hermitian class without reality conditions not realized by the nHSYK model, thus completing our program of finding explicit realizations of
the tenfold classification within the non-Hermitian SYK model.
Interestingly, and contrary to the Hermitian case, the nHWSYK model also realizes classes beyond the tenfold way, illustrating the richer classification of
non-Hermitian random matrix models. More concretely, it also realizes the five pseudo-Hermitian classes AIII$_-$, BDI$^\dagger_{-+}$, CII$^\dagger_{-+}$, BDI$_{+-}$, and CII$_{+-}$. Finally, the complex-fermion nHSYK model realizes non-pseudo-Hermitian classes with reality conditions, namely the Ginibre Orthogonal, Ginibre Symplectic, chiral Ginibre Orthogonal, and chiral Ginibre Symplectic Ensembles.

Additionally, we mention the possibility of realizing further symmetry classes in two-site nHSYK models. These models, relevant for wormhole physics~\cite{maldacena2018,garcia2021,garcia2021b,garcia2019} and also in the exploration of dominant off-diagonal replica configurations~\cite{garcia2021b}, provide an interesting playground to implement not only the universality classes already realized in the nHSYK but also going beyond by tuning the additional parameters appropriately.
However, a potential issue in this case may be that the coupling
needed to lift spectral degeneracies actually determines the RMT symmetry class and it may occur that no new classes are obtained without fine-tuning.

The study of local level statistics presented in this paper will be complemented by a companion work~\cite{usnext} on long-range correlations that explore shorter timescales. Deviations from random matrix universality allow us to use the nHSYK model as a toy model of both generic features of the universal quantum ergodic state reached around the Heisenberg time and non-universal, but still rather generic, properties of quantum interacting systems in its approach to ergodicity. 

Other natural extensions of this work include the analytical calculation of the spectral density of the nHSYK model by combinatorial techniques and a short-time characterization of non-Hermitian many-body quantum chaos in this nHSYK model by the evaluation of the Lyapunov exponent resulting from an out-of-time-order correlation function~\cite{larkin1969,kitaev2015,maldacena2016} for times of the order of the Ehrenfest time. The latter could help dynamically characterize non-Hermitian quantum chaos. For real spectra, we have the BGS conjecture that relates dynamical and spectral correlations. However, despite the heavy use of terminology borrowed from the real-spectrum case, it is still unclear to what extent agreement with non-Hermitian random matrix predictions is related to quantum chaos in the original sense of quantum dynamics of classically chaotic systems. We plan to address some of these issues in the near future.

\appendix

\acknowledgments{
Martin Zirnbauer is thanked for a useful discussion.
AMGG was partially supported by NSFC Grant No.\ 11874259 (AMG), the National Key R$\&$D Program of China (Project ID: 2019YFA0308603), and a Shanghai talent program.
This work was also support by FCT through Grant No.\ SFRH/BD/147477/2019 (LS) as well as by U.S. DOE Grant No. DE-FAG-88FR40388 (JJMV).
}

\bibliography{librarynh.bib}

\end{document}